\documentclass[11pt,reqno,a4paper]{amsart}

\usepackage[
  left=31mm,
  right=31mm,
  top=29mm,
  bottom=31mm,
  headsep=8mm
]{geometry}
\usepackage{amsmath}
\usepackage[T1]{fontenc}
\usepackage{microtype}
\usepackage{mathrsfs}
\usepackage{enumitem}

\usepackage{amsmath,amsxtra,amsfonts,amssymb,amsthm,amscd}
\usepackage{mathtools}
\usepackage{newtxtext,newtxmath}
\usepackage{bm}
\usepackage{graphicx}
\usepackage{tikz}
\usepackage{booktabs}
\usepackage{enumitem}
\usepackage{cite}

\usepackage[dvipsnames]{xcolor}
\definecolor{paperblue}{RGB}{20,67,112}
\definecolor{papergray}{RGB}{75,75,75}
\usepackage[
  colorlinks=true,
  linkcolor=paperblue,
  citecolor=paperblue,
  urlcolor=paperblue,
  pdfborder={0 0 0}
]{hyperref}
\usepackage[nameinlink,noabbrev]{cleveref}

\numberwithin{equation}{section}
\allowdisplaybreaks[2]
\setlist[itemize]{topsep=3pt,itemsep=1pt,parsep=0pt}
\setlist[enumerate]{topsep=3pt,itemsep=1pt,parsep=0pt}

\theoremstyle{plain}
\newtheorem{thm}{Theorem}[section]

\newtheorem{prop}[thm]{Proposition}
\newtheorem{cor}[thm]{Corollary}

\newtheorem{lem}[thm]{Lemma}

\theoremstyle{definition}

\theoremstyle{remark}
\newtheorem{rem}[thm]{Remark}

\newcommand{\eqa}{\begin{eqnarray}}
\newcommand{\eeqa}{\end{eqnarray}}
\newcommand{\beq}{\begin{equation}}
\newcommand{\eeq}{\end{equation}}
\newcommand{\nn}{\nonumber}

\def\res{\mathop{\text{\rm Res}}}

\crefname{thm}{theorem}{theorems}
\crefname{lem}{lemma}{lemmas}
\crefname{prop}{proposition}{propositions}
\crefname{cor}{corollary}{corollaries}
\crefname{de}{definition}{definitions}
\crefname{rem}{remark}{remarks}
\crefname{ex}{example}{examples}

\begin{document}

\title[Generalized Frobenius Structures on Hurwitz Spaces]%
{Generalized Frobenius Structures on Hurwitz Spaces,
	Their Open Extensions and Associated Principal Hierarchies}

\author{Shilin Ma}
\address{School of Mathematics and Statistics, Xi'an Jiaotong University, Xi'an 710049, P. R. China}
\email{mashilin@xjtu.edu.cn}
\date{\today}



\begin{abstract}
We construct generalized Frobenius structures on Hurwitz spaces of arbitrary genus using extended primary differentials. Through contour integrals involving the Cauchy kernel, we derive explicit dispersionless Lax representations for the principal hierarchies of these structures and their almost-dual counterparts. We also construct solutions to the open WDVV equations for both the original and almost-dual structures, and derive explicit principal hierarchies for the resulting rank-one extensions. Finally, we show that these constructions admit Shramchenko deformations.

\end{abstract}

\maketitle
\tableofcontents

\section{Introduction}\label{sec:intro}

The notion of a Frobenius manifold was introduced by Dubrovin \cite{dub1998} as a geometric formulation of the Witten--Dijkgraaf--Verlinde--Verlinde (WDVV) associativity equations arising in two-dimensional topological field theory. A Frobenius manifold carries a commutative and associative multiplication on its tangent spaces, a compatible flat metric, a flat unity vector field, and a potential whose third derivatives determine the multiplication. In the conformal case, the structure is further equipped with an Euler vector field describing its quasi-homogeneity. Frobenius manifolds have found important applications in Gromov--Witten theory, singularity theory, mirror symmetry, and integrable systems; see, for example, \cite{manin1999frobenius,hertling2002frobenius,dubrovin2001normal,givental2005simple} and references therein.

The Dubrovin--Zhang theory establishes a deep connection between Frobenius manifolds and integrable hierarchies \cite{dubrovin2001normal,liu2022variational,liu2023variational}. For a semisimple Frobenius manifold, it constructs a dispersive deformation of the principal hierarchy for which the Virasoro symmetries act linearly on the tau-function. The tau-function of the  topological solution is closely related to generating functions of enumerative invariants. Recently, Liu, Qu and Zhang \cite{liu2022generalized} extended this framework to Frobenius manifolds with non-flat unity, referred to as generalized Frobenius manifolds. Their theory contains, in particular, an example closely related to the Ablowitz--Ladik hierarchy \cite{ablowitz1975nonlinear,ablowitz1976nonlinear}, whose topological deformation is connected with the equivariant Gromov--Witten theory of the resolved conifold \cite{brini2012local,brini2012integrable}.

The purpose of the present paper is to construct a class of generalized Frobenius manifold structures on Hurwitz spaces and to give explicit descriptions of their principal hierarchies. Recall that a Hurwitz space parametrizes compact Riemann surfaces equipped with a meromorphic function with prescribed ramification data over infinity, together with a choice of symplectic basis of cycles. For each choice of a primary differential, Dubrovin constructed a semisimple Frobenius manifold structure on the corresponding Hurwitz space; different choices of primary differentials are related by Legendre transformations (see \cite{dubrovin1999differential} and Lecture~5 of \cite{dub1998}). The corresponding principal hierarchies were also explicitly constructed in Lecture~6 of \cite{dub1998}.
In a related development, Romano studied a broader class of horizontal differentials on double Hurwitz spaces and constructed Frobenius structures of dual type, together with associated Whitham-type hierarchies \cite{Romano2015}.

In the present paper, we investigate the corresponding extended primary differentials on Hurwitz spaces from the perspective of generalized Frobenius manifolds. We prove that each such differential determines a generalized Frobenius manifold structure, whose unity need not be flat, and that the resulting structures are related by generalized Legendre transformations in the sense of \cite{liu2025legendre}.

The constructions of Dubrovin and Romano rely on horizontal differentials and their bilinear pairings, involving residues, period integrals, and principal-value integrals. Although these methods are effective in their respective settings, constructing principal hierarchies for a broader class of primary differentials requires a systematic verification of the corresponding recursion relations. To overcome this technical difficulty, we develop a unified approach based on the Cauchy kernel and its associated decompositions. This approach provides explicit contour-integral representations of the geometric structures and, more importantly, a direct construction of recursion vector fields satisfying the calibration equations. As a result, we obtain explicit dispersionless Lax representations for the principal hierarchies of generalized Frobenius structures on Hurwitz spaces of arbitrary genus, as well as their almost-dual counterparts \cite{dubrovin2004almost}.

Our approach also provides a unified framework for constructing open extensions. The open WDVV equations, introduced in the context of open Gromov--Witten theory by Horev and Solomon \cite{horev2012open}, are closely related to intersection theory on moduli spaces of bordered Riemann surfaces; see, e.g., \cite{buryak2016open,buryak2022open,buryak2024open,pandharipande2014intersection}. As observed by P.~Rossi, solutions of the open WDVV equations give rise to rank-one extensions of the underlying flat \(F\)-manifolds. Alcolado \cite{alcolado2017extended} established a useful criterion for constructing solutions of the open WDVV equations. Basalaev and Buryak \cite{basalaev2019open} constructed genus-zero open descendant potentials for homogeneous solutions of the open WDVV equations and proved the corresponding open Virasoro constraints. They subsequently developed open Saito theory for singularities of types \(A\) and \(D\) \cite{basalaev2021opensaito}, expressing the corresponding open potentials in terms of superpotentials. 
Basalaev \cite{basalaev2022integrable} further showed that these open WDVV solutions satisfy a stabilization condition and constructed the associated integrable hierarchies.

More recently, de Almeida \cite{almeida2026open} constructed solutions of the open WDVV equations for arbitrary Hurwitz Dubrovin--Frobenius manifolds using  the associated Landau--Ginzburg superpotentials. Proserpio and Strachan \cite{proserpio2025duality} investigated the corresponding open extensions of the almost-dual structures and their bi-flat \(F\)-manifold geometry. Basalaev and Rarovskii \cite{basalaev2026hurwitzopen} constructed explicit solutions of the open WDVV equations for genus-zero Hurwitz Frobenius manifolds, realizing the corresponding rank-one extensions by restricting the Frobenius structures on Hurwitz spaces with an additional pole to suitable degeneration loci.

In the present paper, using an Abelian integral of the chosen extended primary differential as a local coordinate, we construct solutions of the open WDVV equations for the generalized Frobenius structures introduced above on Hurwitz spaces of arbitrary genus. When the chosen differential is one of Dubrovin's primary differentials, our construction recovers the solutions obtained by de Almeida \cite{almeida2026open}. An analogous construction for the corresponding almost-dual structures extends the results of Proserpio and Strachan \cite{proserpio2025duality} beyond the setting of Dubrovin's primary differentials. In both cases, we derive explicit formulas for the principal hierarchies of the resulting rank-one extensions. Finally, we show that these constructions also admit Shramchenko deformations \cite{shramchenko2005deformations}, formulated in terms of the deformed canonical bidifferential and its associated Cauchy kernel.

We hope that the Cauchy-kernel approach developed in this paper will provide a useful framework for further investigations. One possible direction is to explore the connection between the Dubrovin--Zhang theory and topological recursion in the Hurwitz setting, motivated by the known relationship between Hurwitz Frobenius manifolds and topological recursion \cite{duninbarkowski2018primary,duninbarkowski2019superpotential}. Another direction is to investigate the construction of infinite-dimensional Frobenius structures associated with higher-genus Whitham hierarchies, building on our previous work in the genus-zero case \cite{ma2024infinite}. In higher genus, the Cauchy-kernel formulation may offer an alternative to approaches based on canonical coordinates. We hope to pursue these directions in future work.

\subsection{Main results}
We first recall the basic ingredients for constructing Frobenius manifold structures on Hurwitz spaces, following \cite{dub1998,Romano2015}.

Let
$
H=H_{g;n_0,\ldots,n_m}
$
be the Hurwitz space parametrizing equivalence classes of data
\[
\bigl(
C;\infty_0,\ldots,\infty_m;
\lambda;
\alpha_1,\ldots,\alpha_g,\beta_1,\ldots,\beta_g
\bigr),
\]
where \(C\) is a compact Riemann surface of genus \(g\),
the points \(\infty_0,\ldots,\infty_m\) are pairwise distinct,
and
$
\alpha_1,\ldots,\alpha_g,\beta_1,\ldots,\beta_g
$
form a symplectic basis of \(H_1(C,\mathbb Z)\).
The function \(\lambda\) is meromorphic on \(C\), with poles precisely at
\(\infty_0,\ldots,\infty_m\) of respective orders
\(n_0,\ldots,n_m\). We further require \(0\) to be a regular value
of \(\lambda\), or equivalently, all zeros of \(\lambda\) to be simple.

Consider the open subset of \(H\) on which all critical points of
\(\lambda\), away from the marked points, are simple and have pairwise
distinct critical values. Let \(P_1,\ldots,P_N\) be the critical points of
\(\lambda\), and set
$
u_i=\lambda(P_i)$ for $ i=1,\ldots,N.
$
By the Riemann--Hurwitz formula,
$
N=2g+m+\sum_{i=0}^m n_i-1.
$
The critical values \(u_1,\ldots,u_N\) form a system of local coordinates
on \(H\), called the canonical coordinates. In these coordinates, the multiplication on \(TH\) is defined by
\begin{equation}\label{multi}
	\frac{\partial}{\partial u_i}\circ
	\frac{\partial}{\partial u_j}
	=
	\delta_{ij}\frac{\partial}{\partial u_i},
	\qquad
	1\leq i,j\leq N.
\end{equation}
The unity and Euler vector fields are
\[
e=\sum_{i=1}^N\frac{\partial}{\partial u_i},\quad E=\sum_{i=1}^N u_{i}\frac{\partial}{\partial u_i}.
\]

Let \(\widetilde H\) be the universal curve over \(H\), with projection \(\pi\), whose fiber over \(u\in H\) is the Riemann surface \(C_u\). Away from the ramification divisor of \(\lambda\), the condition
that \(\lambda\) remain constant along horizontal directions defines a
connection on \(\widetilde H\). Thus, every vector field
\(\partial\in\operatorname{Vect}(H)\) admits a unique horizontal lift,
denoted by \(\partial|_\lambda\), characterized by
\[
\pi_*(\partial|_\lambda)=\partial,
\qquad
\partial|_\lambda(\lambda)=0.
\]
This lift extends meromorphically across the ramification divisor. For a
differential \(\omega\) depending on \(u\in H\), we denote its derivative
along this lift by
$
\partial|_\lambda\omega.
$

We call a differential \(\omega\) on \(\widetilde H\) horizontal if, for
every \(\partial\in\operatorname{Vect}(H)\), the differential
\(\partial|_\lambda\omega\) is meromorphic on the fibers, with possible
poles only at the ramification points of \(\lambda\). We consider a
family of horizontal differentials
$
\{\phi_I\}_{I\in\mathcal I},
$
referred to collectively as \emph{extended primary differentials}, where
\(\mathcal I\) denotes the corresponding index set. Their precise
definitions are given in Subsection~\ref{secextpri}.
Among the extended primary differentials, the following distinguished
subfamily coincides with the primary differentials appearing in
Dubrovin's construction:
\[
\phi_{\infty_0,j},
\qquad
j=1,\ldots,n_0-1,
\]
\[
\phi_{\infty_i,j},
\qquad
i=1,\ldots,m,\quad j=0,\ldots,n_i,
\]
and
\[
\phi_{\alpha_i,0},\ \phi_{\beta_i,0},
\qquad
i=1,\ldots,g.
\]
They form a basis of an \(N\)-dimensional vector space \(\mathcal D\). For each \(u\in H\), let \(\mathcal D_u\) denote the vector space spanned by their restrictions to \(C_u\). The spaces \(\mathcal D\) and \(\mathcal D_u\) are naturally isomorphic, and we shall identify them whenever no confusion arises.

Fix an extended primary differential $\phi_0$. We restrict
our attention to the open subset
\[
H_{\phi_0}^{\mathrm{nd}}
:=
\left\{
u\in H:
\phi_0(P_a)\neq 0,\quad a=1,\ldots,N
\right\},
\]
where $\phi_0(P_a)\neq 0$ means that $\phi_0$ does not
vanish at the ramification point $P_a$.
For simplicity, we continue to denote this open subset
by $H$ throughout the constructions associated with
$\phi_0$.

Let $\Phi_0$ be an Abelian integral of $\phi_0$,
normalized by requiring that its expansion near
$\infty_0$ in the local coordinate
$\zeta_0=\lambda^{-1/n_0}$ have vanishing constant term.
For a vector field \(\partial\) on \(H\), let \(\partial|_{\Phi_0}\) denote its horizontal lift with \(\Phi_0\) held fixed, and set
$
\partial(\lambda)=\partial|_{\Phi_0}\lambda.
$
Following Dubrovin's construction, we identify the tangent space \(T_uH\) with \(\mathcal D_u\) by associating to \(\partial\) the differential \(\partial(\lambda)\phi_0\). This identification is made precise in Lemma~\ref{lem:tangent-admissible} below. For \(\omega_1,\omega_2\in\mathcal D_u\), define the residue pairing
\begin{equation}\label{metric}
	\langle\omega_1,\omega_2\rangle
	:=
	\sum_{d\lambda=0}
	\operatorname{Res}
	\frac{\omega_1\omega_2}{d\lambda},
\end{equation}
where the sum is taken over all ramification points of \(\lambda\).
The nonzero pairings between the Dubrovin primary differentials are
\begin{align*}
	&\left\langle
	\phi_{\infty_0,j},
	\phi_{\infty_0,n_0-j}
	\right\rangle
	=
	\frac{1}{n_0},
	\qquad
	&&j=1,\ldots,n_0-1,
	\\
	&\left\langle
	\phi_{\infty_i,j},
	\phi_{\infty_i,n_i-j}
	\right\rangle
	=
	\frac{1}{n_i},
	\qquad
	&&j=1,\ldots,n_i-1,\quad i=1,\ldots,m,
	\\
	&\left\langle
	\phi_{\infty_i,0},
	\phi_{\infty_i,n_i}
	\right\rangle
	=-\frac{1}{n_i},
	\quad &&i=1,\ldots,m,\\
	&\left\langle
	\phi_{\alpha_i,0},\phi_{\beta_j,0}
	\right\rangle=\delta_{ij},
	\qquad &&i,j=1,\ldots,g.	
\end{align*}
with all remaining pairings equal to zero.
Then the above identification induces a flat metric on \(H\) defined by
\begin{equation}\label{metricphi}
	\eta_{\phi_0}(\partial_1,\partial_2)
	=
	\left\langle
	\partial_1(\lambda)\phi_0,
	\partial_2(\lambda)\phi_0
	\right\rangle .
\end{equation}

The multiplication on \(T_uH\) is compatible with the metric in the sense that
\[
\eta_{\phi_0}
(\partial_1\circ\partial_2,\partial_3)
=
c(\partial_1,\partial_2,\partial_3),
\]
where
\begin{equation}\label{03tensor}
	c(\partial_1,\partial_2,\partial_3)
	:=
	\sum_{d\lambda=0}
	\res
	\frac{\omega_1\omega_2\omega_3}{\phi_0\,d\lambda},
\end{equation}
with
\[
\omega_i=\partial_i(\lambda)\phi_0\in\mathcal D_u,
\qquad
\partial_i\in T_uH,
\qquad
i=1,2,3.
\]

\begin{thm}\label{mainthm-1}
	For every extended primary differential \(\phi_0\), the data
$
	(H,\eta_{\phi_0},\circ,e)
$
	define a generalized Frobenius manifold. The structures corresponding to
	different choices of \(\phi_0\) are related by generalized Legendre
	transformations in the sense of \cite{liu2025legendre}.
	
	In particular, if \(\phi_0\) is of pole type or cycle type, or is
	a logarithmic differential with \(p=0\), then the resulting
	generalized Frobenius manifold is conformal, with Euler vector
	field \(E\). If \(\phi_0\in\mathcal D\), then the unity vector
	field is flat, and hence the resulting structure is an ordinary
	Frobenius manifold.
	
	Moreover, the Legendre-extended principal hierarchy admits the
	dispersionless Lax representation
	\[
	\frac{\partial\lambda}{\partial T_I}\phi_0
	=
	\{\Phi_I,\lambda\},
	\qquad I\in\mathcal I.
	\]
	Here \(d\Phi_I=\phi_I\), and the spatial variable is
	\(x=T_{I_0}\), where \(I_0\in\mathcal I\) is the index
	corresponding to \(\phi_0\). The bracket is defined by
	\begin{equation}\label{poisson}
		\{f,g\}
		:=
		df\,\partial_x g
		-
		dg\,\partial_x f.
	\end{equation}
\end{thm}

We next consider solutions of the open WDVV equations associated with the generalized Frobenius manifolds constructed above. We first recall the following criterion of Alcolado \cite{alcolado2017extended}. Let \(N\) be a Frobenius manifold with flat coordinates
\(\mathbf{t}=(t^{1},\dots,t^{n})\), and let
\(f=f(\mathbf{t},s)\) be a smooth function on \(N\times U\) such that
\(\partial_s f\neq0\). Suppose that
\begin{equation}\label{omegaequ}
	\partial_{\alpha}\partial_{\beta}f
	=
	\partial_s\left(
	\frac{
		\partial_{\alpha}f\,\partial_{\beta}f
		-c_{\alpha\beta}^{\delta}\partial_{\delta}f
	}{
		\partial_s f
	}
	\right),
	\qquad
	\alpha,\beta=1,\dots,n,
\end{equation}
where \(\partial_{\alpha}=\frac{\partial}{\partial t^{\alpha}}\).
Then there exists a function \(F^o=F^o(\mathbf{t},s)\) satisfying
\begin{equation}\label{omegaequ2}
	\partial_{\alpha}\partial_{\beta}F^o
	=
	\frac{
		\partial_{\alpha}f\,\partial_{\beta}f
		-c_{\alpha\beta}^{\delta}\partial_{\delta}f
	}{
		\partial_s f
	},
	\qquad
	\partial_sF^o=f.
\end{equation}
Moreover, \(F^o(\mathbf{t},s)\) is a solution of the open WDVV equations associated with \(N\). The same criterion remains valid for generalized Frobenius manifolds, as can be readily verified.

\begin{thm}\label{openext}
	Fix an extended primary differential \(\phi_0\) defining a
	nondegenerate metric \(\eta_{\phi_0}\), and let \(\Phi_0\) be its
	normalized Abelian integral, so that \(d\Phi_0=\phi_0\).
	For each \(u\in H\), take a domain \(V_u\subset C_u\) on which
	\(\Phi_0\) is biholomorphic onto a domain \(U\subset\mathbb C\)
	and \(d\lambda\neq0\).
	
	Let \(\mathbf t\) be flat coordinates for \(\eta_{\phi_0}\),
	and define \(z=z(\mathbf t,s)\) implicitly by
$
	\Phi_0\bigl(\mathbf t,z(\mathbf t,s)\bigr)=s.
$
	Set
$
	f(\mathbf t,s)
	:=
	\lambda\bigl(\mathbf t,z(\mathbf t,s)\bigr).
$
	Then \(f\) satisfies \eqref{omegaequ}. Consequently,
	there exists a function \(F^o(\mathbf t,s)\) satisfying
	\eqref{omegaequ2}, and \(F^o\) is a solution of the open WDVV
	equations associated with the generalized Frobenius manifold
	\((H,\eta_{\phi_0},\circ,e)\).
	
	Moreover, the principal hierarchy of the corresponding
	rank-one extension is given by
	\begin{equation}\label{open-principal-I}
		\frac{\partial\lambda}{\partial T_I}\phi_0
		=
		\{\Phi_I,\lambda\},
		\qquad
		\frac{\partial s}{\partial T_I}
		=
		\partial_x\Phi_I(s),
		\qquad
		I\in\mathcal I,
	\end{equation}
	where the bracket \(\{\ ,\ \}\) is defined in
	\eqref{poisson}. The flows in the additional \(s\)-direction
	are
	\begin{equation}\label{open-principal-s}
		\frac{\partial\lambda}{\partial T_{s,p}}
		=
		0,
		\qquad
		\frac{\partial s}{\partial T_{s,p}}
		=
		\partial_x
		\left(
		\frac{f(\mathbf t,s)^{p+1}}{(p+1)!}
		\right),
		\qquad
		p\geq0.
	\end{equation}
	Here \(\Phi_I(s)\) denotes
	\(\Phi_I\bigl(\mathbf t,z(\mathbf t,s)\bigr)\).
\end{thm}
When \(\phi_0\in\mathcal D\), our construction recovers the open WDVV solutions obtained by Almeida \cite{almeida2026open}, while providing a more direct proof. Moreover, our approach yields an explicit description of the principal hierarchy of the associated rank-one extension.

We next consider the corresponding almost-dual structure. Let \(\widehat H\subset H\) be the domain on which the Euler vector field \(E\) is invertible with respect to the multiplication \(\circ\) defined in \eqref{multi}. In the canonical coordinates \(u_1,\ldots,u_N\), this is the domain where \(u_i\neq0\) for all \(i\). The almost-dual metric and multiplication are given by
\[
g_{\phi_0}
=
\sum_{i=1}^N
\frac{\eta_i}{u_i}\,du_i^2,
\qquad
\frac{\partial}{\partial u_i}
\ast
\frac{\partial}{\partial u_j}
=
\delta_{ij}\frac{1}{u_i}
\frac{\partial}{\partial u_i},
\]
where
$
\eta_{\phi_0}
=
\sum_{i=1}^N\eta_i\,du_i^2.
$
Under the identification \(T_uH\simeq\mathcal D_u\) determined by \(\phi_0\), the metric \(g_{\phi_0}\) is induced by the bilinear form
\begin{equation}\label{dualmet}
	(\omega_1,\omega_2)
	=
	\sum_{d\lambda=0}
	\res\frac{\omega_1\omega_2}{\lambda\,d\lambda},
	\qquad
	\omega_1,\omega_2\in\mathcal D_u.
\end{equation}
The unity vector field of the almost-dual multiplication \(\ast\) is
$
E=\sum_{i=1}^N u_i\frac{\partial}{\partial u_i}.
$

\begin{thm}\label{thm:dual-principal}
	The principal hierarchy of the almost-dual generalized Frobenius structure
$
	(\widehat H,g_{\phi_0},\ast,E)
$
	admits the dispersionless Lax representation
	\begin{equation}\label{dual-principal-Q}
		\frac{\partial\lambda}{\partial\widehat T_A}\phi_0
		=
		\{\widehat\Phi_A,\lambda\},
		\qquad
		A\in\widehat{\mathcal I}^{\mathrm{ext}},
	\end{equation}
	where the bracket is defined in \eqref{poisson}, and the index set \(\widehat{\mathcal I}^{\mathrm{ext}}\) and the Abelian integrals \(\widehat\Phi_A\) are specified in Subsections~\ref{almprisec} and~\ref{dualprinH}.
	
	In particular, the flow associated with \((\mathrm{un},0)\) is the spatial translation, so that
$
	\frac{\partial}{\partial x}
	=
	\frac{\partial}{\partial\widehat T_{\mathrm{un},0}}.
$
\end{thm}

An analogous result holds for the rank-one open extension.

\begin{thm}\label{dual-openext}
	Fix an extended primary differential \(\phi_0\), and let
	\(\Phi_0\) be its normalized Abelian integral.
	For each \(u\in\widehat H\), choose a domain
	\(V_u\subset C_u\) on which \(\Phi_0\) is biholomorphic
	onto a domain \(U\subset\mathbb C\), with
	\(\lambda\neq0\) and \(d\lambda\neq0\).
	Fix a single-valued branch of \(\log\lambda\) on \(V_u\).
		Let \(\widehat{\mathbf t}\) be flat coordinates for
	\(g_{\phi_0}\), and define \(z=z(\widehat{\mathbf t},s)\)
	implicitly by
$
	\Phi_0\bigl(\widehat{\mathbf t},
	z(\widehat{\mathbf t},s)\bigr)=s.
$
	Set
$
	L(\widehat{\mathbf t},s)
	:=
	\log\lambda\bigl(\widehat{\mathbf t},
	z(\widehat{\mathbf t},s)\bigr).
$
	Then \(L\) satisfies the analogue of \eqref{omegaequ}
	for the almost-dual Frobenius structure
	\((\widehat H,g_{\phi_0},\ast,E)\).
	Consequently, \(L\) determines a solution of the
	corresponding open WDVV equations.
	
	Moreover, the principal hierarchy of the corresponding
	rank-one extension admits the dispersionless Lax
	representation
	\begin{equation}\label{dual-open-principal-I}
		\frac{\partial\lambda}{\partial\widehat T_A}\phi_0
		=
		\{\widehat\Phi_A,\lambda\},
		\qquad
		\frac{\partial s}{\partial\widehat T_A}
		=
		\partial_x\widehat\Phi_A(s),
		\qquad
		A\in\widehat{\mathcal I}^{\mathrm{ext}},
	\end{equation}
	where
$
	\widehat\Phi_A(s)
	=
	\widehat\Phi_A\bigl(
	\widehat{\mathbf t},z(\widehat{\mathbf t},s)
	\bigr).
$
	The flows in the additional \(s\)-direction are
	\begin{equation}\label{dual-open-principal-s}
		\frac{\partial\lambda}{\partial\widehat T_{s,p}}
		=
		0,
		\qquad
		\frac{\partial s}{\partial\widehat T_{s,p}}
		=
		\partial_x
		\left(
		\frac{L(\widehat{\mathbf t},s)^{p+1}}{(p+1)!}
		\right),
		\qquad p\geq0.
	\end{equation}
\end{thm}

When \(\phi_0\in\mathcal D\), our construction recovers
the almost-dual open WDVV solutions obtained by
Proserpio and Strachan \cite{proserpio2025duality},
while providing a direct proof. Moreover, our approach
yields an explicit description of the principal
hierarchy of the corresponding rank-one extension.

In 2005, Shramchenko \cite{shramchenko2005deformations} introduced deformations of Hurwitz Frobenius manifolds by modifying the Bergman bidifferential. We show that our constructions of generalized Frobenius structures, their almost-dual counterparts, associated principal hierarchies, and rank-one open extensions naturally extend to the Shramchenko-deformed setting.

\subsection{Organization of the paper}

The paper is organized as follows.  In Section~\ref{sec:dual-formulation}, we develop the Cauchy-kernel formalism and derive the corresponding representations of the metrics, Levi--Civita connections, and multiplications. Section~\ref{sec-mainthm-1} constructs the generalized Frobenius structures associated with extended primary differentials and derives explicit dispersionless Lax representations for their principal hierarchies. In Section~\ref{sec-openext}, we construct rank-one extensions of these structures, obtain the corresponding open WDVV solutions, and describe their principal hierarchies. Section~\ref{sec-thm:dual-principal} treats the almost-dual structures, including their Cauchy-kernel formulation, principal hierarchies, and rank-one extensions. Finally, Section~\ref{sec:shramchenko} shows that the preceding constructions admit Shramchenko deformations.

\section{Cauchy Decompositions and Dual Formulations}
\label{sec:dual-formulation}
In this section, we describe the cotangent spaces of \(H\) and derive dual representations of the metric, Levi--Civita connection, and multiplication in terms of contour integrals involving the Cauchy kernel. These formulations will be used in the proofs of the main theorems.

\subsection{Cauchy kernel and decompositions}\label{subseccauthy}

Fix \(u\in H\), and let \(C_u\) be the corresponding Riemann surface. A \emph{contour}
\(\gamma\subset C_u\) will mean a finite disjoint union
$
\gamma=\bigsqcup_{a=1}^r\gamma_a
$
of oriented smooth closed curves, each having only finitely many self-intersection points.

Let \(B(z,w)\) be the normalized Bergman bidifferential on \(C_u\), and define the normalized Cauchy kernel by
\[
K(z,w)
=
\int_{\infty_0}^{z}B(\xi,w).
\]
The integral is understood with a choice of integration path from \(\infty_0\) to \(z\), or equivalently, with a choice of local determination of \(K\). For a fixed determination, \(K(z,w)\) is an Abelian differential of the third kind in \(w\), with simple poles at \(w=z\) and \(w=\infty_0\), of residues \(1\) and \(-1\), respectively. The distinguished determination, defined using integration paths in a canonical polygon, has vanishing \(\alpha\)-periods in the second variable and satisfies
$
K(\infty_0,w)=0.
$
Analytic continuation in the first variable gives
\[
K(z+\alpha_i,w)-K(z,w)=0,
\qquad
K(z+\beta_i,w)-K(z,w)=dZ_i(w),
\]
where \(dZ_1,\ldots,dZ_g\) are the normalized holomorphic differentials satisfying
\[
\frac{1}{2\pi\mathrm i}
\oint_{\alpha_j}dZ_i
=
\delta_{ij}.
\]

Let \(\mathcal H_\gamma\) denote the space of branchwise holomorphic germs along \(\gamma\). At a self-intersection point, the germs associated with distinct local branches are regarded as independent. For \(f\in\mathcal H_\gamma\), define its Cauchy transform by
\begin{equation}\label{Cauchy-transform-function}
	C_\gamma f(z)
	:=
	\frac{1}{2\pi\mathrm i}
	\oint_\gamma f(w)K(z,w)
	=
	\frac1{2\pi\mathrm i}\oint_\gamma
	f(w)\left(\int_{\infty_0}^{z}B(\xi,w)\right),
\end{equation}
where the integral over \(\gamma\) is understood as the sum of the integrals over its components, with the corresponding determinations of the kernel fixed.

In general, \(C_\gamma f\) is a multivalued, piecewise holomorphic function on \(C_u\setminus\gamma\). More precisely, on any simply connected open subset of \(C_u\setminus\gamma\), a  choice of determination of \(K\) defines a holomorphic branch of \(C_\gamma f\). The distinguished branch satisfies
$
C_\gamma f(\infty_0)=0.
$
The changes of determination induced by the canonical cycles are
\[
\Delta_{\alpha_i}C_\gamma f=0,
\qquad
\Delta_{\beta_i}C_\gamma f
=
\frac{1}{2\pi\mathrm i}
\oint_\gamma f\,dZ_i,
\qquad
i=1,\ldots,g.
\]
If these increments vanish for all \(i=1,\ldots,g\), then \(C_\gamma f\) is single-valued on \(C_u\setminus\gamma\), although it may still have a nonzero jump across \(\gamma\).

Moreover, along every smooth branch of \(\gamma\), the Cauchy transform satisfies the local jump relation
\begin{equation}\label{Cauchy-jump-function}
	(C_\gamma f)^{\mathrm L}
	-
	(C_\gamma f)^{\mathrm R}
	=
	f,
\end{equation}
where \(\mathrm L\) and \(\mathrm R\) denote boundary values from the left and right sides of the oriented branch, respectively. These boundary values are taken with compatible local determinations of $K(z,w)$ near the branch of \(\gamma\). The jump relation is local and remains valid after a simultaneous change of determination on both sides, since such a change adds the same constant to the two boundary values.

\begin{rem}\label{RHdec}
	If \(\gamma\) is a finite disjoint union of simple closed contours which bounds a domain \(D_\gamma^{\mathrm{in}}\subset C_u\), then the complementary region
$
	D_\gamma^{\mathrm{out}}
	=
	C_u\setminus\overline{D_\gamma^{\mathrm{in}}}
$
	is well defined, and the above Cauchy transform gives the standard additive Riemann--Hilbert decomposition of scalar functions on a compact Riemann surface; see, for example, Chapter~1, Sections~4--5 of \cite{rodin1988riemann}.
\end{rem}

Similarly, for a branchwise holomorphic differential \(\omega\) along \(\gamma\), define
\[
\mathcal C_\gamma\omega(w)
:=
-\frac{1}{2\pi\mathrm i}
\oint_\gamma K(z,w)\omega(z).
\]
The integral is understood with a fixed initial point on each component of \(\gamma\)  and a determination of \(K\) there. Different choices change \(\mathcal C_\gamma\omega\) by a linear combination of \(dZ_1,\ldots,dZ_g\).
The differential \(\mathcal C_\gamma\omega\) is holomorphic away from \(\gamma\) and \(\infty_0\), and has at most a simple pole at \(\infty_0\), with
\[
\operatorname{res}_{\infty_0}\mathcal C_\gamma\omega
=
\frac{1}{2\pi\mathrm i}
\oint_\gamma\omega.
\]
Along every smooth branch of \(\gamma\), its boundary values satisfy
\[
(\mathcal C_\gamma\omega)^{\mathrm L}
-
(\mathcal C_\gamma\omega)^{\mathrm R}
=
\omega.
\]

\begin{lem}\label{lem:Cauchy-d}
	Let \(\gamma\) be represented by a closed contour based at \(P_*\), and fix a determination of \(K(w,z)\) at \(w=P_*\), continued analytically along \(\gamma\). Then, for every \(f\in\mathcal H_\gamma\),
	\begin{equation}\label{Cauchy-d}
		\mathcal C_\gamma(df)(z)
		=
		d_z(C_\gamma f)(z)
		-
		\frac{f(P_*)}{2\pi\mathrm i}\,
		\Delta_\gamma K(P_*,z),
	\end{equation}
	where
$
	\Delta_\gamma K(P_*,z)
$
	denotes the increment of \(K(w,z)\) after analytic continuation of the first variable \(w\) once along \(\gamma\).
\end{lem}

\begin{proof}
	Integration by parts along the based contour \(\gamma\) gives
	\[
	\begin{aligned}
		\mathcal C_\gamma(df)(z)
		&=
		-\frac{1}{2\pi\mathrm i}
		\oint_\gamma K(w,z)\,df(w)\\
		&=
		\frac{1}{2\pi\mathrm i}
		\oint_\gamma f(w)\,d_wK(w,z)
		-
		\frac{f(P_*)}{2\pi\mathrm i}\,
		\Delta_\gamma K(P_*,z).
	\end{aligned}
	\]
	Since
	\[
	d_wK(w,z)=B(w,z)=B(z,w),
	\]
	we obtain
	\[
	\frac{1}{2\pi\mathrm i}
	\oint_\gamma f(w)\,d_wK(w,z)
	=
	\frac{1}{2\pi\mathrm i}
	\oint_\gamma f(w)B(z,w)
	=
	d_z(C_\gamma f)(z),
	\]
	which proves \eqref{Cauchy-d}.
\end{proof}
In particular, if the winding numbers of \(\gamma\) along all the \(\beta\)-cycles vanish, then \(\Delta_\gamma K=0\), and hence
$
\mathcal C_\gamma(df)=d(C_\gamma f).
$

\subsection{Extended primary differentials and cotangent spaces}\label{secextpri}

Let \(H\) be the Hurwitz space introduced above and \(\widetilde H\) its universal curve. We consider the following families of extended primary differentials on \(\widetilde H\).

For each differential \(\phi\) in these families, we require its derivative \(\partial_{u_k}|_\lambda\phi\) to be a single-valued meromorphic differential whose poles are confined to the ramification points of \(\lambda\), with
\[
\oint_{\alpha_j}\partial_{u_k}|_\lambda\phi=0,
\qquad j=1,\ldots,g.
\]
The local singularities and monodromies of the differentials themselves are specified below. 

\begin{enumerate}[label=(\roman*)]
	
	\item \emph{Pole-type differentials.}
	
	For \(j\geq1\) with \(j\not\equiv0\pmod{n_0}\), write \(j=kn_0+r\), where \(k\geq0\) and \(r=1,\ldots,n_0-1\). Let \(\phi_{\infty_0,j}\) be the Abelian differential of the second kind whose only pole is at \(\infty_0\), with principal part
$
	d\left[
	\frac{1}{n_0}
	\frac{\Gamma(r/n_0)}{\Gamma(j/n_0+1)}
	\lambda^{j/n_0}
	\right].
$
	
	For \(i=1,\ldots,m\) and \(j\geq1\), write \(j=kn_i+r\), where \(k\geq0\), \(r=1,\ldots,n_i\), and \(r=n_i\) when \(n_i\mid j\). Let \(\phi_{\infty_i,j}\) be the Abelian differential of the second kind whose only pole is at \(\infty_i\), with principal part
$
	d\left[
	\frac{1}{n_i}
	\frac{\Gamma(r/n_i)}{\Gamma(j/n_i+1)}
	\lambda^{j/n_i}
	\right].
$
	
	For \(q\in\lambda^{-1}(0)\) and \(j\geq1\), let \(\phi_{q,j}\) be the Abelian differential of the second kind whose only pole is at \(q\), with principal part
$
	(-1)^{j+1}(j-1)!\,d\lambda^{-j}.
$
	
	\item \emph{Logarithmic differentials.}
	
	Set \(c_0=0\) and \(c_p=\sum_{r=1}^p\frac1r\) for \(p\geq1\).	
	For \(i=1,\ldots,m\) and \(p\geq0\), let \(\phi^{\log}_{\infty_i,p}\) be the multivalued differential which is holomorphic away from \(\infty_0\) and \(\infty_i\), with logarithmic singular parts
$
	d\left[
	\frac{\lambda^p}{p!}
	\left(\log\lambda^{1/n_0}-\frac{c_p}{n_0}\right)
	\right]
$
	at \(\infty_0\), and
$
	-d\left[
	\frac{\lambda^p}{p!}
	\left(\log\lambda^{1/n_i}-\frac{c_p}{n_i}\right)
	\right]
$
	at \(\infty_i\).
	
	In particular, \(\phi^{\log}_{\infty_i,0}\) is an Abelian differential of the third kind with residues \(-1\) and \(1\) at \(\infty_0\) and \(\infty_i\), respectively. We write
$
	\phi_{\infty_i,0}:=\phi^{\log}_{\infty_i,0}.
$
	
	For \(q\in\lambda^{-1}(0)\) and \(p\geq0\), let \(\phi^{\log}_{q,p}\) be the multivalued differential which is holomorphic away from \(\infty_0\) and \(q\), with logarithmic singular parts
$
	d\left[
	\frac{\lambda^p}{p!}
	\left(\log\lambda^{1/n_0}-\frac{c_p}{n_0}\right)
	\right]
$
	at \(\infty_0\), and
$
	d\left[
	\frac{\lambda^p}{p!}
	\left(\log\lambda-c_p\right)
	\right]
$
	at \(q\).
	
	For \(p=0\), this is an Abelian differential of the third kind with residues \(-1\) and \(1\) at \(\infty_0\) and \(q\), respectively. We write
$
	\phi_{q,0}:=\phi^{\log}_{q,0}.
$

	\item \emph{Cycle-type differentials.}
	
	For \(i=1,\ldots,g\), let \(\phi_{\beta_i,0}=dZ_i\) be the holomorphic differential determined by
$
	\frac{1}{2\pi\mathrm i}
	\oint_{\alpha_j}\phi_{\beta_i,0}
	=\delta_{ij}.
$
		For \(p\geq1\), let \(\phi_{\beta_i,p}\) be the multivalued holomorphic differential with
$
	\Delta_{\alpha_j}\phi_{\beta_i,p}
	=
	2\pi\mathrm i\,\delta_{ij}\,
	d\frac{\lambda^p}{p!},
$ and $
	\Delta_{\beta_j}\phi_{\beta_i,p}=0.
$
	
	For \(p\geq0\), let \(\phi_{\alpha_i,p}\) be the multivalued holomorphic differential with
$
	\Delta_{\alpha_j}\phi_{\alpha_i,p}=0,
$ and $
	\Delta_{\beta_j}\phi_{\alpha_i,p}
	=
	-\delta_{ij}\,
	d\frac{\lambda^{p+1}}{(p+1)!}.
$
	Here \(\Delta_{\alpha_j}\) and \(\Delta_{\beta_j}\) denote the increments obtained by analytic continuation along the corresponding cycles.
	
\end{enumerate}

To fix the remaining holomorphic ambiguities, let
$
D_E=\partial_E|_\lambda+\lambda\partial_\lambda.
$
For the pole-type and cycle-type differentials, we impose
\[
\begin{aligned}
	D_E\phi_{\infty_0,j}
	&=\frac{j}{n_0}\phi_{\infty_0,j},
	&
	D_E\phi_{\infty_i,j}
	&=\frac{j}{n_i}\phi_{\infty_i,j},\\
	D_E\phi_{q,j}
	&=-j\phi_{q,j},
	&
	D_E\phi_{\beta_i,p}
	&=p\phi_{\beta_i,p},\\
	D_E\phi_{\alpha_i,p}
	&=(p+1)\phi_{\alpha_i,p}.
\end{aligned}
\]
For the logarithmic differentials with \(p\geq1\), set \(h_p(\lambda)=\lambda^p/p!\) and impose
\[
D_E\phi^{\log}_{\infty_i,p}
=
p\phi^{\log}_{\infty_i,p}
-
\sum_{s=1}^m
\left(\frac{n_s}{n_0}+\delta_{is}\right)
\phi_{\infty_s,pn_s}
+
\frac1{n_0}dh_p,
\]
and
\[
D_E\phi^{\log}_{q,p}
=
p\phi^{\log}_{q,p}
-
\sum_{s=1}^m
\frac{n_s}{n_0}\phi_{\infty_s,pn_s}
+
\frac1{n_0}dh_p.
\]
The degree-zero logarithmic differentials satisfy
\[
D_E\phi^{\log}_{\infty_i,0}
=
D_E\phi^{\log}_{q,0}=0.
\]
Since these Euler conditions do not remove their holomorphic ambiguities, we additionally require
\[
\oint_{\alpha_j}\phi^{\log}_{\infty_i,0}
=
\oint_{\alpha_j}\phi^{\log}_{q,0}
=0,
\qquad j=1,\ldots,g.
\]

These conditions determine the extended primary differentials uniquely. We denote their index set by \(\mathcal I\) and write
$
\{\phi_I\}_{I\in\mathcal I}.
$
They are the extended primary differentials associated with the standard normalization of the Bergman bidifferential, as in \cite{dub1998,Romano2015,shramchenko2005deformations}.

\begin{lem}\label{lem:tangent-admissible}
	Fix an extended primary differential \(\phi_0\). Then, for every \(u\in H\), the correspondence assigning \(\partial(\lambda)\phi_0\) to each \(\partial\in T_uH\) defines an isomorphism between \(T_uH\) and \(\mathcal D_u\).
\end{lem}

\begin{proof}
	Let \(\Phi_0\) be an Abelian integral of \(\phi_0\), normalized by requiring that its expansion at \(\infty_0\) in the local coordinate \(\zeta_0=\lambda^{-1/n_0}\) has vanishing constant term. In the presence of logarithmic singularities, the constant term is understood as the finite part after subtracting the prescribed singular and logarithmic terms.
	
	Let \(\partial(\lambda)\) denote the derivative of \(\lambda\) along the \(\Phi_0\)-horizontal lift of \(\partial\). Since
$	\partial|_\lambda\Phi_0
	=
	-\frac{\partial(\lambda)}{d\lambda}\,\phi_0,
$
	we obtain
$
	\partial(\lambda)\phi_0
	=
	-\bigl(\partial|_\lambda\Phi_0\bigr)d\lambda
	\in\mathcal D_u.
$
	If \(\partial(\lambda)\phi_0=0\), then \(\partial(\lambda)=0\), and hence
$
	\partial u_a=\partial(\lambda)(P_a)=0
$ for $ a=1,\ldots,N.
$
	Since \(u_1,\ldots,u_N\) are local coordinates on \(H\), it follows that \(\partial=0\). Thus the correspondence is injective. Finally,
$
	\dim T_uH=\dim\mathcal D_u=N,
$
	so it is an isomorphism.
\end{proof}
For \(u\in H\), let \(\gamma\subset C_u\) be a contour. Since a differential \(\phi\in\mathcal D_u\) may be multivalued, we fix an initial point \(P_{\ast}\in\gamma\) and a determination of \(\phi\) at \(P_{\ast}\), which is then analytically continued along \(\gamma\). The pairing
\begin{equation}\label{pair}
	\langle f,\phi\rangle_\gamma
	=
	\frac{1}{2\pi\mathrm{i}}
	\oint_\gamma f\phi,
	\qquad
	f\in\mathcal H_\gamma,\quad
	\phi\in\mathcal D_u,
\end{equation}
depends in general on this choice. More precisely, if \(\widetilde P_{\ast}\in\gamma\) is another initial point, with the determination of \(\phi\) obtained by analytic continuation from \(P_{\ast}\) to \(\widetilde P_{\ast}\) along the oriented arc \(\gamma_{P_{\ast}\widetilde P_{\ast}}\), then the corresponding pairings differ by
$
\frac{1}{2\pi\mathrm{i}}
\int_{\gamma_{P_{\ast}\widetilde P_{\ast}}}
f\Delta_\gamma\phi.
$
For a fixed choice, \eqref{pair} defines a linear map from \(\mathcal H_\gamma\) to \(\mathcal D_u^\ast\). After fixing an extended primary differential \(\phi_0\) and using the identification \(T_uH\simeq\mathcal D_u\), each \(f\in\mathcal H_\gamma\) therefore determines a cotangent vector in \(T_u^\ast H\).

\subsection{A bilinear identity for multivalued differentials}

The following bilinear identity will be used throughout the subsequent analysis.

\begin{lem}\label{bilin}
	Let \(\widehat C\) be the fundamental polygon obtained by cutting
	\(C\) along the canonical homology basis
	$
	\alpha_1,\beta_1,\ldots,\alpha_g,\beta_g,
	$
	and assume that \(d\lambda\) does not vanish on
	\(\partial\widehat C\). Let \(\omega^{(1)}\) and
	\(\omega^{(2)}\) be differentials defined in a neighborhood of
	\(\partial\widehat C\), with jumps
	\[
	\omega^{(a)}(P+\alpha_j)-\omega^{(a)}(P)
	=
	p_j^{(a)}(\lambda)\,d\lambda,
\quad 
	\omega^{(a)}(P+\beta_j)-\omega^{(a)}(P)
	=
	q_j^{(a)}(\lambda)\,d\lambda,
	\]
	for \(a=1,2\) and \(j=1,\ldots,g\), where
	\(p_j^{(a)}\) and \(q_j^{(a)}\) are polynomials in \(\lambda\).
	Set
$
	F^{(a)}=\frac{\omega^{(a)}}{d\lambda}.
$
	Then
	\[
	\begin{aligned}
		\oint_{\partial\widehat C}
		F^{(1)}\omega^{(2)}
		=
		\sum_{j=1}^{g}
		\Bigg[
		&-\int_{\alpha_j}
		\left(
		q_j^{(2)}\omega^{(1)}
		+
		q_j^{(1)}\omega^{(2)}
		\right)
		\\
		&+\int_{\beta_j}
		\left(
		p_j^{(2)}\omega^{(1)}
		+
		p_j^{(1)}\omega^{(2)}
		\right)
		\Bigg].
	\end{aligned}
	\]
\end{lem}

\begin{proof}
	The jump relations imply
$
	\Delta_{\alpha_j}F^{(a)}=p_j^{(a)}(\lambda)
$ and $
	\Delta_{\beta_j}F^{(a)}=q_j^{(a)}(\lambda).
$
	Pairing the oppositely oriented sides of the canonical cuts gives
	\[
	\oint_{\partial\widehat C}F^{(1)}\omega^{(2)}
	=
	\sum_{j=1}^g
	\left[
	-\int_{\alpha_j}
	\Delta_{\beta_j}\bigl(F^{(1)}\omega^{(2)}\bigr)
	+
	\int_{\beta_j}
	\Delta_{\alpha_j}\bigl(F^{(1)}\omega^{(2)}\bigr)
	\right].
	\]
	Using \(F^{(1)}\omega^{(2)}=F^{(2)}\omega^{(1)}\), we obtain
	\[
	\Delta_{\beta_j}\bigl(F^{(1)}\omega^{(2)}\bigr)
	=
	q_j^{(2)}\omega^{(1)}
	+
	q_j^{(1)}\omega^{(2)}
	+
	q_j^{(1)}q_j^{(2)}\,d\lambda,
	\]
	and similarly
	\[
	\Delta_{\alpha_j}\bigl(F^{(1)}\omega^{(2)}\bigr)
	=
	p_j^{(2)}\omega^{(1)}
	+
	p_j^{(1)}\omega^{(2)}
	+
	p_j^{(1)}p_j^{(2)}\,d\lambda.
	\]
	The last terms have vanishing integrals along the corresponding
	closed cycles, since they are exact differentials in \(\lambda\).
	The stated identity follows.
\end{proof}
\subsection{Dual representation of the metric }

Let \(\gamma\subset C_u\) be a contour, and choose a path from \(\infty_0\) to an initial point \(P_{\ast}\in\gamma\). This path determines the corresponding branch of \(K(z,w)\) along \(\gamma\) by analytic continuation from the standard determination at \(\infty_0\), characterized by
$
K(\infty_0,w)=0.
$

Using the Cauchy decompositions introduced above, define
\begin{equation}\label{dulmet}
	\eta^{\sharp}_\gamma(f)(w)
	=
	-C_\gamma(f)(w)\,d\lambda(w)
	+
	\mathcal C_\gamma(f\,d\lambda)(w),
\end{equation}
for \(f\in\mathcal H_\gamma\).
The jumps of the two terms in \(\eta_\gamma^\sharp(f)\) cancel across \(\gamma\). Moreover, the resulting differential has the prescribed singularities and monodromies of an admissible differential. Hence
$
\eta_\gamma^\sharp(f)\in\mathcal D_u.
$

\begin{cor}\label{dulmecimm}
	For every \(f\in\mathcal H_\gamma\) and
	\(\phi\in\mathcal D_u\), one has
	\begin{equation}\label{metpair}
		\bigl\langle\eta^{\sharp}_\gamma(f),\phi\bigr\rangle
		=
		\langle f,\phi\rangle_\gamma,
	\end{equation}
	where \(\langle\,\cdot,\cdot\,\rangle\) and
	\(\langle\,\cdot,\cdot\,\rangle_\gamma\) are defined in
	\eqref{metric} and \eqref{pair}, respectively.	
	Here the standard determination of \(\phi\) near \(\infty_0\) is fixed by
$
	\frac{\phi}{d\lambda}|_{\infty_{0}}=0.	
$
	The determination of \(\phi\) at the initial point \(P^{\ast}\in\gamma\) is obtained by analytic continuation along the same path from \(\infty_0\) to \(P^{\ast}\) used in the definition of \(\eta_\gamma^\sharp\).
\end{cor}

\begin{proof}
	Since \(C_\gamma(f)\) is holomorphic at the ramification points and
	every \(\phi\in\mathcal D_u\) is also holomorphic there, the term
	\(-C_\gamma(f)d\lambda\) does not contribute to the pairing
	\(\langle\,\cdot,\cdot\,\rangle\). Hence
	\begin{align*}
		\bigl\langle\eta^{\sharp}_\gamma(f),\phi\bigr\rangle
		&=
		\sum_{d\lambda(w)=0}
		\operatorname{res}_{w}
		\frac{
			\mathcal C_\gamma(f\,d\lambda)(w)\phi(w)
		}{
			d\lambda(w)
		}
		\\
		&=
		-\frac{1}{2\pi\mathrm{i}}
		\oint_\gamma
		f(z)d\lambda(z)
		\sum_{d\lambda(w)=0}
		\operatorname{res}_{w}
		\left(
		K(z,w)\frac{\phi(w)}{d\lambda(w)}
		\right).
	\end{align*}
	
 For fixed \(\xi\),
	consider
$
	B(\xi,w)\frac{\phi(w)}{d\lambda(w)}
$
	as a differential in \(w\). Besides the poles at the ramification points, it has a pole at \(w=\xi\), with
	\[
	\operatorname{res}_{w=\xi}
	\left(
	B(\xi,w)\frac{\phi(w)}{d\lambda(w)}
	\right)
	=
	d_\xi\left(\frac{\phi(\xi)}{d\lambda(\xi)}\right).
	\]
	By Lemma~\ref{bilin}, together with the vanishing \(\alpha\)-periods of \(B(\xi,w)\), the boundary contributions along the canonical cuts vanish. Therefore,
	\begin{equation}\label{B-res-identity}
		\sum_{d\lambda(w)=0}
		\operatorname{res}_{w}
		\left(
		B(\xi,w)\frac{\phi(w)}{d\lambda(w)}
		\right)
		=
		-d_\xi\left(\frac{\phi(\xi)}{d\lambda(\xi)}\right).
	\end{equation}
	Both sides of \eqref{B-res-identity} are independent of the determination of \(\phi\).
	
	For \(z\in\gamma\), let \(\rho_z\) be the path from \(\infty_0\) to \(z\) used to determine \(K(z,w)\), so that
$
	K(z,w)=\int_{\rho_z}B(\xi,w).
$
	Integrating \eqref{B-res-identity} along \(\rho_z\), we obtain
	\begin{align*}
		\sum_{d\lambda(w)=0}
		\operatorname{res}_{w}
		\left(
		K(z,w)\frac{\phi(w)}{d\lambda(w)}
		\right)
		&=
		\int_{\rho_z}
		\sum_{d\lambda(w)=0}
		\operatorname{res}_{w}
		\left(
		B(\xi,w)\frac{\phi(w)}{d\lambda(w)}
		\right)
		\\
		&=
		-\int_{\rho_z}
		d_\xi\left(\frac{\phi(\xi)}{d\lambda(\xi)}\right)
		\\
		&=
		-\frac{\phi(z)}{d\lambda(z)},
	\end{align*}
	where \(\phi(z)\) is the determination obtained by analytic continuation of its standard determination along \(\rho_z\), and we have used
$
	\frac{\phi}{d\lambda}=0
$
	at \(\infty_0\).
	Substituting this into the preceding expression gives
	\begin{align*}
		\bigl\langle\eta^{\sharp}_\gamma(f),\phi\bigr\rangle
		&=
		\frac{1}{2\pi\mathrm{i}}
		\oint_\gamma
		f(z)d\lambda(z)
		\frac{\phi(z)}{d\lambda(z)}
		\\
		&=
		\langle f,\phi\rangle_\gamma,
	\end{align*}
	as claimed.
\end{proof}

After fixing an extended primary differential \(\phi_0\), let
\(dF\in T_u^*H\) be represented by \(f\in\mathcal H_\gamma\), in the sense that
\[
dF(\partial)
=
\frac{1}{2\pi\mathrm{i}}
\oint_\gamma
f\,\partial(\lambda)\phi_0,
\qquad
\partial\in T_uH.
\]
By Corollary~\ref{dulmecimm},
\[
dF(\partial)
=
\left\langle
\eta^{\sharp}_\gamma(f),
\partial(\lambda)\phi_0
\right\rangle.
\]
Consequently, under the identification
\(T_uH\simeq\mathcal D_u\), the metric dual
\(\eta_{\phi_0}^{\sharp}(dF)\) is represented by the admissible differential
$
\eta^{\sharp}_\gamma(f).
$
\subsection{Levi--Civita connection}\label{subsec-Levi}

Let \(\gamma\subset C_u\) be the disjoint union of sufficiently small positively oriented circles surrounding the ramification points \(P_1,\ldots,P_N\), chosen to vary smoothly with \(u\in H\). Denote by \(D_\gamma^+\) the union of the discs bounded by the components of \(\gamma\), and by \(D_\gamma^-\) its complement in \( C_u\).

For a function or differential \(\omega\) defined in a neighborhood of \(\gamma\), we denote by \(\mathcal P_+(\omega)\) and \(\mathcal P_-(\omega)\) the corresponding Cauchy transforms restricted to \(D_\gamma^+\) and \(D_\gamma^-\), respectively. These are the two components of the Cauchy decomposition described above. We choose the determination of \(K(z,w)\) along each component of \(\gamma\) defined by paths contained in the canonical polygon, so that \(\mathcal C_\gamma(\omega)\) has vanishing \(\alpha\)-periods on \(D_\gamma^-\).

\begin{lem}\label{lem:var-positive} 
	For any \(\omega\in\mathcal D\) and  vector field 
	\(\partial\in\operatorname{Vect}(H)\), one has 
	$
	\mathcal P_+\bigl(\partial|_\lambda\omega\bigr)\in\mathcal D.
	$ 
\end{lem} 

\begin{proof} 
	Expand \(\omega\) in the basis of Dubrovin primary differentials as 
	$
	\omega=\sum_I c_I\phi_I.
	$ 
	Then 
	\[ 
	\partial|_\lambda\omega 
	= 
	\sum_I\partial(c_I)\phi_I 
	+ 
	\sum_I c_I\,\partial|_\lambda\phi_I. 
	\] 
	
	By the defining properties of the Dubrovin primary differentials, their \(\lambda\)-horizontal derivatives have vanishing positive projections, whereas the differentials themselves are holomorphic at the ramification points. Thus
$
	\mathcal P_+(\partial|_\lambda\phi_I)=0,
$ and $
	\mathcal P_+(\phi_I)=\phi_I.
$
We obtain 
	\[ 
	\mathcal P_+\bigl(\partial|_\lambda\omega\bigr) 
	= 
	\sum_I\partial(c_I)\phi_I\in\mathcal D. 
	\] 
	This proves the lemma.
\end{proof}

Fix an extended primary differential \(\phi_0\), and identify
\(T_uH\) with \(\mathcal D_u\). For
\(\partial_1,\partial_2\in\operatorname{Vect}(H)\), set
$
\omega
=
\partial_2(\lambda)\phi_0.
$ Since
\[
\partial_1|_\lambda f
=
\partial_1(f)
-
\frac{\partial_1(\lambda)}{d\lambda}\,df
\]
and
\[
\partial_1|_\lambda\phi_0
=
-d\left(
\frac{\partial_1(\lambda)}{d\lambda}\phi_0
\right),
\]
we obtain
\begin{align*}
	\partial_1|_\lambda
	\bigl(\partial_2(\lambda)\phi_0\bigr)
	={}&
	\left(
	\partial_1\partial_2(\lambda)
	-
	\frac{\partial_1(\lambda)}{d\lambda}
	d\bigl(\partial_2(\lambda)\bigr)
	\right)\phi_0-
	\partial_2(\lambda)\,
	d\left(
	\frac{\partial_1(\lambda)}{d\lambda}\phi_0
	\right)
	\\
	={}&
	\partial_1\partial_2(\lambda)\phi_0
	-
	d\left(
	\frac{\partial_1(\lambda)\partial_2(\lambda)}
	{d\lambda}\phi_0
	\right).
\end{align*}
Hence
\begin{equation}\label{LC-pre}
	\mathcal P_+
	\left(
	\partial_1|_\lambda
	\bigl(\partial_2(\lambda)\phi_0\bigr)
	\right)
	=
	\partial_1\partial_2(\lambda)\phi_0
	-
	\mathcal P_+
	\left[
	d\left(
	\frac{\partial_1(\lambda)\partial_2(\lambda)}
	{d\lambda}\phi_0
	\right)
	\right].
\end{equation}

\begin{prop}\label{prop:LC}
	The Levi--Civita connection \(\nabla\) of the metric
	\(\eta_{\phi_0}\) is characterized, under the identification
	\(T_uH\simeq\mathcal D_u\), by
	\begin{equation}\label{LC}
		\bigl(\nabla_{\partial_1}\partial_2\bigr)(\lambda)\phi_0
		=
		\mathcal P_+
		\left(
		\partial_1|_\lambda
		\bigl(\partial_2(\lambda)\phi_0\bigr)
		\right).
	\end{equation}
	Equivalently,
	\begin{equation}\label{LC-explicit}
		\bigl(\nabla_{\partial_1}\partial_2\bigr)(\lambda)\phi_0
		=
		\partial_1\partial_2(\lambda)\phi_0
		-
		\mathcal P_+
		\left[
		d\left(
		\frac{\partial_1(\lambda)\partial_2(\lambda)}
		{d\lambda}\phi_0
		\right)
		\right].
	\end{equation}
\end{prop}

\begin{proof}
By Lemma~\ref{lem:var-positive}, the right-hand side of
\eqref{LC} belongs to \(\mathcal D_u\), and satisfies the Leibniz rule in \(\partial_2\).
Thus \eqref{LC} defines a connection on \(TH\).
	
	We first verify torsion-freeness. Since the term involving \(\mathcal P_+\) in
	\eqref{LC-explicit} is symmetric in \(\partial_1\) and \(\partial_2\), one has
	\begin{align*}
		&
		\bigl(
		\nabla_{\partial_1}\partial_2
		-
		\nabla_{\partial_2}\partial_1
		-
		[\partial_1,\partial_2]
		\bigr)(\lambda)\phi_0=
		\left(
		\partial_1\partial_2(\lambda)
		-
		\partial_2\partial_1(\lambda)
		-
		[\partial_1,\partial_2](\lambda)
		\right)\phi_0
		=
		0.
	\end{align*}
	Thus the connection is torsion-free.
	
	We next verify compatibility with the residue metric. For \(\omega_1,\omega_2\in\mathcal D\), one has
	\begin{align*}
		\partial\langle\omega_1,\omega_2\rangle
		&=
		\partial
		\left(
		\sum_{a=1}^N
		\res_{P_a}
		\frac{\omega_1\omega_2}{d\lambda}
		\right)
		\\
		&=
		\sum_{a=1}^N
		\res_{P_a}
		\left(
		\frac{(\partial|_\lambda\omega_1)\omega_2}{d\lambda}
		+
		\frac{\omega_1(\partial|_\lambda\omega_2)}{d\lambda}
		\right).
	\end{align*}
Since \(\mathcal P_-(\partial|_\lambda\omega_1)\) is single-valued and has vanishing \(\alpha\)-periods, the residue theorem and Lemma~\ref{bilin} imply that
	\[
	\sum_{a=1}^N
	\res_{P_a}
	\frac{
		\mathcal{P}_{-}(\partial|_\lambda\omega_1)\omega_2
	}{d\lambda}
	=
	0.
	\]
	Consequently,
	\begin{align*}
		\partial\langle\omega_1,\omega_2\rangle
		={}&
		\sum_{a=1}^N
		\res_{P_a}
		\frac{
			\mathcal P_+
			(\partial|_\lambda\omega_1)\,\omega_2
		}{d\lambda}
		+
		\sum_{a=1}^N
		\res_{P_a}
		\frac{
			\omega_1\,
			\mathcal P_+
			(\partial|_\lambda\omega_2)
		}{d\lambda}
		\\
		={}&
		\left\langle
		\mathcal P_+
		(\partial|_\lambda\omega_1),
		\omega_2
		\right\rangle
		+
		\left\langle
		\omega_1,
		\mathcal P_+
		(\partial|_\lambda\omega_2)
		\right\rangle.
	\end{align*}
	Under the identification \(T_uH\simeq\mathcal D_u\), this is precisely
	\[
	\partial\,
	\eta_{\phi_0}(\partial_1,\partial_2)
	=
	\eta_{\phi_0}
	(\nabla_\partial\partial_1,\partial_2)
	+
	\eta_{\phi_0}
	(\partial_1,\nabla_\partial\partial_2).
	\]
	Thus the connection is compatible with \(\eta_{\phi_0}\).
	It is therefore the Levi--Civita connection.
\end{proof}
The formula above admits an alternative expression in terms of the multiplication on \(TH\), which will be used in the proof of Theorem~\ref{openext}.

\begin{cor}
	The positive projection satisfies
	\begin{align}
		\mathcal P_+\left(
		\frac{
			\partial_1(\lambda)\partial_2(\lambda)\phi_0
		}{d\lambda}
		\right)
		&=
		\mathcal P_+\left(
		\frac{
			\partial_1(\lambda)\partial_2(\lambda)\phi_0
		}{d\lambda}
		-
		\frac{
			(\partial_1\circ\partial_2)(\lambda)\phi_0
		}{d\lambda}
		\right)
		\nonumber\\
		&=
		\frac{
			\partial_1(\lambda)\partial_2(\lambda)\phi_0
		}{d\lambda}
		-
		\frac{
			(\partial_1\circ\partial_2)(\lambda)\phi_0
		}{d\lambda}.
		\label{LC-product-proj}
	\end{align}
	Hence
	\begin{equation}\label{LC-product}
		\bigl(\nabla_{\partial_1}\partial_2\bigr)(\lambda)\phi_0
		=
		\partial_1\partial_2(\lambda)\phi_0
		-
		d\left(
		\frac{
			\partial_1(\lambda)\partial_2(\lambda)\phi_0
		}{d\lambda}
		-
		\frac{
			(\partial_1\circ\partial_2)(\lambda)\phi_0
		}{d\lambda}
		\right).
	\end{equation}
\end{cor}

\begin{proof}
	The function
$
	\frac{(\partial_1\circ\partial_2)(\lambda)\phi_0}{d\lambda}
$
	is holomorphic on \(D_\gamma^-\), has vanishing \(\alpha\)-increments, and vanishes at \(\infty_0\). Hence its positive projection is zero.
	
	On the other hand, one has
	\[
	\partial_1(\lambda)(P_a)\partial_2(\lambda)(P_a)
	=
	(\partial_1\circ\partial_2)(\lambda)(P_a),
	\qquad a=1,\ldots,N.
	\]
	Consequently, the difference
$
	\frac{
		\partial_1(\lambda)\partial_2(\lambda)\phi_0
	}{d\lambda}
	-
	\frac{
		(\partial_1\circ\partial_2)(\lambda)\phi_0
	}{d\lambda}
$
	is holomorphic on \(D_\gamma^+\) and is therefore fixed by \(\mathcal P_+\). This proves \eqref{LC-product-proj}.
	
	Since each component of \(\gamma\) is contractible, Lemma~\ref{lem:Cauchy-d} gives
$
	\mathcal P_+(dh)=d\mathcal P_+(h)
$
	for functions \(h\in\mathcal{H}_{\gamma}\) defined near \(\gamma\). Applying this identity to \eqref{LC-explicit} and using \eqref{LC-product-proj}, we obtain \eqref{LC-product}.
\end{proof}

We next derive a dual representation of the Levi--Civita connection. Let \(\omega\) be a covector field on \(H\), represented by
$
f\in\mathcal H_\gamma,
$
where both the contour \(\gamma\) and the germ \(f\) vary smoothly with \(u\in H\).  Let \(\nabla^*\) denote the connection on \(T^*H\) dual to the Levi--Civita connection \(\nabla\).

For 
\(\partial\in\operatorname{Vect}(H)\), define
\[
\{f,\lambda\}_\partial
:=
df\,\partial(\lambda)-\partial(f)\,d\lambda,
\]
where \(\partial(f)\) denotes differentiation along the
\(\phi_0\)-horizontal lift.

\begin{lem}\label{lem:dulconn}
	Fix an extended primary differential \(\phi_0\). Let \(\omega\) be a covector field on \(H\) represented by \(f\in\mathcal H_\gamma\), with a fixed initial determination of \(K(z,w)\) at an initial point \(P_*\in\gamma\). We assume that \(P_*\) varies with \(u\) while \(\lambda(P_*)\) remains constant. Then, for every \(\partial\in\operatorname{Vect}(H)\),
	\begin{equation}\label{dual-LC}
		\bigl(
		\eta_{\phi_0}^{\sharp}(\nabla_\partial^*\omega)
		\bigr)(\lambda)\phi_0
		=
		\{C_\gamma(f),\lambda\}_\partial
		-
		\mathcal C_\gamma
		\bigl(\{f,\lambda\}_\partial\bigr).
	\end{equation}
	In particular, the right-hand side belongs to \(\mathcal D_u\).
\end{lem}

\begin{proof}
	By the compatibility of \(\nabla\) with \(\eta_{\phi_0}\), one has
	\[
	\eta_{\phi_0}^{\sharp}(\nabla_\partial^*\omega)
	=
	\nabla_\partial
	\bigl(\eta_{\phi_0}^{\sharp}(\omega)\bigr).
	\]
	Hence Proposition~\ref{prop:LC} gives
	\begin{equation}\label{dual-LC-proof1}
		\bigl(
		\eta_{\phi_0}^{\sharp}(\nabla_\partial^*\omega)
		\bigr)(\lambda)\phi_0
		=
		\mathcal P_+
		\left(
		\partial|_\lambda\eta_\gamma^\sharp(f)
		\right).
	\end{equation}	
	Using
$
	-\bigl(\partial|_\lambda C_\gamma(f)\bigr)d\lambda
	=
	\{C_\gamma(f),\lambda\}_\partial
$
	and \eqref{dulmet}, we obtain
	\begin{align*}
		\partial|_\lambda\eta_\gamma^\sharp(f)
		={}&
		\{C_\gamma(f),\lambda\}_\partial
		+
		\partial|_\lambda\mathcal C_\gamma(f\,d\lambda)
		\\
		={}&
		\{C_\gamma(f),\lambda\}_\partial
		-
		\mathcal C_\gamma
		\bigl(\{f,\lambda\}_\partial\bigr)
		+
		R,
	\end{align*}
	where
$
	R=
	\partial|_\lambda\mathcal C_\gamma(f\,d\lambda)
	-
	\mathcal C_\gamma
	\bigl(\partial|_\lambda(f\,d\lambda)\bigr).
$
	
	We now show that \(\mathcal P_+(R)=0\). Since the initial point is transported with \(\lambda(P_*)\) fixed, differentiation of the contour integral gives
	\[
	R
	=
	-\frac{1}{2\pi\mathrm i}
	\oint_\gamma
	f(z)d\lambda(z)\,
	\partial|_\lambda K(z,w).
	\]	
	For fixed \(z\), the differential \(\partial|_\lambda K(z,w)\) has vanishing \(\alpha\)-periods in \(w\).
Moreover, near the diagonal \(w=z\), away from the ramification points, the singular part of \(K(z,w)\) is
$
\frac{d\lambda(w)}{\lambda(w)-\lambda(z)}.
$
Hence \(\partial|_\lambda K(z,w)\) is regular along the diagonal, so \(R\) extends analytically across \(\gamma\). Consequently, its periods are well defined even when \(\gamma\) intersects the canonical cycles. We obtain
	\[
	\begin{aligned}
		\oint_{\alpha_j}R(w)
		&=
		-\frac{1}{2\pi\mathrm i}
		\oint_\gamma f(z)d\lambda(z)
		\oint_{\alpha_j}
		\partial|_\lambda K(z,w)
=0.
	\end{aligned}
	\]
	Furthermore, \(R\) is single-valued and holomorphic away from the ramification points, except possibly for a simple pole at \(\infty_0\). Hence \(R\) belongs to the negative subspace of the Cauchy decomposition associated with the small contours surrounding the ramification points. Consequently,
$
	\mathcal P_+(R)=0.
$
	
	On the other hand, the jumps across \(\gamma\) of
$
	\{C_\gamma(f),\lambda\}_\partial
$ and $
	\mathcal C_\gamma
	\bigl(\{f,\lambda\}_\partial\bigr)
$
	are both equal to \(\{f,\lambda\}_\partial\). Their difference therefore extends across \(\gamma\), and the properties of the Cauchy transforms imply that
$
	\{C_\gamma(f),\lambda\}_\partial
	-
	\mathcal C_\gamma
	\bigl(\{f,\lambda\}_\partial\bigr)
	\in\mathcal D_u.
$
	In particular, it is holomorphic at the ramification points and is therefore fixed by \(\mathcal P_+\). Applying \(\mathcal P_+\) to the preceding identity and using \eqref{dual-LC-proof1} yields \eqref{dual-LC}.
\end{proof}

\subsection{Dual representation of the multiplication}
\label{subsec:dual-mul}

Let \(\gamma\subset C_u\) be a contour and \(f\in\mathcal H_\gamma\). Fix an initial point \(P_*\in\gamma\) and a determination of \(K(z,w)\) near \(z=P_*\). For \(\partial\in T_uH\), assuming that \(\partial(\lambda)\) is holomorphic near \(\gamma\), define
\begin{equation}\label{dual-mul-def}
	\mathcal M_{\partial,\gamma}(f)
	:=
	-C_\gamma\bigl(f\,\partial(\lambda)\bigr)\,d\lambda
	+
	\partial(\lambda)\,
	\mathcal C_\gamma(f\,d\lambda).
\end{equation}
The jumps of the two terms cancel across \(\gamma\), so \(\mathcal M_{\partial,\gamma}(f)\) extends analytically across \(\gamma\). Moreover, it is holomorphic near the ramification points, and hence the residue pairing \eqref{metric} remains well defined when one of its arguments is \(\mathcal M_{\partial,\gamma}(f)\). However, it need not belong to \(\mathcal D_u\), since \(\partial(\lambda)\) may have poles at the zeros of \(\phi_0\).

\begin{lem}\label{lem:dual-mul}
	Let \(\partial_1,\partial_2,\partial_3\in T_uH\), and suppose that
$
	\partial_2(\lambda)\phi_0
	=
	\eta_\gamma^\sharp(f)
$
	for some \(f\in\mathcal H_\gamma\). Then
	\[
	c(\partial_1,\partial_2,\partial_3)
	=
	\left\langle
	\mathcal M_{\partial_1,\gamma}(f),
	\partial_3(\lambda)\phi_0
	\right\rangle,
	\]
	where \(c\) and \(\langle\cdot,\cdot\rangle\) are defined in
	\eqref{03tensor} and \eqref{metric}, respectively.
\end{lem}

\begin{proof}
	By the definitions of \(\mathcal M_{\partial_1,\gamma}(f)\) and \(\eta_\gamma^\sharp(f)\), one has
	\[
	\mathcal M_{\partial_1,\gamma}(f)
	-
	\partial_1(\lambda)\eta_\gamma^\sharp(f)
	=
	\left[
	\partial_1(\lambda)C_\gamma(f)
	-
	C_\gamma\bigl(f\,\partial_1(\lambda)\bigr)
	\right]d\lambda.
	\]
	The coefficient of \(d\lambda\) on the right-hand side is holomorphic at the ramification points, so this term does not contribute to the residue pairing. Consequently,
	\begin{align*}
		\left\langle
		\mathcal M_{\partial_1,\gamma}(f),
		\partial_3(\lambda)\phi_0
		\right\rangle
		&=
		\sum_{d\lambda=0}
		\res
		\frac{
			\partial_1(\lambda)
			\eta_\gamma^\sharp(f)
			\partial_3(\lambda)\phi_0
		}{d\lambda}
		\\
		&=
		\sum_{d\lambda=0}
		\res
		\frac{
			\partial_1(\lambda)
			\partial_2(\lambda)
			\partial_3(\lambda)\phi_0^2
		}{d\lambda}
		\\
		&=
		c(\partial_1,\partial_2,\partial_3),
	\end{align*}
	as claimed.
\end{proof}

In summary, we have established a contour representation
of cotangent vectors on the Hurwitz space $H$.
More precisely, a cotangent vector can be represented
by contour data consisting of a contour $\gamma$,
a function $f\in\mathcal H_\gamma$, an initial point
$P_*$, and an initial determination of the
Cauchy kernel $K$.
  We have also expressed the metric, its Levi--Civita connection, and the multiplication in terms of these contour data. In the next section, we construct recursion vector fields as the metric duals of covectors represented by explicit contour data. 
  We then use these vector fields to establish the
  generalized Frobenius structure and construct
  the associated Legendre-extended principal hierarchy.

\section{Generalized Frobenius structures and principal hierarchies}\label{sec-mainthm-1}

In this section, we prove that every choice of an extended primary differential determines a generalized Frobenius manifold structure on \(H\), and we give an explicit construction of its principal hierarchy.

\subsection{Recursion vector fields}

We first associate a vector field on \(H\) with each extended primary differential. We then show that these vector fields satisfy the recursion relations characterizing the gradients of the Hamiltonian densities of the principal hierarchy.

Recall that the index set \(\mathcal I\) of the extended primary differentials is decomposed as
\[
\mathcal I
=
\mathcal I_{\mathrm{pole}}
\sqcup
\mathcal I_{\log}
\sqcup
\mathcal I_{\mathrm{cyc}},
\]
where
\begin{align*}
	\mathcal I_{\mathrm{pole}}
	={}&
	\left\{
	(\infty_0,j)
	\mid
	j\geq1,\ 
	j\not\equiv0\pmod{n_0}
	\right\}
	\\
	&\sqcup
	\left\{
	(\infty_i,j)
	\mid
	i=1,\ldots,m,\ j\geq1
	\right\}
	\\
	&\sqcup
	\left\{
	(q,j)
	\mid
	q\in\lambda^{-1}(0),\ j\geq1
	\right\},
\end{align*}
\[
\mathcal I_{\log}
=
\left\{
(\infty_i,p)_{\log}
\mid
i=1,\ldots,m,\ p\geq0
\right\}
\sqcup
\left\{
(q,p)_{\log}
\mid
q\in\lambda^{-1}(0),\ p\geq0
\right\},
\]
and
\[
\mathcal I_{\mathrm{cyc}}
=
\left\{
(\alpha_i,p),(\beta_i,p)
\mid
i=1,\ldots,g,\ p\geq0
\right\}.
\]

For each \(I\in\mathcal I\), we associate a contour
\(\gamma_I\subset C_u\) and a function
\(Q_I\in\mathcal H_{\gamma_I}\) as follows. For the Cauchy transforms associated with these contours, we choose the initial determinations of \(K(z,w)\) using paths in the canonical polygon \(\widehat C_u\), and continue them analytically along the oriented contours. 

\begin{enumerate}[label=(\roman*)]
	
	\item \emph{Pole-type indices.}
	
	For \(I=(\infty_0,j)\), write
	\[
	j=kn_0+r,
	\qquad
	k\geq0,\quad r=1,\ldots,n_0-1.
	\]
	Let \(\gamma_I=\gamma_{0}\) be a sufficiently small
	positively oriented contour around \(\infty_0\), and define
	\[
	Q_{I}
	=
	-\frac{1}{n_0}
	\frac{\Gamma(r/n_0)}{\Gamma(j/n_0)}
	\lambda^{j/n_0-1}.
	\]
	
	For \(I=(\infty_i,j)\), \(i=1,\ldots,m\), write
	\[
	j=kn_i+r,
	\qquad
	k\geq0,\quad r=1,\ldots,n_i,
	\]
	with \(r=n_i\) when \(n_i\mid j\). Let \(\gamma_I=\gamma_{i}\) be a sufficiently small
	positively oriented contour around \(\infty_i\) and define
	\[
	Q_{\infty_i,j}
	=
	-\frac{1}{n_i}
	\frac{\Gamma(r/n_i)}{\Gamma(j/n_i)}
	\lambda^{j/n_i-1}.
	\]
	
	For \(I=(q,j)\), where \(q\in\lambda^{-1}(0)\) and \(j\geq1\), let
	\(\gamma_I=\gamma_q\), with \(\gamma_q\) a sufficiently small
	positively oriented contour around \(q\), and define
	\[
	Q_{q,j}
	=
	(-1)^{j+1}j!\lambda^{-j-1}.
	\]
	
	\item \emph{Logarithmic indices.}
	
For \(i=1,\ldots,m\), let \(dw_{0i}\) be the normalized Abelian differential of the third kind characterized by
\[
\oint_{\alpha_j}dw_{0i}=0,
\qquad j=1,\ldots,g,
\]
and
\[
\res_{\infty_i}dw_{0i}=1,
\qquad
\res_{\infty_0}dw_{0i}=-1.
\]
Choose an Abelian integral \(w_{0i}\) of \(dw_{0i}\).

For \(I=(\infty_i,p)_{\log}\), \(i=1,\ldots,m\), and \(p\geq1\), let
$
\gamma_I
=
\gamma_0\cup\gamma_i\cup\gamma_{0i},
$
where \(\gamma_0\) and \(\gamma_i\) are sufficiently small positively oriented contours around \(\infty_0\) and \(\infty_i\), respectively, and \(\gamma_{0i}\) is a positively oriented contour enclosing both \(\gamma_0\) and \(\gamma_i\) and no other marked points. Define \(Q_{(\infty_i,p)_{\log}}\) on the components of \(\gamma_I\) by
\begin{align*}
	\left.Q_{(\infty_i,p)_{\log}}\right|_{\gamma_0}
	&=
	-\frac{\lambda^{p-1}}{(p-1)!}
	\left(
	\log\lambda^{1/n_0}
	-w_{0i}
	-\frac{c_{p-1}}{n_0}
	\right),
	\\
	\left.Q_{(\infty_i,p)_{\log}}\right|_{\gamma_i}
	&=
	\frac{\lambda^{p-1}}{(p-1)!}
	\left(
	\log\lambda^{1/n_i}
	+w_{0i}
	-\frac{c_{p-1}}{n_i}
	\right),
	\\
	\left.Q_{(\infty_i,p)_{\log}}\right|_{\gamma_{0i}}
	&=
	-\frac{\lambda^{p-1}}{(p-1)!}w_{0i}.
\end{align*}
	For \(p=0\), let
	$
	\gamma_I=\gamma_0\cup\gamma_i
	$
	and define
	\[
	\left.Q_{(\infty_i,0)_{\log}}\right|_{\gamma_0}
	=
	-\frac{1}{n_0\lambda},
	\qquad
	\left.Q_{(\infty_i,0)_{\log}}\right|_{\gamma_i}
	=
	\frac{1}{n_i\lambda}.
	\]
	
For \(q\in\lambda^{-1}(0)\), let \(dw_{0q}\) be the normalized Abelian differential of the third kind characterized by
\[
\oint_{\alpha_j}dw_{0q}=0,
\qquad j=1,\ldots,g,
\]
and
\[
\res_qdw_{0q}=1,
\qquad
\res_{\infty_0}dw_{0q}=-1.
\]
Choose an Abelian integral \(w_{0q}\) of \(dw_{0q}\).

For \(I=(q,p)_{\log}\) and \(p\geq1\), let
$
\gamma_I
=
\gamma_0\cup\gamma_q\cup\gamma_{0q},
$
where \(\gamma_{0q}\) is the positively oriented boundary of a sufficiently small simply connected neighborhood containing \(\infty_0\) and \(q\), but no other marked points. Define \(Q_{(q,p)_{\log}}\) on the components of \(\gamma_I\) by
\begin{align*}
	\left.Q_{(q,p)_{\log}}\right|_{\gamma_0}
	&=
	-\frac{\lambda^{p-1}}{(p-1)!}
	\left(
	\log\lambda^{1/n_0}
	-w_{0q}
	-\frac{c_{p-1}}{n_0}
	\right),
	\\
	\left.Q_{(q,p)_{\log}}\right|_{\gamma_q}
	&=
	-\frac{\lambda^{p-1}}{(p-1)!}
	\left(
	\log\lambda
	-w_{0q}
	-c_{p-1}
	\right),
	\\
	\left.Q_{(q,p)_{\log}}\right|_{\gamma_{0q}}
	&=
	-\frac{\lambda^{p-1}}{(p-1)!}w_{0q}.
\end{align*}

For \(p=0\), let
$
\gamma_I=\gamma_0\cup\gamma_q
$
and define
\[
\left.Q_{(q,0)_{\log}}\right|_{\gamma_0}
=
-\frac{1}{n_0\lambda},
\qquad
\left.Q_{(q,0)_{\log}}\right|_{\gamma_q}
=
-\frac{1}{\lambda}.
\]

\item \emph{Cycle-type indices.}

For \(i=1,\ldots,g\) and \(p\geq 0\), define
\[
\gamma_{(\alpha_i,p)}=\alpha_i,
\qquad
Q_{(\alpha_i,p)}
=
\frac{\lambda^p}{p!}.
\]

For the $\beta$-type indices, choose representatives of the
cycles $\beta_i$ based at a common point
$P_*\in\lambda^{-1}(0)$, and fix a determination of
$K(z,w)$ near $z=P_*$. 
For \(p\geq1\), set
\[
\gamma_{(\beta_i,p)}=\beta_i,
\qquad
Q_{(\beta_i,p)}
=
2\pi\mathrm{i}\frac{\lambda^{p-1}}{(p-1)!}.
\]
For \(p=0\), let \(\gamma_{(\beta_i,0)}\) be a positively oriented simple closed contour contained in the canonical polygon \(\widehat C_u\) and sufficiently close to its boundary, and set
\[
Q_{(\beta_i,0)}
=
\frac{\phi_{\beta_i,0}}{d\lambda},
\qquad i=1,\ldots,g.
\]
\end{enumerate}

For an index \(I\in\mathcal I\) admitting a successor, we define its successor \(I^+\) by
\[
\begin{aligned}
	(\infty_0,j)^+&=(\infty_0,j+n_0),\\
	(\infty_i,j)^+&=(\infty_i,j+n_i),\\
	(q,j)^+&=(q,j-1),\qquad j\geq2,\\
	(q,1)^+&=(q,0)_{\log},\\
	\bigl((\infty_i,p)_{\log}\bigr)^+
	&=(\infty_i,p+1)_{\log},\\
	\bigl((q,p)_{\log}\bigr)^+
	&=(q,p+1)_{\log},\\
	(\alpha_i,p)^+&=(\alpha_i,p+1),\\
	(\beta_i,p)^+&=(\beta_i,p+1).
\end{aligned}
\]
Whenever \(I\) is not an initial element of its recursion family, we denote by \(I^-\) its predecessor.

We now establish the relation between the contour data and the extended primary differentials.

\begin{lem}\label{lem:cauchy-Phi}
	For every index \(I\in\mathcal I\), the Cauchy transforms satisfy
	\begin{equation}\label{cauchy-Phi}
		C_{\gamma_I}(Q_I)=\Phi_{I^-},
		\qquad
		\mathcal C_{\gamma_I}(Q_I\,d\lambda)=\phi_I
	\end{equation}
	on suitable local branches. In other words, the Abelian integral \(\Phi_{I^-}\) and the differential \(\phi_I\) are obtained by analytic continuation of the corresponding local branches of these Cauchy transforms, respectively.
	For initial indices \(I\), we adopt the convention \(\Phi_{I^-}=0\).
\end{lem}

\begin{proof}
We first consider the pole-type indices \(I=(\infty_0,j)\) or \(I=(\infty_i,j)\). The local expansion of the Cauchy kernel shows that
$
\mathcal C_{\gamma_I}(Q_I\,d\lambda)
$
has precisely the principal part prescribed for \(\phi_I\), with no other poles. Its \(\lambda\)-horizontal derivatives are single-valued meromorphic differentials with poles only at the ramification points and vanishing \(\alpha\)-periods.

Moreover, one has
\[
D_E\mathcal C_{\gamma_I}(Q_I\,d\lambda)
=
\frac{j}{n_a}\mathcal C_{\gamma_I}(Q_I\,d\lambda).
\]
Thus this Cauchy transform satisfies all the defining conditions of \(\phi_I\). By uniqueness,
$
\mathcal C_{\gamma_I}(Q_I\,d\lambda)=\phi_I.
$

On the other hand, the definitions of the weights give
$
dQ_{I^+}=Q_I\,d\lambda,
$ and $
\gamma_{I^+}=\gamma_I.
$
Since the contours are contractible, the integration-by-parts identity \eqref{Cauchy-d} yields
\[
dC_{\gamma_{I^+}}(Q_{I^+})
=
\mathcal C_{\gamma_I}(Q_I\,d\lambda)
=
\phi_I.
\]
The normalization at \(\infty_0\) therefore gives, on the corresponding branch,
$
C_{\gamma_{I^+}}(Q_{I^+})=\Phi_I.
$
For an initial index \(I\), the weight \(Q_I\) is holomorphic at the corresponding pole. When the pole is \(\infty_0\), it also satisfies \(Q_I(\infty_0)=0\). Hence the exterior Cauchy transform vanishes, giving
$
C_{\gamma_I}(Q_I)=0=\Phi_{I^-}.
$
This proves both identities for the pole-type indices under consideration.

The case \(I=(q,j)\) can be treated similarly.
	
We next consider the logarithmic indices. Let \(I=(\infty_i,p)_{\log}\), \(p\geq1\), and \(I^+=(\infty_i,p+1)_{\log}\). Denote by \(D_0,D_i,D_{0i}\) the domains enclosed by \(\gamma_0,\gamma_i,\gamma_{0i}\), respectively. Consider the piecewise function
\[
F=
\begin{cases}
	\displaystyle
	\Phi_I-\frac{\lambda^p}{p!}
	\left(\log\lambda^{1/n_0}-\frac{c_p}{n_0}\right),
	&\text{in }D_0,\\[6pt]
	\displaystyle
	\Phi_I+\frac{\lambda^p}{p!}
	\left(\log\lambda^{1/n_i}-\frac{c_p}{n_i}\right),
	&\text{in }D_i,\\[6pt]
	\displaystyle
	\Phi_I-\frac{\lambda^p}{p!}w_{0i},
	&\text{in }D_{0i}\setminus(\overline D_0\cup\overline D_i),\\[6pt]
	\Phi_I,
	&\text{outside }\overline D_{0i},
\end{cases}
\]
where the branches are chosen consistently with the prescribed determinations of the logarithms and Abelian integrals. By the defining local expansions of \(\Phi_I\), the function \(F\) is holomorphic at \(\infty_0\) and \(\infty_i\). Moreover, the logarithmic monodromies cancel in the intermediate region. Across the three positively oriented contours, its jumps are
\[
\begin{aligned}
	\left.F^L-F^R\right|_{\gamma_0}
	&=
	-\frac{\lambda^p}{p!}
	\left(
	\log\lambda^{1/n_0}-w_{0i}-\frac{c_p}{n_0}
	\right),\\
	\left.F^L-F^R\right|_{\gamma_i}
	&=
	\frac{\lambda^p}{p!}
	\left(
	\log\lambda^{1/n_i}+w_{0i}-\frac{c_p}{n_i}
	\right),\\
	\left.F^L-F^R\right|_{\gamma_{0i}}
	&=-\frac{\lambda^p}{p!}w_{0i}.
\end{aligned}
\]
These coincide with the prescribed values of \(Q_{I^+}\). Furthermore, \(F\) satisfies the normalization conditions of the scalar Cauchy transform. By uniqueness of the normalized Cauchy jump problem, we obtain, on the exterior branch,
$
C_{\gamma_{I^+}}(Q_{I^+})=\Phi_I.
$

For the initial index \(I=(\infty_i,0)_{\log}\), the weights are
\[
Q_I=
\begin{cases}
	\displaystyle-\frac{1}{n_0\lambda},
	&\text{on }\gamma_0,\\[6pt]
	\displaystyle\frac{1}{n_i\lambda},
	&\text{on }\gamma_i.
\end{cases}
\]
Both weights are holomorphic and vanish at their respective poles. Thus, on the exterior branch, the residue theorem gives
$
C_{\gamma_I}(Q_I)=0=\Phi_{I^-}.
$

We now prove the differential Cauchy transform identity. Let \(I=(\infty_i,p)_{\log}\), \(p\geq1\). Using \(c_p-c_{p-1}=1/p\), direct differentiation gives
\[
dQ_{I^+}=
\begin{cases}
	\displaystyle
	Q_I\,d\lambda+\frac{\lambda^p}{p!}\,dw_{0i},
	&\text{on }\gamma_0\cup\gamma_i,\\[6pt]
	\displaystyle
	Q_I\,d\lambda-\frac{\lambda^p}{p!}\,dw_{0i},
	&\text{on }\gamma_{0i}.
\end{cases}
\]
Consequently,
\[
\begin{aligned}
	\mathcal C_{\gamma_I}(Q_I\,d\lambda)
	={}&
	\mathcal C_{\gamma_I}(dQ_{I^+})
	-\mathcal C_{\gamma_0\cup\gamma_i}
	\left(\frac{\lambda^p}{p!}\,dw_{0i}\right)+\mathcal C_{\gamma_{0i}}
	\left(\frac{\lambda^p}{p!}\,dw_{0i}\right).
\end{aligned}
\]
The differential \(\frac{\lambda^p}{p!}\,dw_{0i}\) is single-valued and meromorphic, with no poles in the region between \(\gamma_0\cup\gamma_i\) and \(\gamma_{0i}\). Hence the last two terms cancel by contour deformation on the common exterior region. Since \(\gamma_I=\gamma_{I^+}\) and \(Q_{I^+}\) is single-valued along each contour component, the integration-by-parts identity \eqref{Cauchy-d} yields
\[
\begin{aligned}
	\mathcal C_{\gamma_I}(Q_I\,d\lambda)
	&=\mathcal C_{\gamma_I}(dQ_{I^+})\\
	&=dC_{\gamma_{I^+}}(Q_{I^+})\\
	&=d\Phi_I\\
	&=\phi_I.
\end{aligned}
\]

 For \(I=(\infty_i,0)_{\log}\), the definitions give
\[
Q_I\,d\lambda=
\begin{cases}
	\displaystyle-\frac{d\lambda}{n_0\lambda},
	&\text{on }\gamma_0,\\[6pt]
	\displaystyle\frac{d\lambda}{n_i\lambda},
	&\text{on }\gamma_i.
\end{cases}
\]
The two differentials have residues \(1\) at \(\infty_0\) and \(-1\) at \(\infty_i\), respectively. Since \(K(\infty_0,w)=0\), the residue theorem gives
\[
\mathcal C_{\gamma_I}(Q_I\,d\lambda)(w)
=K(\infty_i,w)
=dw_{0i}(w)
=\phi_I(w).
\]

The case \(I=(q,p)_{\log}\) is analogous.

Finally, we consider the cycle-type indices. 
First, let \(I=(\alpha_i,p)\), \(p\geq0\). The vanishing \(\alpha\)-periods of the Cauchy kernel give
\[
C_{\alpha_i}(1)
=
\frac{1}{2\pi\mathrm i}\oint_{\alpha_i}K(z,w)
=0.
\]
Thus the first identity holds for \(p=0\), since \(\Phi_{I^-}=0\) by convention.

For \(p\geq1\), consider
\[
dC_{\alpha_i}\left(\frac{\lambda^p}{p!}\right)
=
\frac{1}{2\pi\mathrm i}
\oint_{\alpha_i}\frac{\lambda(w)^p}{p!}B(z,w).
\]
By the normalization of \(B\) and the Cauchy jump relation, this differential has vanishing \(\alpha\)-periods and satisfies
\[
\Delta_{\alpha_j}
dC_{\alpha_i}\left(\frac{\lambda^p}{p!}\right)=0,
\qquad
\Delta_{\beta_j}
dC_{\alpha_i}\left(\frac{\lambda^p}{p!}\right)
=
-\delta_{ij}\,d\frac{\lambda^p}{p!}.
\]
These are precisely the defining conditions of \(\phi_{\alpha_i,p-1}\). Hence
$
dC_{\alpha_i}\left(\frac{\lambda^p}{p!}\right)
=
\phi_{\alpha_i,p-1}.
$
Since both \(C_{\alpha_i}(\lambda^p/p!)\) and \(\Phi_{\alpha_i,p-1}\) vanish at \(\infty_0\) on the normalized determination, we obtain
$
C_{\alpha_i}\left(\frac{\lambda^p}{p!}\right)
=
\Phi_{\alpha_i,p-1}.
$
Therefore \(C_{\gamma_I}(Q_I)=\Phi_{I^-}\) for all \(\alpha\)-type indices.

Moreover,
$
Q_I\,d\lambda
=
d\left(\frac{\lambda^{p+1}}{(p+1)!}\right).
$
Since \(\Delta_{\alpha_i}K=0\), the integration-by-parts identity \eqref{Cauchy-d} gives
\[
\begin{aligned}
	\mathcal C_{\alpha_i}(Q_I\,d\lambda)
	=
	dC_{\alpha_i}\left(\frac{\lambda^{p+1}}{(p+1)!}\right)=
	\phi_{\alpha_i,p}.
\end{aligned}
\]

Next, let \(I=(\beta_i,p)\), \(p\geq1\). We first observe that
\[
dC_{\beta_i}(2\pi\mathrm i)
=
\oint_{\beta_i}B(z,w)
=
dZ_i(z)
=
\phi_{\beta_i,0}(z).
\]
For \(p\geq2\), one has
\[
dC_{\beta_i}\left(
2\pi\mathrm i\,\frac{\lambda^{p-1}}{(p-1)!}
\right)
=
\oint_{\beta_i}
\frac{\lambda(w)^{p-1}}{(p-1)!}B(z,w).
\]
The Cauchy jump relation gives
\[
\begin{aligned}
	\Delta_{\alpha_j}
	dC_{\beta_i}\left(
	2\pi\mathrm i\,\frac{\lambda^{p-1}}{(p-1)!}
	\right)
	&=
	2\pi\mathrm i\,\delta_{ij}
	\,d\frac{\lambda^{p-1}}{(p-1)!},\\
	\Delta_{\beta_j}
	dC_{\beta_i}\left(
	2\pi\mathrm i\,\frac{\lambda^{p-1}}{(p-1)!}
	\right)
	&=0.
\end{aligned}
\]
Moreover, since the Euler flow rescales \(\lambda\) while preserving the normalized Bergman bidifferential and the canonical cycles, we have
\[
D_E\left(
\oint_{\beta_i}
\frac{\lambda(w)^{p-1}}{(p-1)!}B(z,w)
\right)
=
(p-1)
\oint_{\beta_i}
\frac{\lambda(w)^{p-1}}{(p-1)!}B(z,w).
\]
Together with the prescribed monodromy, these conditions identify the differential with \(\phi_{\beta_i,p-1}\).
The normalization at \(\infty_0\) therefore yields
$
C_{\beta_i}\left(
2\pi\mathrm i\,\frac{\lambda^{p-1}}{(p-1)!}
\right)
=
\Phi_{\beta_i,p-1}.
$

Furthermore,
$
Q_I\,d\lambda
=
d\left(2\pi\mathrm i\,\frac{\lambda^p}{p!}\right).
$
Applying \eqref{Cauchy-d} and using
\[
\Delta_{\beta_i}K(P_*,z)=dZ_i(z),
\]
we obtain
\[
\begin{aligned}
	\mathcal C_{\beta_i}(Q_I\,d\lambda)
	&=
	dC_{\beta_i}\left(
	2\pi\mathrm i\,\frac{\lambda^p}{p!}
	\right)
	-\frac{\lambda(P_*)^p}{p!}\,dZ_i\\
	&=
	dC_{\beta_i}\left(
	2\pi\mathrm i\,\frac{\lambda^p}{p!}
	\right)\\
	&=
	d\Phi_{\beta_i,p}
	=
	\phi_{\beta_i,p},
\end{aligned}
\]
where we used \(p\geq1\) and \(\lambda(P_*)=0\).

Finally, consider \(I=(\beta_i,0)\). In this case, \(\gamma_I\) is the positively oriented contour lying inside and sufficiently close to the boundary of the canonical polygon, and
$
Q_I=\frac{\phi_{\beta_i,0}}{d\lambda}.
$
For fixed \(z\) in the region enclosed by \(\gamma_I\), the differential
$
\frac{\phi_{\beta_i,0}(w)}{d\lambda(w)}K(z,w)
$
is meromorphic in \(w\) on the canonical polygon. Since \(\gamma_I\) surrounds all its poles, the residue theorem gives
\[
C_{\gamma_I}(Q_I)(z)=0=\Phi_{I^-}.
\]

Moreover, \(Q_I\,d\lambda=\phi_{\beta_i,0}\). For fixed \(w\) in the same region, the differential
$
K(z,w)\phi_{\beta_i,0}(z)
$
has a simple pole at \(z=w\), with
\[
\operatorname{Res}_{z=w}
\bigl(K(z,w)\phi_{\beta_i,0}(z)\bigr)
=
-\phi_{\beta_i,0}(w).
\]
Hence
\[
\begin{aligned}
	\mathcal C_{\gamma_I}(Q_I\,d\lambda)(w)
	&=
	-\frac{1}{2\pi\mathrm i}
	\oint_{\gamma_I}
	K(z,w)\phi_{\beta_i,0}(z)\\
	&=
	\phi_{\beta_i,0}(w).
\end{aligned}
\]
This proves \eqref{cauchy-Phi} for all cycle-type indices on the corresponding determinations.

\end{proof}

Combining Lemma~\ref{lem:cauchy-Phi} and \eqref{dulmet}, using the compatible determinations of the Cauchy transforms, we immediately obtain the following.
\begin{lem}\label{lem:Q-primary}
	For every index
	$
	I\in\mathcal I
	$, one has
	\begin{equation}
\eta_{\gamma_I}^{\sharp}(Q_I)
=
\phi_I-\Phi_{I^-}\,d\lambda,
	\end{equation}
	where  $\eta^{\sharp}_{\gamma}$ is defined in \eqref{dulmet}.
\end{lem}

Fix an extended primary differential \(\phi_0\). For each
$
I\in\mathcal I,
$
let \(\omega_I\) be the covector field on \(H\) represented by the pair
\((\gamma_I,Q_I)\), namely,
\[
\omega_I(\partial)
=
\frac{1}{2\pi\mathrm i}
\oint_{\gamma_I}
Q_I\,\partial(\lambda)\phi_0,
\qquad
\partial\in T_uH.
\]
Define
$
\xi_I:=\eta_{\phi_0}^{\sharp}(\omega_I).
$
Equivalently, under the identification \(T_uH\simeq\mathcal D_u\),
$
\xi_I(\lambda)\phi_0
=
\eta_{\gamma_I}^{\sharp}(Q_I).
$
Then Lemma~\ref{lem:Q-primary} gives
\begin{equation}\label{eq:Q-primary}
	\xi_I(\lambda)\phi_0
	=
	\phi_I-\Phi_{I^-}\,d\lambda.
\end{equation}

\begin{lem}
	Let \(I\in\mathcal I\) admit a successor \(I^+\). Then, for every vector field \(\partial\) on \(H\), one has
	\begin{equation}\label{eq:xi-Phi}
		(\nabla_{\partial}\xi_{I^{+}})(\lambda)\phi_{0}
		=
		\{\Phi_{I},\lambda\}_{\partial},
	\end{equation}
	where
$
	\{\Phi_{I},\lambda\}_{\partial}
	=
	d\Phi_{I}\,\partial(\lambda)
	-
	\partial(\Phi_{I})\,d\lambda.
$
\end{lem}

\begin{proof}
	The identity can  be derived directly from \eqref{dual-LC}. Here we give an alternative proof using \eqref{eq:Q-primary}. By \eqref{LC},
	\[
	\begin{aligned}
		(\nabla_{\partial}\xi_{I^{+}})(\lambda)\phi_{0}
		&=
		\mathcal{P}_{+}\!\left(
		\partial|_{\lambda}
		\bigl(\xi_{I^{+}}(\lambda)\phi_0\bigr)
		\right)\\
		&=
			\mathcal{P}_{+}\!\left(
		\partial|_{\lambda}
		\bigl(\phi_{I^{+}}-\Phi_I\,d\lambda\bigr)
		\right).
	\end{aligned}
	\]
	Since \(\phi_{I^{+}}\) is horizontal, its \(\lambda\)-horizontal derivative has vanishing positive part.
 Hence
	\[
	(\nabla_{\partial}\xi_{I^{+}})(\lambda)\phi_{0}
	=
	-	\mathcal{P}_{+}\!\left(
	(\partial|_{\lambda}\Phi_I)\,d\lambda
	\right).
	\]
	Using
	\[
	-(\partial|_{\lambda}\Phi_I)\,d\lambda
	=
	d\Phi_I\,\partial(\lambda)
	-
	\partial(\Phi_I)\,d\lambda
	=
	\{\Phi_I,\lambda\}_{\partial},
	\]
	and observing that \(\{\Phi_I,\lambda\}_{\partial}\) is holomorphic at the ramification points and is therefore fixed by \(\mathcal{P}_{+}\), we obtain
\eqref{eq:xi-Phi}.
\end{proof}

\begin{lem}\label{cor:Q-vector-recursion}
	For every \(I\in\mathcal I\) admitting a successor \(I^+\) and every vector field \(\partial\) on \(H\), one has
	\begin{equation}\label{xi-rec}
			\nabla_{\partial}\xi_{I^+}
		=
		\partial\circ\xi_I.
	\end{equation}
\end{lem}

\begin{proof}
	By \eqref{eq:xi-Phi} and the dual representation of the multiplication, it suffices to show that
	\begin{equation}\label{eq:rec-cauchy-proof}
		\left\langle
		\{\Phi_I,\lambda\}_{\partial_1},
		\partial_2(\lambda)\phi_0
		\right\rangle
		=
		\left\langle
		\mathcal M_{\partial_1,\gamma_I}(Q_I),
		\partial_2(\lambda)\phi_0
		\right\rangle
	\end{equation}
	for all vector fields \(\partial_1,\partial_2\) on \(H\).
	
	By the definition of \(\mathcal M_{\partial,\gamma}\) and Lemma~\ref{lem:cauchy-Phi},
	\[
	\begin{aligned}
		\mathcal M_{\partial_1,\gamma_I}(Q_I)
		&=
		-C_{\gamma_I}\bigl(Q_I\partial_1(\lambda)\bigr)\,d\lambda
		+
		\partial_1(\lambda)\phi_I,\\
		\{\Phi_I,\lambda\}_{\partial_1}
		&=
		\partial_1(\lambda)\phi_I
		-
		\partial_1(\Phi_I)\,d\lambda.
	\end{aligned}
	\]
	Consequently,
	\[
	\mathcal M_{\partial_1,\gamma_I}(Q_I)
	-
	\{\Phi_I,\lambda\}_{\partial_1}
	=
	\left[
	\partial_1(\Phi_I)
	-
	C_{\gamma_I}\bigl(Q_I\partial_1(\lambda)\bigr)
	\right]d\lambda.
	\]
	The coefficient in brackets is holomorphic at the ramification points of \(\lambda\), so the difference above does not contribute to the residue pairing. Thus \eqref{eq:rec-cauchy-proof} holds, proving \eqref{xi-rec}.
\end{proof}

\subsection{Generalized Frobenius structure on $H$}

We now prove the first part of Theorem~\ref{mainthm-1}, namely, that
$
(H,\eta_{\phi_0},\circ,e)
$
defines a generalized Frobenius manifold.

The critical values \(u_1,\ldots,u_N\) are canonical coordinates for the multiplication \(\circ\). Since the metric \(\eta_{\phi_0}\) is flat and invariant with respect to \(\circ\), it remains to verify the potentiality condition. In canonical coordinates,
$
\eta_{\phi_0}(e,\cdot)
=
\sum_{a=1}^N\eta_a\,du_a.
$
Thus the potentiality condition
$
\partial_{u_b}\eta_a
=
\partial_{u_a}\eta_b
$
is equivalent to the closedness of \(\eta_{\phi_0}(e,\cdot)\).

Let \(I_0\) be the index corresponding to the chosen extended primary differential \(\phi_0\). By Lemma~\ref{lem:Q-primary},
\[
\xi_{I_0}(\lambda)\phi_0
=
\phi_0-\Phi_{I_0^-}\,d\lambda,
\]
where \(\Phi_{I_0^-}=0\) if \(I_0\) is initial. Evaluating at each ramification point \(P_a\), where \(d\lambda=0\) and \(\phi_0\) is nonzero on the nondegenerate locus, gives
\[
\xi_{I_0}(u_a)=1,
\qquad a=1,\ldots,N.
\]
Hence \(\xi_{I_0}=e\), and therefore
\[
\eta_{\phi_0}(e,\cdot)
=
\eta_{\phi_0}(\xi_{I_0},\cdot)
=
\omega_{I_0}.
\]

By metric compatibility and torsion-freeness of \(\nabla\), for any vector fields \(X,Y\) on \(H\),
\[
d\omega_{I_0}(X,Y)
=
\eta_{\phi_0}(\nabla_X\xi_{I_0},Y)
-
\eta_{\phi_0}(\nabla_Y\xi_{I_0},X).
\]
If \(I_0\) admits a predecessor \(I_0^-\), Lemma~\ref{cor:Q-vector-recursion} gives
$
\nabla_X\xi_{I_0}=X\circ\xi_{I_0^-}.
$
Consequently,
\[
\begin{aligned}
	d\omega_{I_0}(X,Y)
	&=
	\eta_{\phi_0}(X\circ\xi_{I_0^-},Y)
	-
	\eta_{\phi_0}(Y\circ\xi_{I_0^-},X)=0,
\end{aligned}
\]
by the commutativity of \(\circ\) and the invariance of \(\eta_{\phi_0}\).

If \(I_0\) is initial, then
$
\xi_{I_0}(\lambda)\phi_0=\phi_0.
$
By \eqref{LC} and the horizontality of \(\phi_0\),
\[
(\nabla_X\xi_{I_0})(\lambda)\phi_0
=
\mathcal P_+\bigl(X|_\lambda\phi_0\bigr)
=
0.
\]
Thus \(\nabla\xi_{I_0}=0\), which also implies
$
d\omega_{I_0}=0.
$
Therefore, in all cases,
$
d\bigl(\eta_{\phi_0}(e,\cdot)\bigr)=0.
$
The potentiality condition follows, and hence
$
(H,\eta_{\phi_0},\circ,e)
$
is a generalized Frobenius manifold.
\begin{rem}\label{density}
	When the chosen differential \(\phi_0\) has no logarithmic singularities, in particular when \(\phi_0\) is an ordinary Dubrovin primary differential, the functions \(F_I\) satisfying
	$
	dF_I=\omega_I,
	$
	or equivalently,
	\begin{equation}\label{density deriv}
		X(F_I)
		=
		\omega_I(X)
		=
		\frac{1}{2\pi\mathrm{i}}
		\oint_{\gamma_I}
		Q_I\,X(\lambda)\phi_0
	\end{equation}
	for any vector field \(X\) on \(H\), can be written explicitly, up to additive constants, as follows.
	
	For the pole-type indices,
	\[
	\begin{alignedat}{2}
		F_{\infty_0,j}
		={}&
		-\frac{1}{n_0}
		\frac{\Gamma(r/n_0)}{\Gamma(j/n_0+1)}
		\operatorname{Res}_{\infty_0}
		\lambda^{j/n_0}\phi_0,
		&\qquad&
		j=kn_0+r,\quad k\geq0,\quad r=1,\ldots,n_0-1,
		\\[1mm]
		F_{\infty_i,j}
		={}&
		-\frac{1}{n_i}
		\frac{\Gamma(r/n_i)}{\Gamma(j/n_i+1)}
		\operatorname{Res}_{\infty_i}
		\lambda^{j/n_i}\phi_0,
		&&
		j=kn_i+r,\quad k\geq0,\quad r=1,\ldots,n_i,
		\\[1mm]
		F_{q,j}
		={}&
		(-1)^j(j-1)!
		\operatorname{Res}_{q}\lambda^{-j}\phi_0,
		&&
		q\in\lambda^{-1}(0),\quad j\geq1.
	\end{alignedat}
	\]
	In the second formula, \(r=n_i\) whenever \(n_i\mid j\).
	
	For the logarithmic indices,
	\[
	\begin{aligned}
		F_{(\infty_i,p)_{\log}}
		={}&
		-\operatorname{Res}_{\infty_0}
		\frac{\lambda^p}{p!}
		\left(
		\log\lambda^{1/n_0}
		-w_{0i}
		-\frac{c_p}{n_0}
		\right)\phi_0
		\\
		&+
		\operatorname{Res}_{\infty_i}
		\frac{\lambda^p}{p!}
		\left(
		\log\lambda^{1/n_i}
		+w_{0i}
		-\frac{c_p}{n_i}
		\right)\phi_0
		\\
		&-
		\frac{1}{2\pi\mathrm{i}}
		\oint_{\gamma_{0i}}
		\frac{\lambda^p}{p!}w_{0i}\phi_0,
		\qquad
		i=1,\ldots,m,\quad p\geq0,
	\end{aligned}
	\]
	and
	\[
	\begin{aligned}
		F_{(q,p)_{\log}}
		={}&
		-\operatorname{Res}_{\infty_0}
		\frac{\lambda^p}{p!}
		\left(
		\log\lambda^{1/n_0}
		-w_{0q}
		-\frac{c_p}{n_0}
		\right)\phi_0
		\\
		&-
		\operatorname{Res}_{q}
		\frac{\lambda^p}{p!}
		\left(
		\log\lambda
		-w_{0q}
		-c_p
		\right)\phi_0
		\\
		&-
		\frac{1}{2\pi\mathrm{i}}
		\oint_{\gamma_{0q}}
		\frac{\lambda^p}{p!}w_{0q}\phi_0,
		\qquad
		q\in\lambda^{-1}(0),\quad p\geq0.
	\end{aligned}
	\]
	
	For the cycle-type indices,
	\[
	\begin{alignedat}{2}
		F_{\beta_i,p}
		={}&
		\oint_{\beta_i}
		\frac{\lambda^p}{p!}\phi_0,
		&\qquad&
		i=1,\ldots,g,\quad p\geq0,
		\\[1mm]
		F_{\alpha_i,p}
		={}&
		\frac{1}{2\pi\mathrm{i}}
		\oint_{\alpha_i}
		\frac{\lambda^{p+1}}{(p+1)!}\phi_0,
		&&
		i=1,\ldots,g,\quad p\geq0.
	\end{alignedat}
	\]
	These formulas follow by direct verification from
	\eqref{density deriv} and Lemma~\ref{bilin}.
\end{rem}
We next consider the conformal case. Suppose that the chosen
extended primary differential \(\phi_0\) is homogeneous with
respect to
$
D_E=\partial_E|_\lambda+\lambda\partial_\lambda,
$
that is,
$
D_E\phi_0=\nu_0\phi_0.
$
This holds for the pole-type and cycle-type differentials,
as well as for the logarithmic differentials of degree zero.
More precisely, the corresponding Euler degrees are
\[
\begin{aligned}
	\nu_{\infty_0,j}&=\frac{j}{n_0},
	&
	\nu_{\infty_i,j}&=\frac{j}{n_i},
	&
	\nu_{q,j}&=-j,
	\\
	\nu_{\beta_i,p}&=p,
	&
	\nu_{\alpha_i,p}&=p+1,
	&
	\nu_{(\infty_i,0)_{\log}}
	&=\nu_{(q,0)_{\log}}=0.
\end{aligned}
\]
Thus, if \(\phi_0=\phi_I\) is one of these homogeneous
differentials, then \(\nu_0=\nu_I\).

Since
$
D_E\lambda=\lambda
$
and
$
D_E\bigl(X(\lambda)\bigr)
=
[E,X](\lambda)+X(\lambda)
$
for every vector field \(X\) on \(H\), the residue
representation of the metric gives
\[
\begin{aligned}
	E\bigl(\eta_{\phi_0}(X,Y)\bigr)
	={}&
	\eta_{\phi_0}([E,X],Y)
	+
	\eta_{\phi_0}(X,[E,Y])+
	(2\nu_0+1)\eta_{\phi_0}(X,Y).
\end{aligned}
\]
Consequently,
\[
\mathcal L_E\eta_{\phi_0}
=
(2\nu_0+1)\eta_{\phi_0}
=
(2-d)\eta_{\phi_0},
\]
where $
d=1-2\nu_0.
$
Hence
$
(H,\eta_{\phi_0},\circ,e,E)
$
is a conformal generalized Frobenius manifold.

For completeness, the logarithmic differentials with
\(p\geq1\) satisfy nonhomogeneous Euler relations.
Writing \(h_p(\lambda)=\lambda^p/p!\), one has
\[
\begin{aligned}
	D_E\phi^{\log}_{\infty_i,p}
	={}&
	p\phi^{\log}_{\infty_i,p}
	-
	\sum_{s=1}^m
	\left(
	\frac{n_s}{n_0}+\delta_{is}
	\right)
	\phi_{\infty_s,pn_s}
	+
	\frac{1}{n_0}dh_p,
	\\
	D_E\phi^{\log}_{q,p}
	={}&
	p\phi^{\log}_{q,p}
	-
	\sum_{s=1}^m
	\frac{n_s}{n_0}\phi_{\infty_s,pn_s}
	+
	\frac{1}{n_0}dh_p.
\end{aligned}
\]
Thus these differentials are not homogeneous in general,
and the preceding conformality argument does not apply
to the corresponding generalized Frobenius structures.

\begin{rem}
	Recall that the Abelian integrals are normalized
	to have vanishing constant terms in their expansions
	at \(\infty_0\), as specified in the Introduction.	
	Suppose that the chosen extended primary
	differential \(\phi_0\) is homogeneous, namely,
	$
	D_E\phi_0=\nu_0\phi_0.
	$
	Then
	$
	D_E\Phi_0=\nu_0\Phi_0+r_0,
	$
	where the constant \(r_0\) is determined by the
	prescribed normalization. More precisely,
	\[
	r_0=
	\begin{cases}
		\dfrac{1}{n_0},
		& \phi_0=\phi_{(\infty_i,0)_{\log}}
		\text{ or }\phi_{(q,0)_{\log}},\\[6pt]
		0,
		& \text{for all other homogeneous
			extended primary differentials}.
	\end{cases}
	\]
	
	For the non-homogeneous logarithmic differentials,
	set
	$
	\Phi_{W_j}
	=-\sum_{s=1}^m a_s\Phi_{\infty_s,jn_s},
	$
	where
	\[
	a_s=
	\begin{cases}
		\dfrac{n_s}{n_0}+\delta_{is},
		& \phi_0=\phi_{(\infty_i,j)_{\log}},\\[6pt]
		\dfrac{n_s}{n_0},
		& \phi_0=\phi_{(q,j)_{\log}}.
	\end{cases}
	\]
	Then
	$
	D_E\phi_0
	=
	j\phi_0+
	d\left(\Phi_{W_j}+\frac{\lambda^j}{n_0j!}\right).
	$
	The normalization at \(\infty_0\) determines
	the integration constant and gives
	$
	D_E\Phi_0
	=
	j\Phi_0+\Phi_{W_j}
	+\frac{\lambda^j}{n_0j!}.
	$
	In particular, no additional constant term occurs
	in the non-homogeneous logarithmic cases.
	
	Finally, using the reciprocity relation
	$
	\partial_E(\lambda)\phi_0
	=
	-\bigl(\partial_E|_\lambda\Phi_0\bigr)d\lambda,
	$
	we obtain
	\begin{equation}\label{Eurform}
		\partial_E(\lambda)\phi_0
		=
		\begin{cases}
			\lambda\phi_0-(\nu_0\Phi_0+r_0)d\lambda,
			& D_E\phi_0=\nu_0\phi_0,\\[4pt]
			\lambda\phi_0-
			\left(j\Phi_0+\Phi_{W_j}
			+\dfrac{\lambda^j}{n_0j!}\right)d\lambda,
			& \phi_0=\phi_{(\infty_i,j)_{\log}}
			\text{ or }\phi_{(q,j)_{\log}},\ j\ge1.
		\end{cases}
	\end{equation}
\end{rem}

\begin{rem}
	Assume that the chosen extended primary differential
	\(\phi_0\) is homogeneous and has no logarithmic
	singularities. Then the Hamiltonian densities in
	Remark~\ref{density} satisfy quasi-homogeneity
	relations with respect to the Euler vector field \(E\).
	
	For the pole-type densities,
	\begin{align*}
		E(F_{\infty_0,j})
		&=
		\left(\nu_0+\frac{j}{n_0}\right)
		F_{\infty_0,j},
		\quad
		E(F_{\infty_i,j})
		=
		\left(\nu_0+\frac{j}{n_i}\right)
		F_{\infty_i,j},
		\\
		E(F_{q,j})
		&=
		(\nu_0-j)F_{q,j}.
	\end{align*}
	
	For the logarithmic densities with \(p=0\),
	\[
	\begin{aligned}
		E\bigl(F_{(\infty_i,0)_{\log}}\bigr)
		=
		\nu_0 F_{(\infty_i,0)_{\log}},
		\quad 
		E\bigl(F_{(q,0)_{\log}}\bigr)
		=
		\nu_0 F_{(q,0)_{\log}}.
	\end{aligned}
	\]
	
For \(p\geq1\), one has
\[
\begin{aligned}
	E\bigl(F_{(\infty_i,p)_{\log}}\bigr)
	={}&
	(\nu_0+p)F_{(\infty_i,p)_{\log}}
	-
	\sum_{s=1}^m
	\left(\frac{n_s}{n_0}+\delta_{is}\right)
	F_{\infty_s,pn_s},
	\\[2mm]
	E\bigl(F_{(q,p)_{\log}}\bigr)
	={}&
	(\nu_0+p)F_{(q,p)_{\log}}
	-
	\sum_{s=1}^m
	\frac{n_s}{n_0}F_{\infty_s,pn_s}.
\end{aligned}
\]
These identities hold without correction unless
\(\phi_0=\phi_{q,p}\). In this exceptional case, the right-hand sides acquire, the additive constants
$
\frac{(-1)^p}{n_0}
$ and $
(-1)^p\left(\frac{1}{n_0}-1\right) 
$, respectively.
	
	For the cycle-type densities,
	\[
	\begin{aligned}
		E(F_{\beta_i,p})
		&=
		(\nu_0+p)F_{\beta_i,p},
		\quad
		E(F_{\alpha_i,p})
		=
		(\nu_0+p+1)F_{\alpha_i,p}.
	\end{aligned}
	\]
	These identities follow by applying \(E\) to the
	explicit expressions in Remark~\ref{density},
	using the Euler homogeneity of \(\phi_0\) and
	the prescribed normalization of the Abelian
	integrals.
\end{rem}
\subsection{Principal hierarchy}\label{prinH}

We finally derive the principal hierarchy associated with the generalized Frobenius structure constructed above. Let
$
v\in LH=C^\infty(S^1,H)
$
be a smooth loop. The derivative \(v_x\) may be regarded pointwise as a tangent vector in \(T_{v(x)}H\), and, by a slight abuse of notation, we write
$
\partial_x:=v_x.
$
When acting on functions on the universal curve, \(\partial_x\) denotes the corresponding \(\Phi_0\)-horizontal lift.

Since
$
d\omega_I=0,
$
 there exist local functions \(F_I\), unique up to additive constants, satisfying
$
dF_I=\omega_I,
$ and $
\eta_{\phi_0}^{\sharp}(dF_I)=\xi_I.
$

For each \(I\in\mathcal I\), consider the Hamiltonian
$
H_I[v]
=
\int_{S^1}F_{I^+}(v(x))\,dx.
$
Let \(X_I\) be its Hamiltonian vector field with respect to the Poisson structure determined by the flat metric \(\eta_{\phi_0}\). Then
\[
X_I
=
\eta_{\phi_0}^{\sharp}
\nabla_{\partial_x}^{*}dF_{I^+}
=
\nabla_{\partial_x}\xi_{I^+}
=
\partial_x\circ\xi_I.
\]
Under the identification \(T_{v(x)}H\simeq\mathcal D_{v(x)}\), where a vector field \(X\) corresponds to the differential \(X(\lambda)\phi_0\), equality \eqref{eq:xi-Phi} gives
\[
X_I(\lambda)\phi_0
=
(\partial_x\circ\xi_I)(\lambda)\phi_0
=
\{\Phi_I,\lambda\},
\]
where
$
\{f,g\}
:=
df\,\partial_x g
-
dg\,\partial_x f.
$
Writing
$
\frac{\partial\lambda}{\partial T_I}
=
X_I(\lambda),
$
we obtain the dispersionless Lax representation
\begin{equation}\label{principal-lax}
	\frac{\partial\lambda}{\partial T_I}\phi_0
	=
	\{\Phi_I,\lambda\},
	\qquad
	I\in\mathcal I.
\end{equation}

In particular, let \(I_0\) be the index corresponding to the chosen differential \(\phi_0\). Since \(\xi_{I_0}=e\), we have
\[
X_{I_0}
=
\partial_x\circ\xi_{I_0}
=
\partial_x\circ e
=
\partial_x.
\]
Therefore the \(T_{I_0}\)-flow coincides with spatial translation,
$
\frac{\partial}{\partial T_{I_0}}
=
\frac{\partial}{\partial x}.
$

Moreover, the recursion relations imply that each \(\xi_I\) is a Legendre field. Indeed, for all vector fields \(X,Y\) on \(H\),
$
X\circ\nabla_Y\xi_I
=
Y\circ\nabla_X\xi_I,
$
by commutativity and associativity of \(\circ\), with both sides vanishing when \(I\) is initial. Consequently, the flows
\[
\frac{\partial v}{\partial T_I}
=
v_x\circ\xi_I
\]
commute pairwise.

The initial indices give the flat vector fields associated with the ordinary Dubrovin primary differentials. The remaining indices provide their recursive descendants and additional Legendre-field families, including the family containing the unity vector field \(e\). Thus these flows constitute a Legendre-extended principal hierarchy in the sense of \cite{liu2025legendre}, with the explicit Lax representation \eqref{principal-lax}.

This establishes the dispersionless Lax representation stated in Theorem~\ref{mainthm-1} and completes the proof.
\section{Open extensions and their principal hierarchies}\label{sec-openext}

In this section, we prove Theorem~\ref{openext}. We begin by recalling some basic facts about flat \(F\)-manifolds, open WDVV equations, and their principal hierarchies.

\subsection{Flat \(F\)-manifolds and principal hierarchies}

We first recall the notion of a flat \(F\)-manifold, following
\cite{manin2005f,arsie2018flat,alcolado2017extended}. A flat
\(F\)-manifold is a quadruple
$
(\mathcal N,\nabla,\circ,e),
$
where \(\mathcal N\) is an analytic manifold, \(\nabla\) is a flat
torsion-free connection on \(T\mathcal N\), \(\circ\) is a commutative
and associative multiplication on the tangent spaces, and \(e\) is a
unit vector field satisfying
$
\nabla e=0.
$
The multiplication is required to admit a vector potential
\(\mathcal V\), in the sense that
$
X\circ Y=[X,[Y,\mathcal V]]
$
for any \(\nabla\)-flat vector fields \(X\) and \(Y\).

An important class of flat \(F\)-manifolds is obtained from solutions of
the open WDVV equations. Let \(N\) be a Frobenius manifold with flat
coordinates
$
\mathbf t=(t^1,\ldots,t^n),
$
flat metric \(\eta\), prepotential \(F(\mathbf t)\), and structure
constants \(c_{\alpha\beta}^{\delta}\). A function
$
F^{o}=F^{o}(\mathbf t,s)
$
satisfies the open WDVV equations if
\[
c_{\alpha\beta}^{\delta}
\frac{\partial^2 F^{o}}
{\partial t^{\delta}\partial t^{\gamma}}
+
\frac{\partial^2 F^{o}}
{\partial t^{\alpha}\partial t^{\beta}}
\frac{\partial^2 F^{o}}
{\partial t^{\gamma}\partial s}
=
c_{\beta\gamma}^{\delta}
\frac{\partial^2 F^{o}}
{\partial t^{\delta}\partial t^{\alpha}}
+
\frac{\partial^2 F^{o}}
{\partial t^{\beta}\partial t^{\gamma}}
\frac{\partial^2 F^{o}}
{\partial t^{\alpha}\partial s},
\]
and
\[
c_{\alpha\beta}^{\delta}
\frac{\partial^2 F^{o}}
{\partial t^{\delta}\partial s}
+
\frac{\partial^2 F^{o}}
{\partial t^{\alpha}\partial t^{\beta}}
\frac{\partial^2 F^{o}}
{\partial s^2}
=
\frac{\partial^2 F^{o}}
{\partial t^{\alpha}\partial s}
\frac{\partial^2 F^{o}}
{\partial t^{\beta}\partial s},
\]
where summation over repeated indices is understood.

As observed by Rossi, these equations describe rank-one extensions
of Frobenius manifolds  through compatible flat connections and
associative multiplications. More precisely, let \(U\subset\mathbb C\)
be a domain with coordinate \(s\). The extended multiplication on
\(N\times U\), equipped with the product flat connection, admits
the vector potential
\begin{equation}\label{vecpot2}
	\mathcal V
	=
	\frac{\partial F(\mathbf t)}{\partial t^{\alpha}}
	\eta^{\alpha\beta}
	\frac{\partial}{\partial t^{\beta}}
	+
	F^{o}(\mathbf t,s)\frac{\partial}{\partial s}.
\end{equation}
If \(e=\partial/\partial t^1\) and the standard normalization
\[
\frac{\partial^2 F^o}
{\partial t^1\partial t^\alpha}=0,
\qquad
\frac{\partial^2 F^o}
{\partial t^1\partial s}=1
\]
is imposed, the extended multiplication has the flat unity
\(\partial/\partial t^1\), and hence defines a flat \(F\)-manifold.
The same vector-potential construction also applies to generalized
Frobenius manifolds with non-flat unity, although the resulting
unity need not be flat.

We next recall the principal hierarchy associated with a flat
\(F\)-manifold, following
\cite{arsie2018flat,basalaev2019open}. Let \(\mathcal N\) have
dimension \(d\), with flat coordinates
$
\mathbf y=(y^1,\ldots,y^d).
$
A calibration is a collection of vector fields
\[
\bigl\{\Theta_{y^a,p}
\mid a=1,\ldots,d,\ p\geq0\bigr\},
\]
satisfying
\begin{equation}\label{recur}
	\nabla_X\Theta_{y^a,p+1}
	=
	\Theta_{y^a,p}\circ X
\end{equation}
for every vector field \(X\) on \(\mathcal N\), together with the
initial conditions
$
\Theta_{y^a,0}
=
\frac{\partial}{\partial y^a}.
$
Writing
\[
\Theta_{y^b,p}
=
\Theta_{y^b,p}^{c}
\frac{\partial}{\partial y^c},
\qquad
\frac{\partial}{\partial y^c}
\circ
\frac{\partial}{\partial y^d}
=
c_{cd}^{a}
\frac{\partial}{\partial y^a},
\]
the associated principal hierarchy is the system of flows on the
loop space of \(\mathcal N\) given by
\begin{equation}\label{fpridef}
	\frac{\partial y^a(x)}
	{\partial T^{y^b,p}}
	=
	c_{cd}^{a}
	\Theta_{y^b,p}^{c}(\mathbf y(x))
	\frac{\partial y^d(x)}{\partial x}.
\end{equation}

\subsection{Open extensions of generalized Frobenius structures}

We now prove the first part of Theorem \ref{openext}, namely, the existence of the corresponding solution of the open WDVV equations. It suffices to show that the function
$
f(\mathbf t,s)
:=
\lambda\bigl(\mathbf t,z(\mathbf t,s)\bigr),
$
defined in Theorem \ref{openext}, satisfies Alcolado's criterion \eqref{omegaequ}.

For brevity, write
$
\lambda:=\lambda\bigl(\mathbf t,z(\mathbf t,s)\bigr).
$
By the definition of \(z(\mathbf t,s)\), we have
$
\Phi_{0} \bigl(z(\mathbf t,s)\bigr)=s.
$
Hence
$
\frac{\partial \Phi_{0}}{\partial z}
\frac{\partial z}{\partial s}
=
1
$
and
\begin{equation}\label{openlem1}
	\partial_s\lambda
	=
	\frac{\partial\lambda}{\partial z}
	\frac{\partial z}{\partial s}
	=
	\frac{\partial\lambda/\partial z}
	{\partial\Phi_{0}/\partial z}=\frac{d\lambda}{\phi_0}.
\end{equation}
For flat vector fields \(\partial_\alpha\) and \(\partial_\beta\) with respect to \(\eta_{\phi_0}\), we have
$
\nabla_{\partial_\alpha}\partial_\beta=0.
$
Combining \eqref{openlem1} and \eqref{LC-product}, we obtain
\[
\partial_\alpha\partial_\beta\lambda
=
\partial_s
\left(
\frac{
	\partial_\alpha\lambda\,
	\partial_\beta\lambda
	-
	(\partial_\alpha\circ\partial_\beta)\lambda
}{
	\partial_s\lambda
}
\right),
\]
which is precisely \eqref{omegaequ} for
$
f(\mathbf t,s)
:=
\lambda\bigl(\mathbf t,z(\mathbf t,s)\bigr).
$
Therefore, by the criterion recalled above, \(f\) locally determines a solution \(F^o(\mathbf t,s)\) of the open WDVV equations associated with
$
(H,\eta_{\phi_0},\circ,e).
$
This proves the first part of Theorem \ref{openext}.
\subsection{Principal hierarchy of the open extension}

We now construct the principal hierarchy of the rank-one extension and complete the proof of Theorem \ref{openext}. 

We first enlarge the index set by setting
$
\mathcal I^{o}
:=
\mathcal I
\sqcup
\bigl\{(s,p)\mid p\geq0\bigr\}.
$
For each \(A\in\mathcal I^{o}\), we define a vector field \(\Theta_A\) on \(H\times U\) as follows. For \(I\in\mathcal I\), set
$
\Theta_I
=
\xi_I+\Phi_{I^-}(s)\partial_s,
$
while for \(A=(s,p)\), \(p\geq0\), set
$
\Theta_{(s,p)}
=
\frac{\lambda^p}{p!}\partial_s.
$

Let \(\nabla^{\mathrm{ext}}\) denote the flat connection on \(H\times U\) obtained by extending \(\nabla\) trivially in the \(s\)-direction, namely,
\[
\nabla^{\mathrm{ext}}_X Y=\nabla_XY,
\qquad
\nabla^{\mathrm{ext}}_X\partial_s
=
\nabla^{\mathrm{ext}}_{\partial_s}X
=
\nabla^{\mathrm{ext}}_{\partial_s}\partial_s
=
0,
\]
for vector fields \(X,Y\) on \(H\) viewed as independent of \(s\).
For \(A\in\mathcal I^o\), we define its successor \(A^+\) by
\[
A^+
=
\begin{cases}
	I^+, & A=I\in\mathcal I,\\[2mm]
	(s,p+1), & A=(s,p).
\end{cases}
\]
We now show that
\begin{equation}\label{open-calibration}
	\nabla^{\mathrm{ext}}_{\widetilde X}\Theta_{A^+}
	=
	\widetilde X\star\Theta_A,
	\qquad
	A\in\mathcal I^o,
\end{equation}
for every vector field \(\widetilde X\) on \(H\times U\).
Here the multiplication \(\star\) on \(H\times U\) is given by
\begin{align}
	X\star Y
	={}&
	X\circ Y
	+
	\frac{
		X(\lambda)Y(\lambda)
		-
		(X\circ Y)(\lambda)
	}{
		\partial_s\lambda
	}
	\partial_s,
	\label{open-mult1}
	\\
	X\star\partial_s
	={}&
	X(\lambda)\partial_s,
	\qquad
	\partial_s\star\partial_s
	=
	\partial_s\lambda\,\partial_s,
	\label{open-mult2}
\end{align}
where \(X,Y\) are vector fields on \(H\).

We first consider the case \(A=I\in\mathcal I\). Write
$
\widetilde X=X+X^s\partial_s,
$
where \(X\) is tangent to \(H\). It follows from \eqref{eq:Q-primary} -- \eqref{xi-rec} that
\begin{align}
	\xi_I(\lambda)
	&=
	\partial_s\Phi_I(s)
	-
	\Phi_{I^-}(s)\partial_s\lambda,
	\label{open-xi1}
	\\
	(X\circ\xi_I)(\lambda)
	&=
	X(\lambda)\partial_s\Phi_I(s)
	-
	\partial_s\lambda\,X\bigl(\Phi_I(s)\bigr).
	\label{open-xi2}
\end{align}
Since
$
\Theta_{I^+}
=
\xi_{I^+}+\Phi_I(s)\partial_s,
$
one has
\[
\nabla^{\mathrm{ext}}_{\widetilde X}\Theta_{I^+}
=
\nabla_X\xi_{I^+}
+
\left(
X\bigl(\Phi_I(s)\bigr)
+
X^s\partial_s\Phi_I(s)
\right)\partial_s.
\]
Using the recursion relation
\eqref{xi-rec}, 
the component of $\nabla^{\mathrm{ext}}_{\widetilde X}\Theta_{I^+}$ tangent to \(H\) is therefore \(X\circ\xi_I\).

On the other hand, by \eqref{open-mult1}--\eqref{open-mult2},
\begin{align*}
	\widetilde X\star\Theta_I
	={}&
	\left(X+X^s\partial_s\right)
	\star
	\left(\xi_I+\Phi_{I^-}(s)\partial_s\right)
	\\
	={}&
	X\circ\xi_I
	+
	\Bigg[
	\frac{
		X(\lambda)\xi_I(\lambda)
		-
		(X\circ\xi_I)(\lambda)
	}{
		\partial_s\lambda
	}
	+
	\Phi_{I^-}(s)X(\lambda)
	+
	X^s\xi_I(\lambda)
	+
	X^s\Phi_{I^-}(s)\partial_s\lambda
	\Bigg]\partial_s.
\end{align*}
By \eqref{open-xi1} and \eqref{open-xi2},
\begin{align*}
	&
	\frac{
		X(\lambda)\xi_I(\lambda)
		-
		(X\circ\xi_I)(\lambda)
	}{
		\partial_s\lambda
	}
	+
	\Phi_{I^-}(s)X(\lambda)
	\\
	\qquad =&
	\frac{
		X(\lambda)
		\left(
		\partial_s\Phi_I(s)
		-
		\Phi_{I^-}(s)\partial_s\lambda
		\right)
		-
		X(\lambda)\partial_s\Phi_I(s)
		+
		\partial_s\lambda\,X\bigl(\Phi_I(s)\bigr)
	}{
		\partial_s\lambda
	}
	+
	\Phi_{I^-}(s)X(\lambda)
	\\
\qquad =	&
	X\bigl(\Phi_I(s)\bigr),
\end{align*}
while
\begin{align*}
	X^s\xi_I(\lambda)
	+
	X^s\Phi_{I^-}(s)\partial_s\lambda
	&=
	X^s
	\left(
	\partial_s\Phi_I(s)
	-
	\Phi_{I^-}(s)\partial_s\lambda
	\right)
	+
	X^s\Phi_{I^-}(s)\partial_s\lambda
	\\
	&=
	X^s\partial_s\Phi_I(s).
\end{align*}
Therefore,
\[
\widetilde X\star\Theta_I
=
X\circ\xi_I
+
\left(
X\bigl(\Phi_I(s)\bigr)
+
X^s\partial_s\Phi_I(s)
\right)\partial_s
=
\nabla^{\mathrm{ext}}_{\widetilde X}\Theta_{I^+},
\]
which proves \eqref{open-calibration} for \(A=I\in\mathcal I\).

We next consider the case \(A=(s,p)\), \(p\geq0\). By definition,
$
\Theta_{(s,p+1)}
=
\frac{\lambda^{p+1}}{(p+1)!}\partial_s.
$
For
$
\widetilde X=X+X^s\partial_s,
$
we have
\begin{align*}
	\nabla^{\mathrm{ext}}_{\widetilde X}\Theta_{(s,p+1)}
	&=
	\widetilde X
	\left(
	\frac{\lambda^{p+1}}{(p+1)!}
	\right)\partial_s
	\\
	&=
	\frac{\lambda^p}{p!}
	\left(
	X(\lambda)+X^s\partial_s\lambda
	\right)\partial_s.
\end{align*}
On the other hand, using \eqref{open-mult2},
\begin{align*}
	\widetilde X\star\Theta_{(s,p)}
	&=
	\left(
	X+X^s\partial_s
	\right)
	\star
	\frac{\lambda^p}{p!}\partial_s
	\\
	&=
	\frac{\lambda^p}{p!}
	\left(
	X(\lambda)+X^s\partial_s\lambda
	\right)\partial_s.
\end{align*}
Therefore,
\[
\nabla^{\mathrm{ext}}_{\widetilde X}\Theta_{(s,p+1)}
=
\widetilde X\star\Theta_{(s,p)},
\]
which proves \eqref{open-calibration} for \(A=(s,p)\). 

For \(p=0\), we have
$
\Theta_{(s,0)}=\partial_s.
$
Moreover, if \(I\in\mathcal I\) is an initial index corresponding to a Dubrovin primary differential, then \(\Phi_{I^-}=0\), and hence
$
\Theta_I=\xi_I
$
is the flat vector field corresponding to that primary differential, namely, the coordinate vector field associated with the corresponding flat coordinate on \(H\). 
Together with \eqref{open-calibration}, this shows that the family \(\{\Theta_A\}_{A\in\mathcal I^o}\) gives an extension of the calibration to the rank-one extension.

We now derive the explicit form of the corresponding principal hierarchy. Let
$
(v(x),s(x))\in L(H\times U),
$
and regard
$
\partial_x
=
v_x+s_x\partial_s
$
as a tangent vector to \(H\times U\). By the definition of the principal hierarchy recalled above, the flow associated with \(A\in\mathcal I^o\) is
\[
\frac{\partial (v,s)}{\partial T_A}
=
\nabla^{\mathrm{ext}}_{\partial_x}\Theta_{A^+}
=
\partial_x\star\Theta_A.
\]

We first consider \(A=I\in\mathcal I\). Since
$
\Theta_{I^+}
=
\xi_{I^+}+\Phi_I(s)\partial_s,
$
we have
\[
\nabla^{\mathrm{ext}}_{\partial_x}\Theta_{I^+}
=
\nabla_{v_x}\xi_{I^+}
+
\partial_x\Phi_I(s)\partial_s.
\]
By the recursion relation,
$
\nabla_{v_x}\xi_{I^+}
=
v_x\circ\xi_I,
$
and 
$$
(X\circ\xi_I)(\lambda)\phi_0
=
\{\Phi_I,\lambda\}_X
$$
with \(X=v_x\), we obtain
\begin{equation}
	\frac{\partial\lambda}{\partial T_I}\phi_0
	=
	\{\Phi_I,\lambda\},
	\qquad
	\frac{\partial s}{\partial T_I}
	=
	\partial_x\Phi_I(s),
	\qquad
	I\in\mathcal I.
\end{equation}
Here \(\lambda=\lambda(v,z)\) denotes the Hurwitz function, whose evolution is induced by the \(H\)-component of each flow, whereas \(\lambda(s)=\lambda(v,z(v,s))\) denotes its evaluation at the point determined by \(s\).

For \(A=(s,p)\), \(p\geq0\), we have
$
\Theta_{(s,p+1)}
=
\frac{\lambda(s)^{p+1}}{(p+1)!}\partial_s.
$
Hence
\[
\nabla^{\mathrm{ext}}_{\partial_x}\Theta_{(s,p+1)}
=
\partial_x
\left(
\frac{\lambda(s)^{p+1}}{(p+1)!}
\right)\partial_s,
\]
and the corresponding flows are
\begin{equation}
	\frac{\partial\lambda}{\partial T_{(s,p)}}
	=
	0,
	\qquad
	\frac{\partial s}{\partial T_{(s,p)}}
	=
	\partial_x
	\left(
	\frac{\lambda(s)^{p+1}}{(p+1)!}
	\right),
	\qquad
	p\geq0.
\end{equation}
Equations \eqref{open-principal-I} and \eqref{open-principal-s} give the extended principal hierarchy of the rank-one extension and complete the proof of Theorem \ref{openext}.

\begin{rem}
	The family \(\{\Theta_A\}_{A\in\mathcal I^o}\) contains the usual calibration associated with the flat coordinate vector fields on \(H\times U\), together with additional recursive families generated by the non-flat unity vector field \(e\) on \(H\).
		All flows in this extended hierarchy commute. Indeed, writing \(w=(v,s)\), the recursion relation \eqref{open-calibration} gives
	\[
	\frac{\partial w}{\partial T_A}
	=
	\partial_x\Theta_{A^+}
	=
	w_x\star\Theta_A.
	\]
	Hence, for any \(A,B\in\mathcal I^o\),
	\[
	\begin{aligned}
		\frac{\partial^2w}{\partial T_B\partial T_A}
		&=
		\partial_x\left(
		\nabla^{\mathrm{ext}}_{w_{T_B}}\Theta_{A^+}
		\right)\\
		&=
		\partial_x\left(
		(w_x\star\Theta_B)\star\Theta_A
		\right)\\
		&=
		\partial_x\left(
		(w_x\star\Theta_A)\star\Theta_B
		\right)
		=
		\frac{\partial^2w}{\partial T_A\partial T_B},
	\end{aligned}
	\]
	where we have used the commutativity and associativity of \(\star\).
\end{rem}

\section{Almost-dual structures and their principal hierarchies}\label{sec-thm:dual-principal}

In this section, we derive an explicit form of the principal hierarchy associated with the almost-dual structure of \(H\), which we denote by \(\widehat H\).

\subsection{Review of almost duality}

We first recall the principal hierarchy associated with the almost duality of a Frobenius manifold. Let \(M\) be a Frobenius manifold and let
\[
\Sigma
=
\left\{
q\in M
\ \middle|\
\det\bigl(g^{\alpha\beta}(q)\bigr)=0
\right\}
\]
be its discriminant, where \(g^{\alpha\beta}\) denotes the intersection form. On
$
\widehat M=M\setminus\Sigma,
$
the Euler vector field \(E\) is invertible with respect to the multiplication \(\circ\), and one defines
\[
u\ast v
=
E^{-1}\circ u\circ v,
\qquad
u,v\in T_q\widehat M.
\]
Together with the intersection form, the multiplication \(\ast\) defines the almost-dual structure of \(M\) \cite{dubrovin2004almost}, whose unit vector field is \(E\), which is in general not flat.

Let
$
\widehat{\mathbf t}
=
(\widehat t^1,\ldots,\widehat t^n)
$
be flat coordinates for the intersection form, and denote by
\(\widehat c_{\alpha\beta}^{\gamma}\) the structure constants of \(\ast\) in these coordinates. A calibration is given by
\[
\widehat\theta_\alpha(\widehat{\mathbf t},\zeta)
=
\sum_{p\geq0}
\widehat\theta_{\alpha,p}(\widehat{\mathbf t})\zeta^p,
\qquad
\alpha=1,\ldots,n,
\]
satisfying
\[
\frac{\partial^2\widehat\theta_{\alpha,p+1}}
{\partial\widehat t^\beta\partial\widehat t^\gamma}
=
\widehat c_{\beta\gamma}^{\epsilon}
\frac{\partial\widehat\theta_{\alpha,p}}
{\partial\widehat t^\epsilon},
\qquad
\widehat\theta_{\alpha,0}
=
g_{\alpha\beta}\widehat t^\beta.
\]
Equivalently, the differentials
\(d\widehat\theta_\alpha(\widehat{\mathbf t},\zeta)\)
are flat sections of the dual connection induced by the deformed flat connection
\[
\widehat\nabla^{\mathrm{def}}_u v
=
\widehat\nabla_u v
+
\zeta\,u\ast v,
\]
where \(\widehat\nabla\) denotes the Levi--Civita connection of the intersection form.

Let \(\{\cdot,\cdot\}_2\) denote the hydrodynamic Poisson bracket associated with the intersection form, which in its flat coordinates is given by
\[
\bigl\{
\widehat t^\alpha(x),
\widehat t^\beta(y)
\bigr\}_2
=
g^{\alpha\beta}\delta'(x-y).
\]
The corresponding principal hierarchy is given by
\[
\frac{\partial\widehat t^\gamma}
{\partial\widehat T^{\alpha,p}}
=
\left\{
\widehat t^\gamma(x),
\int
\widehat\theta_{\alpha,p+1}(\widehat{\mathbf t})\,dx
\right\}_2
=
g^{\gamma\beta}
\partial_x
\left(
\frac{\partial\widehat\theta_{\alpha,p+1}}
{\partial\widehat t^\beta}
\right),
\qquad
p\geq0.
\]
\subsection{Dual formulation of the almost-dual structure}

As in the case of \(H\), we derive dual formulations of the intersection form and its Levi--Civita connection on \(\widehat H\).

Under the identification \(T_u\widehat H\simeq\mathcal D_u\), where a vector field \(X\) corresponds to the differential \(X(\lambda)\phi_0\), the intersection form is the pullback of the bilinear form
\begin{equation}\label{dual-residue-pairing}
	(\phi_1,\phi_2)
	=
	\sum_{d\lambda=0}
	\operatorname{Res}
	\frac{\phi_1\phi_2}{\lambda\,d\lambda},
	\qquad
	\phi_1,\phi_2\in\mathcal D_u.
\end{equation}
We first give a dual representation of \eqref{dual-residue-pairing}. Let \(\gamma\subset C_u\) be a contour, and choose a path from \(\infty_0\) to an initial point \(P_*\in\gamma\). As in the definition of \(\eta^\sharp_\gamma\), this path determines the corresponding branch of \(K(z,w)\) along \(\gamma\) by analytic continuation from the standard determination at \(\infty_0\). For \(f\in\mathcal H_\gamma\), define
\begin{equation}\label{dual-gsharp}
	g^\sharp_\gamma(f)
	:=
	-C_\gamma(\lambda f)\,d\lambda
	+
	\lambda\,\mathcal C_\gamma(f\,d\lambda)
	+
	\frac{d\lambda}{n_0}
	\frac{1}{2\pi\mathrm i}
	\oint_\gamma f\,d\lambda.
\end{equation}
The jumps of the first two terms cancel, so \(g^\sharp_\gamma(f)\) extends across every smooth branch of \(\gamma\). Moreover, since
\[
\operatorname{Res}_{\infty_0}
\mathcal C_\gamma(f\,d\lambda)
=
\frac{1}{2\pi\mathrm i}
\oint_\gamma f\,d\lambda,
\]
the last term in \eqref{dual-gsharp} cancels the leading pole at \(\infty_0\). The resulting differential satisfies the admissibility conditions defining \(\mathcal D_u\), and hence
$
g^\sharp_\gamma(f)\in\mathcal D_u.
$

\begin{lem}\label{dual-gsharp-pairing}
	For any \(f\in\mathcal H_\gamma\) and \(\phi\in\mathcal D_u\), one has
	\[
	\bigl(g^\sharp_\gamma(f),\phi\bigr)
	=
	\langle f,\phi\rangle_\gamma,
	\]
	where the determination of \(K\) along \(\gamma\) is obtained by analytic continuation along the same path from \(\infty_0\) to \(P_*\) used in the definition of \(g^\sharp_\gamma\).
\end{lem}

\begin{proof}
	Comparing \eqref{dual-gsharp} with \eqref{dulmet}, we have
	\[
	g^\sharp_\gamma(f)-\lambda\,\eta^\sharp_\gamma(f)
	=
	\left[
	\lambda C_\gamma(f)
	-
	C_\gamma(\lambda f)
	+
	\frac{1}{n_0}
	\frac{1}{2\pi\mathrm i}
	\oint_\gamma f\,d\lambda
	\right]d\lambda.
	\]
	The coefficient on the right-hand side is holomorphic at the critical points of \(\lambda\). Since the critical values of \(\lambda\) are nonzero on \(\widehat H\), this term does not contribute to the residue pairing \eqref{dual-residue-pairing}. Hence
	\begin{align*}
		\bigl(g^\sharp_\gamma(f),\phi\bigr)
		&=
		\sum_{d\lambda=0}
		\operatorname{Res}
		\frac{g^\sharp_\gamma(f)\phi}{\lambda\,d\lambda}
		\\
		&=
		\sum_{d\lambda=0}
		\operatorname{Res}
		\frac{\eta^\sharp_\gamma(f)\phi}{d\lambda}
		\\
		&=
		\bigl\langle\eta^\sharp_\gamma(f),\phi\bigr\rangle.
	\end{align*}
	By Corollary~\ref{dulmecimm},
	$
	\bigl\langle\eta^\sharp_\gamma(f),\phi\bigr\rangle
	=
	\langle f,\phi\rangle_\gamma,
	$
	which proves the claim.
\end{proof}

We next derive an explicit expression for the Levi--Civita connection of the intersection form. Set
\[
\lambda^{-1}\mathcal D
:=
\left\{
\frac{\omega}{\lambda}
\ \middle|
\omega\in\mathcal D
\right\}.
\]

\begin{lem}\label{dual-projection-lemma}
	For any \(\omega\in\mathcal D\) and \(X\in\operatorname{Vect}(\widehat H)\), one has
	\[
	\mathcal{P}_+
	\left(
	\frac{X|_\lambda\omega}{\lambda}
	\right)
	\in
	\lambda^{-1}\mathcal D,
	\]
	where \(\mathcal{P}_+\) denotes the positive projection introduced in Subsection~\ref{subsec-Levi}.
\end{lem}

\begin{proof}
	Since \(\lambda\) does not vanish at the ramification points on \(\widehat H\), division by \(\lambda\) introduces no additional singularities there. By the Cauchy decomposition,
	\[
	\lambda
	\mathcal P_+
	\left(
	\frac{X|_\lambda\omega}{\lambda}
	\right)
	=
	X|_\lambda\omega
	-
	\lambda
	\mathcal P_-
	\left(
	\frac{X|_\lambda\omega}{\lambda}
	\right).
	\]
	The left-hand side is holomorphic near the ramification points, while the right-hand side is defined on their complement. These expressions agree near the contours and therefore determine a globally defined differential.
	
	We verify its behavior at \(\infty_0\). Since \(X|_\lambda\omega\) has at most double poles with vanishing residues at the ramification points, and \(\lambda\) is a nonzero even function of the local coordinate \(\sqrt{\lambda-u_a}\), we have
$
	\operatorname{Res}_{P_a}
	\frac{X|_\lambda\omega}{\lambda}=0.
$
	Consequently, the negative projection has vanishing residue at \(\infty_0\) and is therefore holomorphic there. It follows that multiplication by \(\lambda\) introduces no forbidden leading term proportional to \(d\lambda\) at \(\infty_0\).
	
	At the remaining points, the resulting differential has only the prescribed singularities and monodromies of an element of \(\mathcal D\). Hence
$
	\lambda
	\mathcal P_+
	\left(
	\frac{X|_\lambda\omega}{\lambda}
	\right)
	\in\mathcal D,
$
	and therefore
$
	\mathcal P_+
	\left(
	\frac{X|_\lambda\omega}{\lambda}
	\right)
	\in\lambda^{-1}\mathcal D.
$
\end{proof}

\begin{cor}\label{cor:dual-LC}
Under the identification of \(T\widehat H\) with \(\lambda^{-1}\mathcal D\), where \(X\) corresponds to \(X(\log\lambda)\phi_0\),
	the Levi--Civita connection \(\widehat\nabla\) of the intersection form
	\((\, ,\,)_{\phi_0}\) is characterized by
	\begin{align}
		\bigl(\widehat\nabla_{\partial_1}\partial_2\bigr)
		(\log\lambda)\phi_0
		&=
		\mathcal{P}_+
		\left(
		\left.\partial_1\right|_\lambda
		\bigl(\partial_2(\log\lambda)\phi_0\bigr)
		\right)
		\nonumber\\
		&=
		\partial_1\partial_2(\log\lambda)\phi_0
		-
		\mathcal{P}_+
		\left[
		d\left(
		\frac{
			\partial_1(\log\lambda)\,
			\partial_2(\log\lambda)
		}{
			d\log\lambda
		}
		\phi_0
		\right)
		\right].
		\label{intconn}
	\end{align}
\end{cor}
\begin{proof}
	By Lemma~\ref{dual-projection-lemma}, the right-hand side of the first identity in \eqref{intconn} belongs to \(\lambda^{-1}\mathcal D\), and hence defines a connection on \(T\widehat H\). The second identity follows by the same computation as \eqref{LC-pre}.
	
	The symmetry of the correction term implies that this connection is torsion-free. To verify metric compatibility, observe that the intersection form can be written as
	\[
	g_{\phi_0}(X,Y)
	=
	\sum_{d\lambda=0}
	\operatorname{Res}
	\frac{
		X(\log\lambda)\phi_0\,
		Y(\log\lambda)\phi_0
	}{
		d\log\lambda
	}.
	\]
	Since \(d\log\lambda\) is fixed under \(\lambda\)-horizontal differentiation, the same Cauchy decomposition and residue argument as in the proof of Proposition~\ref{prop:LC} shows that the connection is compatible with \(g_{\phi_0}\). Hence it is the Levi--Civita connection.
\end{proof}
\begin{cor}\label{dual-LC-product-proj}
	The positive projection satisfies
	\begin{align*}
		\mathcal P_+\left(
		\frac{\partial_1(\log\lambda)\partial_2(\log\lambda)\phi_0}{d\log\lambda}
		\right)
		&=
		\mathcal P_+\left(
		\frac{\partial_1(\log\lambda)\partial_2(\log\lambda)\phi_0}{d\log\lambda}
		-
		\frac{(\partial_1*\partial_2)(\log\lambda)\phi_0}{d\log\lambda}
		\right)
		\\
		&=
		\frac{\partial_1(\log\lambda)\partial_2(\log\lambda)\phi_0}{d\log\lambda}
		-
		\frac{(\partial_1*\partial_2)(\log\lambda)\phi_0}{d\log\lambda}.
	\end{align*}
	Consequently,
	\begin{equation}\label{intconn2}
		\bigl(\widehat\nabla_{\partial_1}\partial_2\bigr)
		(\log\lambda)\phi_0
		=
		\partial_1\partial_2(\log\lambda)\phi_0
		-
		d\left(
		\frac{\partial_1(\log\lambda)\partial_2(\log\lambda)\phi_0}{d\log\lambda}
		-
		\frac{(\partial_1*\partial_2)(\log\lambda)\phi_0}{d\log\lambda}
		\right).
	\end{equation}
\end{cor}

\begin{proof}
	Observe that
$
	\frac{(\partial_1*\partial_2)(\log\lambda)\phi_0}{d\log\lambda}
	=
	\frac{(\partial_1*\partial_2)(\lambda)\phi_0}{d\lambda}.
$
	As in the case of the original Frobenius structure, this function is holomorphic in the exterior domain, has vanishing \(\alpha\)-increments, and vanishes at \(\infty_0\). Hence its positive projection is zero.
	
	On the other hand, at each ramification point \(P_a\), the definition of the almost-dual multiplication gives
$
	\frac{
		\partial_1(\lambda)(P_a)\,
		\partial_2(\lambda)(P_a)
	}{
		\lambda(P_a)
	}
	=
	(\partial_1*\partial_2)(\lambda)(P_a).
$
	Since \(\lambda(P_a)\ne0\) on \(\widehat H\), the difference
	is holomorphic at every ramification point. It is therefore fixed by \(\mathcal P_+\). This proves the projection identity.
	
	Finally, since the contours surrounding the ramification points are contractible, Lemma~2.2 gives \(\mathcal P_+(dh)=d\mathcal P_+(h)\) for functions \(h\) defined near these contours. Combining this identity with \eqref{intconn} yields \eqref{intconn2}.
\end{proof}
We shall use this Corollary below to construct the open extension of \(\widehat H\).
\begin{lem}\label{intdulconn}
	Fix an extended primary differential \(\phi_0\), and denote by
	\(g_{\phi_0}\) the  intersection form defined by \(\phi_0\).
	Let \(\omega\) be a local section of \(T^*\widehat H\) represented by \(f\in\mathcal H_\gamma\), with the same choice of initial point and determination of \(K\) as in the definition of \(g^\sharp_\gamma\). We assume that the initial point \(P_*\) varies with the parameters while \(\lambda(P_*)\) remains constant. Then, for any
	\(\partial\in\operatorname{Vect}(\widehat H)\),
	\begin{equation}\label{intdulconnformula}
		\begin{aligned}
			\bigl(g_{\phi_0}^\sharp(\widehat\nabla_\partial^*\omega)\bigr)(\lambda)\phi_0
			={}&
			\bigl\{C_\gamma(\lambda f),\lambda\bigr\}_\partial
			-
			\lambda\,
			\mathcal C_\gamma
			\bigl(\{f,\lambda\}_\partial\bigr)
			-
			\frac{d\lambda}{n_0}
			\frac{1}{2\pi\mathrm i}
			\oint_\gamma
			\{f,\lambda\}_\partial .
		\end{aligned}
	\end{equation}
	Here
$
	\{f,\lambda\}_\partial
	=
	df\,\partial(\lambda)
	-
	\partial(f)\,d\lambda,
$
	and \(\partial(f)\) denotes differentiation along the \(\phi_0\)-horizontal lift.
\end{lem}

\begin{proof}
	By the compatibility of \(\widehat\nabla\) with the intersection form and Lemma~\ref{dual-gsharp-pairing}, one has
$
	g_{\phi_0}^\sharp(\widehat\nabla_\partial^*\omega)
	=
	\widehat\nabla_\partial
	\bigl(g_{\phi_0}^\sharp(\omega)\bigr).
$
	Using \eqref{intconn} and the representation
	\(g_{\phi_0}^\sharp(\omega)(\lambda)\phi_0=g^\sharp_\gamma(f)\), we obtain
	\[
	\bigl(g_{\phi_0}^\sharp(\widehat\nabla_\partial^*\omega)\bigr)(\lambda)\phi_0
	=
	\lambda\mathcal P_+
	\left(
	\left.\partial\right|_\lambda
	\frac{g^\sharp_\gamma(f)}{\lambda}
	\right).
	\]
	Substituting \eqref{dual-gsharp} gives
	\begin{align*}
		\bigl(g_{\phi_0}^\sharp(\widehat\nabla_\partial^*\omega)\bigr)(\lambda)\phi_0
		=
		\lambda\mathcal P_+\Biggl(
		&
		-\left(
		\left.\partial\right|_\lambda
		C_\gamma(\lambda f)
		\right)\frac{d\lambda}{\lambda}
		+
		\left.\partial\right|_\lambda
		\mathcal C_\gamma(f\,d\lambda)
+
		\frac{d\lambda}{n_0\lambda}
		\partial\left(
		\frac{1}{2\pi\mathrm i}
		\oint_\gamma f\,d\lambda
		\right)
		\Biggr).
	\end{align*}
	
	As in the proof of the corresponding dual-connection formula for \(\eta_{\phi_0}\), the difference
$
	\left.\partial\right|_\lambda
	\mathcal C_\gamma(f\,d\lambda)
	-
	\mathcal C_\gamma
	\left(
	\left.\partial\right|_\lambda(f\,d\lambda)
	\right)
$
	belongs to the negative part of the Cauchy decomposition and is therefore annihilated by \(\mathcal P_+\).
	Moreover, since the contour is transported with \(\lambda\) fixed,
$
	\partial\left(
	\frac{1}{2\pi\mathrm i}
	\oint_\gamma f\,d\lambda
	\right)
	=
	-\frac{1}{2\pi\mathrm i}
	\oint_\gamma\{f,\lambda\}_\partial.
$
	Using
$
	\left.\partial\right|_\lambda(f\,d\lambda)
	=
	-\{f,\lambda\}_\partial,
$ 
	we obtain
	\begin{align*}
		\bigl(g_{\phi_0}^\sharp(\widehat\nabla_\partial^*\omega)\bigr)(\lambda)\phi_0
		=
		\lambda\mathcal P_+\Biggl(
		&
		\frac{1}{\lambda}
		\{C_\gamma(\lambda f),\lambda\}_\partial
		-
		\mathcal C_\gamma
		\bigl(\{f,\lambda\}_\partial\bigr)
-
		\frac{d\lambda}{n_0\lambda}
		\frac{1}{2\pi\mathrm i}
		\oint_\gamma
		\{f,\lambda\}_\partial
		\Biggr).
	\end{align*}
	The differential inside \(\mathcal P_+\) is holomorphic near the ramification points, since both Poisson bracket and Cauchy transform terms are holomorphic there and the critical values of \(\lambda\) are nonzero on \(\widehat H\). It is therefore fixed by \(\mathcal P_+\), yielding \eqref{intdulconnformula}.
\end{proof}

\subsection{Recursion vector fields}\label{almprisec}

We now construct the recursion vector fields of the almost-dual structure \(\widehat H\) using contour representatives of covectors.

Let \(\mathcal S=\lambda^{-1}(\{0,\infty\})\), and choose a smoothly varying regular base point \(P^\ast\in C_u\setminus\mathcal S\) such that \(\lambda(P^\ast)=1\). Fix an initial determination of the Cauchy kernel \(K(z,w)\) at \(z=P^\ast\), together with a branch of \(\log\lambda\) satisfying \(\log\lambda(P^\ast)=0\).
Let \(\mathcal G\) be a collection of closed contours based at \(P^\ast\), each satisfying the zero-winding condition
\[
\operatorname{wind}_\lambda(\gamma)
:=
\frac{1}{2\pi\mathrm i}
\oint_\gamma\frac{d\lambda}{\lambda}
=0,
\qquad \gamma\in\mathcal G.
\]

 For each \(\gamma\in\mathcal G\), write
$
[\gamma]
=
\sum_{i=1}^g
\bigl(a_i[\alpha_i]+b_i[\beta_i]\bigr)
$ in $H_1(C_u,\mathbb Z)$.
By the normalization of the Bergman bidifferential,
\[
dC_\gamma(1)
=
\frac{1}{2\pi\mathrm i}
\sum_{i=1}^g b_i\,dZ_i.
\]
Hence \(dC_\gamma(1)=0\) if and only if all the \(\beta\)-components \(b_i\) vanish.

Set
$
\mathcal G_\beta
=
\{\gamma\in\mathcal G\mid dC_\gamma(1)\neq0\}.
$
We introduce the index set
\[
\widehat{\mathcal I}
=
\{(\gamma,p)\mid\gamma\in\mathcal G,\ p\ge0\}
\sqcup
\{(\gamma,-1)\mid\gamma\in\mathcal G_\beta\}.
\]
For each \(I=(\gamma,p)\in\widehat{\mathcal I}\), define its contour representative by
\[
\gamma_I=
\begin{cases}
	\gamma, & p\ge0,\\[1mm]
	\partial\widehat C_u^\varepsilon, & p=-1,
\end{cases}
\]
where \(\widehat C_u\) is a fundamental polygon associated with the canonical symplectic basis, and \(\partial\widehat C_u^\varepsilon\) denotes a sufficiently small inward deformation of its positively oriented boundary.

For each \(I=(\gamma,p)\in\widehat{\mathcal I}\), define
$
I^+=(\gamma,p+1),
$ and $
I^-=(\gamma,p-1).
$
 We define \(\widehat\Phi_{I^-}\) as a chosen holomorphic branch of
$
C_{\gamma_I}(\lambda\widehat Q_I),
$
and set \(\widehat\phi_I=d\widehat\Phi_I\). In particular, for \(p=0\),
$
\widehat\Phi_{\gamma,-1}=C_\gamma(1)
$
and $
\widehat\phi_{\gamma,-1}=dC_\gamma(1).
$
The latter differential vanishes when \(\gamma\notin\mathcal G_\beta\). For \(p=-1\), the formal predecessor \(\widehat\Phi_{\gamma,-2}\) vanishes.

We next verify that
\begin{equation}\label{eq:hatQ-cauchy}
	\widehat\phi_I
	=
	\mathcal C_{\gamma_I}
	\bigl(\widehat Q_I\,d\lambda\bigr),
	\qquad I\in\widehat{\mathcal I}.
\end{equation}
For \(I=(\gamma,p)\) with \(p\ge0\), we have
$
\gamma_{I^+}=\gamma_I=\gamma
$ and $
d(\lambda\widehat Q_{I^+})
=
\widehat Q_I\,d\lambda.
$
The zero-winding condition ensures that \(\lambda\widehat Q_{I^+}\) is single-valued along \(\gamma\). Moreover, since \(\log\lambda(P^\ast)=0\), one has
$
(\lambda\widehat Q_{I^+})(P^\ast)=0.
$
Thus, integration by parts gives
\[
\begin{aligned}
	\mathcal C_{\gamma_I}
	\bigl(\widehat Q_I\,d\lambda\bigr)
	=
	dC_\gamma(\lambda\widehat Q_{I^+})=
	d\widehat\Phi_I
	=
	\widehat\phi_I.
\end{aligned}
\]

For \(I=(\gamma,-1)\), we instead have
$
\widehat Q_I\,d\lambda
=
dC_\gamma(1)
=
\widehat\phi_I,
$
which is a holomorphic differential. Since
\(\gamma_I=\partial\widehat C_u^\varepsilon\), the Cauchy reproducing formula on the fundamental polygon yields
$
\mathcal C_{\gamma_I}
\bigl(\widehat Q_I\,d\lambda\bigr)
=
\widehat\phi_I.
$
This proves \eqref{eq:hatQ-cauchy} for every \(I\in\widehat{\mathcal I}\).

Moreover, in all cases,
\[
\oint_{\gamma_I}\widehat Q_I\,d\lambda=0.
\]
Indeed, for \(p\ge0\) this follows from the zero-winding condition, while for \(p=-1\) it follows from the holomorphicity of \(\widehat\phi_I\) and the polygon-boundary contour.

Substituting these identities into \eqref{dual-gsharp}, we obtain
\[
g^\sharp_{\gamma_I}(\widehat Q_I)
=
\lambda\widehat\phi_I
-
\widehat\Phi_{I^-}\,d\lambda.
\]
Let \(\widehat\omega_I\) denote the covector represented by \((\gamma_I,\widehat Q_I)\), and define
\[
\widehat\xi_I
:=
g_{\phi_0}^{\sharp}(\widehat\omega_I).
\]
Under the identification of tangent vectors with admissible differentials, we therefore obtain
\begin{equation}\label{eq:hatQ-primary}
	\widehat\xi_I(\lambda)\phi_0
	=
	\lambda\widehat\phi_I
	-
	\widehat\Phi_{I^-}\,d\lambda.
\end{equation}


\begin{lem}\label{lem:dual-xi-Phi}
	For every \(I\in\widehat{\mathcal I}\) and every vector field \(\partial\) on \(\widehat H\), one has
	\begin{equation}\label{eq:dual-xi-Phi}
		(\widehat\nabla_{\partial}\widehat\xi_{I^+})(\lambda)\phi_0
		=
		\{\widehat\Phi_I,\lambda\}_{\partial},
	\end{equation}
	where
$
	\{\widehat\Phi_I,\lambda\}_{\partial}
	=
	d\widehat\Phi_I\,\partial(\lambda)
	-
	\partial(\widehat\Phi_I)\,d\lambda.
$
\end{lem}

\begin{proof}
	By the Levi--Civita formula \eqref{intconn} for the intersection form and \eqref{eq:hatQ-primary},
	\[
	\begin{aligned}
		(\widehat\nabla_{\partial}\widehat\xi_{I^+})(\lambda)\phi_0
		&=
		\lambda \mathcal{P}_+\left(
		\partial|_\lambda
		\frac{\widehat\xi_{I^+}(\lambda)\phi_0}{\lambda}
		\right)\\
		&=
		\lambda\mathcal{P}_+\left(
		\partial|_\lambda
		\left(
		\widehat\phi_{I^+}
		-
		\widehat\Phi_I\,d\log\lambda
		\right)
		\right).
	\end{aligned}
	\]	
We first observe that
$
\mathcal{P}_+\bigl(\partial|_\lambda\widehat\phi_{I^+}\bigr)=0,
$
since \(\partial|_\lambda\widehat\phi_{I^+}\) is a normalized meromorphic differential with possible poles only at the critical points of \(\lambda\). By using
\[
-(\partial|_\lambda\widehat\Phi_I)\,d\lambda
=
d\widehat\Phi_I\,\partial(\lambda)
-
\partial(\widehat\Phi_I)\,d\lambda,
\]
	one has
	\[
	\begin{aligned}
		(\widehat\nabla_{\partial}\widehat\xi_{I^+})(\lambda)\phi_0
		&=
		\lambda \mathcal{P}_+\left(
		\frac{\{\widehat\Phi_I,\lambda\}_{\partial}}{\lambda}
		\right).
	\end{aligned}
	\]
	Since \(\frac{\{\widehat\Phi_I,\lambda\}_{\partial}}{\lambda}\) is holomorphic at the ramification points, the positive projection leaves it unchanged. Thus
$
	(\widehat\nabla_{\partial}\widehat\xi_{I^+})(\lambda)\phi_0
	=
	\{\widehat\Phi_I,\lambda\}_{\partial},
$
	as required.
\end{proof}
\begin{lem}\label{cor:dual-Q-vector-recursion}
	For every \(I\in\widehat{\mathcal I}\) and
	every vector field \(\partial\) on \(\hat{H}\), one has
	\begin{equation}\label{dual-xi-rec}
		\widehat\nabla_{\partial}\widehat\xi_{I^+}
		=
		\partial*\widehat\xi_I.
	\end{equation}
\end{lem}
\begin{proof}
	Since
$
	g_{\phi_0}(X,Y)
	=
	\eta_{\phi_0}(E^{-1}\circ X,Y),
$
	we have
$
	\widehat\xi_I
	=
	E\circ\eta_{\phi_0}^{\sharp}(\widehat\omega_I).
$
	Consequently,
$$
	\partial_1*\widehat\xi_I
	=
	\partial_1\circ
	\eta_{\phi_0}^{\sharp}(\widehat\omega_I).
$$
	By Lemma~\ref{lem:dual-xi-Phi} and the dual representation of the original multiplication, it therefore suffices to show that
	\begin{equation}\label{eq:dual-rec-cauchy-proof}
		\left\langle
		\mathcal M_{\partial_1,\gamma_I}(\widehat Q_I),
		\partial_2(\lambda)\phi_0
		\right\rangle
		=
		\left\langle
		\{\widehat\Phi_I,\lambda\}_{\partial_1},
		\partial_2(\lambda)\phi_0
		\right\rangle
	\end{equation}
	for every vector field \(\partial_2\) on \(\widehat H\).
	
	Recall that
$
	\mathcal C_{\gamma_I}
	(\widehat Q_I\,d\lambda)
	=
	\widehat\phi_I
	=
	d\widehat\Phi_I.
$
	By the definition of \(\mathcal M_{\partial,\gamma}\),
	\[
	\begin{aligned}
		\mathcal M_{\partial_1,\gamma_I}(\widehat Q_I)
		&=
		-C_{\gamma_I}
		\bigl(\widehat Q_I\partial_1(\lambda)\bigr)
		\,d\lambda
		+
		\partial_1(\lambda)
		\mathcal C_{\gamma_I}
		(\widehat Q_I\,d\lambda)\\
		&=
		-C_{\gamma_I}
		\bigl(\widehat Q_I\partial_1(\lambda)\bigr)
		\,d\lambda
		+
		\partial_1(\lambda)\,d\widehat\Phi_I.
	\end{aligned}
	\]
	Consequently,
	\[
	\mathcal M_{\partial_1,\gamma_I}(\widehat Q_I)
	-
	\{\widehat\Phi_I,\lambda\}_{\partial_1}
	=
	\left(
	\partial_1(\widehat\Phi_I)
	-
	C_{\gamma_I}
	\bigl(\widehat Q_I\partial_1(\lambda)\bigr)
	\right)d\lambda.
	\]
	The coefficient in parentheses is holomorphic at the ramification points of \(\lambda\). Therefore the difference does not contribute to the residue pairing, and \eqref{eq:dual-rec-cauchy-proof} follows. Since the residue pairing is nondegenerate, this proves
$
	\widehat\nabla_{\partial_1}\widehat\xi_{I^+}
	=
	\partial_1*\widehat\xi_I.
$
\end{proof}
\subsection{Recursion vector fields associated with \(E\)}\label{sec-Erec}

Besides the recursion families \(\{\widehat\xi_I\}_{I\in\widehat{\mathcal I}}\) constructed above, we seek an additional recursion family containing the unity vector field \(E\) of the almost-dual multiplication.

First, consider the conformal case, where
$
D_E\phi_0=\nu_0\phi_0.
$
It follows from \eqref{Eurform} that
\[
E(\log\lambda)\phi_0
=
\phi_0-(\nu_0\Phi_0+r_0)\,d\log\lambda.
\]
For any vector field \(X\) on \(\widehat H\), using
$
\partial_X|_\lambda\Phi_0
=
-X(\lambda)\frac{\phi_0}{d\lambda},
$
we obtain
\[
\begin{aligned}
	\partial_X|_\lambda
	\bigl(E(\log\lambda)\phi_0\bigr)
	&=
	\partial_X|_\lambda\phi_0
	-
	\nu_0(\partial_X|_\lambda\Phi_0)\,d\log\lambda\\
	&=
	\partial_X|_\lambda\phi_0
	+
	\nu_0X(\log\lambda)\phi_0.
\end{aligned}
\]
Recall that
$
\mathcal P_+(\partial_X|_\lambda\phi_0)=0
$ and $
\mathcal P_+\bigl(X(\log\lambda)\phi_0\bigr)
=
X(\log\lambda)\phi_0.
$
Applying the  formula \eqref{intconn} yields
\[
\begin{aligned}
	(\widehat\nabla_XE)(\log\lambda)\phi_0
	&=
	\mathcal P_+\left(
	\partial_X|_\lambda
	\bigl(E(\log\lambda)\phi_0\bigr)
	\right)\\
	&=
	\nu_0X(\log\lambda)\phi_0.
\end{aligned}
\]
Consequently,
$
\widehat\nabla_XE=\nu_0X.
$

When \(\nu_0=0\), the identity \(\widehat\nabla_XE=\nu_0X\) shows that \(E\) is \(\widehat\nabla\)-flat. As we shall show in the next subsection, the initial recursion fields span the space of \(\widehat\nabla\)-flat vector fields. Hence \(E\) is a constant linear combination of these fields. Taking the same linear combination of the corresponding recursion families, with their initial indices aligned, gives a recursion family \(\{\widehat\xi_{\mathrm{un},p}\}_{p\ge0}\) satisfying \(\widehat\xi_{\mathrm{un},0}=E\).

Let \(\widehat\Phi_{\mathrm{un},p}^{\mathrm{lin}}\) denote the corresponding constant linear combinations of Abelian integrals, with \(\widehat\Phi_{\mathrm{un},-1}^{\mathrm{lin}}=0\). By linearity, these satisfy the recursion identities \begin{equation}\label{E-dul-recu}
	\widehat\xi_{\mathrm{un},p}(\lambda)\phi_0
	=
	\lambda\,d\widehat\Phi_{\mathrm{un},p}^{\mathrm{lin}}
	-
	\widehat\Phi_{\mathrm{un},p-1}^{\mathrm{lin}}\,d\lambda,
\end{equation}
and
\begin{equation}\label{E-dul-recu-con}
	(\widehat\nabla_X\widehat\xi_{\mathrm{un},p+1})(\lambda)\phi_0
	=
	\{\widehat\Phi_{\mathrm{un},p}^{\mathrm{lin}},\lambda\}_X.
\end{equation} In particular, by \eqref{eq:hatQ-primary} and \eqref{Eurform},
\[
d\widehat\Phi_{\mathrm{un},0}^{\mathrm{lin}}
=
\phi_0-r_0\,d\log\lambda.
\]
Therefore,
\[
\widehat\Phi_{\mathrm{un},0}^{\mathrm{lin}}
=
\Phi_0-r_0\log\lambda+c(u).
\]
Since \(\widehat\xi_{\mathrm{un},0}=E\), the recursion relation gives
\[
\{\widehat\Phi_{\mathrm{un},0}^{\mathrm{lin}},\lambda\}_X
=
(X\ast E)(\lambda)\phi_0
=
X(\lambda)\phi_0.
\]
On the other hand,
\[
\{\widehat\Phi_{\mathrm{un},0}^{\mathrm{lin}},\lambda\}_X
=
X(\lambda)\phi_0-X(c)\,d\lambda.
\]
Hence \(X(c)=0\) for every vector field \(X\), so \(c\in\mathbb C\).

We now define
\[
\widehat\Phi_{\mathrm{un},p}
=
\widehat\Phi_{\mathrm{un},p}^{\mathrm{lin}}
+
r_0\frac{(\log\lambda)^{p+1}}{(p+1)!},
\qquad p\ge0,
\]
and set \(\widehat\Phi_{\mathrm{un},-1}=r_0\). The corrected Abelian integrals satisfy the same recursion identities as (5.11) and (5.12), with \(\widehat\Phi_{\mathrm{un},p}^{\mathrm{lin}}\) replaced by \(\widehat\Phi_{\mathrm{un},p}\), for all \(p\ge0\). In particular,
$
\widehat\Phi_{\mathrm{un},0}=\Phi_0+c$ with $
c\in\mathbb C.
$

Now suppose that \(\nu_0\neq0\). For each \(p\in\mathbb Z\), define
$
\widehat\xi_{\mathrm{un},p}
=
\nu_0^{-p}E.
$
Since \(E\) is the unity vector field of the almost-dual multiplication, we have
\begin{equation}\label{E-recursion-con}
	\begin{aligned}
		\widehat\nabla_X\widehat\xi_{\mathrm{un},p+1}
		&=
		\nu_0^{-(p+1)}\widehat\nabla_XE\\
		&=
		\nu_0^{-p}X\\
		&=
		X*\widehat\xi_{\mathrm{un},p}.
	\end{aligned}
\end{equation}

Thus \(\{\widehat\xi_{\mathrm{un},p}\}_{p\in\mathbb Z}\) forms an additional recursion family containing \(E\).
Moreover, since \(d\Phi_0=\phi_0\) and \(\partial_X|_{\Phi_0}\Phi_0=0\), we obtain
\[
\begin{aligned}
	(\widehat\nabla_X\widehat\xi_{\mathrm{un},p+1})
	(\lambda)\phi_0
	&=
	\nu_0^{-p}X(\lambda)\phi_0\\
	&=
	\{\nu_0^{-p}\Phi_0,\lambda\}_X\\
	&=
	\{\widehat\Phi_{\mathrm{un},p},\lambda\}_X,
\end{aligned}
\]
where
\[
\widehat\Phi_{\mathrm{un},p}
=
\nu_0^{-p}\Phi_0
+
\nu_0^{-(p+1)}r_0,
\qquad p\in\mathbb Z.
\]
Finally, by \eqref{Eurform}, these Abelian integrals satisfy
\[
\widehat\xi_{\mathrm{un},p}(\lambda)\phi_0
=
\lambda\,d\widehat\Phi_{\mathrm{un},p}
-
\widehat\Phi_{\mathrm{un},p-1}\,d\lambda.
\]

For the non-conformal case, where
\(\phi_0=\phi_{(\infty_i,j)_{\log}}\) or
\(\phi_0=\phi_{(q,j)_{\log}}\) with \(j\geq1\), define
$
W_j=-\sum_{s=1}^m a_s\xi_{\infty_s,jn_s},
$
where
\[
a_s=
\begin{cases}
	\dfrac{n_s}{n_0}+\delta_{is},
	& \phi_0=\phi_{(\infty_i,j)_{\log}},\\[6pt]
	\dfrac{n_s}{n_0},
	& \phi_0=\phi_{(q,j)_{\log}}.
\end{cases}
\]
We also write
$
\Phi_{W_j}
=
-\sum_{s=1}^m a_s\Phi_{\infty_s,jn_s}.
$
By \eqref{Eurform}, we have
\begin{equation}\label{nonhomEu}
	E(\log\lambda)\phi_0
	=
	\phi_0-
	\left(
	j\Phi_0+\Phi_{W_j}
	+\frac{\lambda^j}{n_0j!}
	\right)d\log\lambda.
\end{equation}

\begin{lem}\label{scalarrec}
	For every vector field \(X\) on \(\widehat H\), one has
	\[
	\widehat\nabla_XE=jX+X\circ W_j,
	\]
	and
	\[
	\widehat\nabla_X(E\circ W_j)=jX\circ W_j.
	\]
\end{lem}

\begin{proof}
	By \eqref{nonhomEu} and the identity
	\[
	\{\Phi_{W_j},\lambda\}_X
	=
	(X\circ W_j)(\lambda)\phi_0,
	\]
	which follows from \eqref{cor:Q-vector-recursion} and
	\eqref{xi-rec}, we obtain
	\[
	\begin{aligned}
		\partial_X|_\lambda
		\bigl(E(\log\lambda)\phi_0\bigr)
		&=
		\partial_X|_\lambda\phi_0
		+jX(\log\lambda)\phi_0+
		(X\circ W_j)(\log\lambda)\phi_0.
	\end{aligned}
	\]
	Here we used the fact that
	\(\lambda^j/(n_0j!)\) depends only on \(\lambda\).
	Applying the Levi--Civita formula
	\eqref{intconn} and using
$
	\mathcal P_+(\partial_X|_\lambda\phi_0)=0,
$
	we obtain
$
	\widehat\nabla_XE=jX+X\circ W_j.
$
	
	For the second identity, let
	\(I=(\infty_s,jn_s)\), where \(1\leq s\leq m\).
	Recall that
$
	g_{\phi_0}^{\sharp}
	=
	E\circ\eta_{\phi_0}^{\sharp}.
$
	Thus the \(g_{\phi_0}\)-dual of the covector
	represented by \((\gamma_I,Q_I)\) is
	\(E\circ\xi_I\).
	
By \eqref{intdulconnformula}, one has
	\[
	\begin{aligned}
		\bigl(\widehat\nabla_X(E\circ\xi_I)\bigr)
		(\lambda)\phi_0
		&=
		\{C_{\gamma_I}(\lambda Q_I),\lambda\}_X\\
		&=
		j\{\Phi_I,\lambda\}_X\\
		&=
		j(X\circ\xi_I)(\lambda)\phi_0.
	\end{aligned}
	\]
	Taking the constant linear combination defining
	\(W_j\), we conclude that
	\[
	\widehat\nabla_X(E\circ W_j)
	=
	jX\circ W_j.
	\]
\end{proof}

Motivated by the preceding lemma, define
\[
\widehat\xi_{\mathrm{un},p}
=
\frac{1}{j^p}E
-
\frac{p}{j^{p+1}}E\circ W_j,
\qquad p\in\mathbb Z.
\]
In particular,
\(\widehat\xi_{\mathrm{un},0}=E\).
By Lemma~\ref{scalarrec}, we have
\begin{equation}\label{E-recursion-noncon}
	\begin{aligned}
		\widehat\nabla_X\widehat\xi_{\mathrm{un},p+1}
		&=
		\frac{1}{j^{p+1}}(jX+X\circ W_j)
		-
		\frac{p+1}{j^{p+1}}X\circ W_j\\
		&=
		\frac{1}{j^p}X
		-
		\frac{p}{j^{p+1}}X\circ W_j\\
		&=
		X*\widehat\xi_{\mathrm{un},p}.
	\end{aligned}
\end{equation}
Thus
\(\{\widehat\xi_{\mathrm{un},p}\}_{p\in\mathbb Z}\)
forms a bi-infinite recursion family containing
the unity vector field \(E\) of the almost-dual multiplication.
The corresponding Abelian integrals are defined by
\[
\widehat\Phi_{\mathrm{un},p}
=
\frac{1}{j^p}\Phi_0
-
\frac{p}{j^{p+1}}
\left(
\Phi_{W_j}
+
\frac{\lambda^j}{n_0j!}
\right),
\qquad p\in\mathbb Z.
\]
Indeed, since
$
\left\{
\frac{\lambda^j}{n_0j!},\lambda
\right\}_X=0
$
and
$
\{\Phi_{W_j},\lambda\}_X
=
(X\circ W_j)(\lambda)\phi_0,
$
we obtain
\begin{align}
	\{\widehat\Phi_{\mathrm{un},p},\lambda\}_X
	&=
	\frac{1}{j^p}X(\lambda)\phi_0
	-
	\frac{p}{j^{p+1}}
	(X\circ W_j)(\lambda)\phi_0
	\nn\\
	&=
	(X*\widehat\xi_{\mathrm{un},p})(\lambda)\phi_0.
	\label{Erecpoisson}
\end{align}

On the other hand, for
\(I=(\infty_s,jn_s)\), one has
$
\lambda Q_I=jQ_{I^+}
$ and $
\oint_{\gamma_I}Q_I\,d\lambda=0.
$
By the dual representation \eqref{dual-gsharp} and
\(g_{\phi_0}^{\sharp}=E\circ\eta_{\phi_0}^{\sharp}\),
we obtain
\[
\begin{aligned}
	(E\circ\xi_I)(\lambda)\phi_0
	&=
	g_{\gamma_I}^{\sharp}(Q_I)\\
	&=
	\lambda\,d\Phi_I-j\Phi_I\,d\lambda.
\end{aligned}
\]
Taking the linear combination defining \(W_j\) gives
\[
(E\circ W_j)(\lambda)\phi_0
=
\lambda\,d\Phi_{W_j}
-
j\Phi_{W_j}\,d\lambda.
\]
Combining this identity with \eqref{nonhomEu}
and the definition of \(\widehat\xi_{\mathrm{un},p}\),
we find
\[
\begin{aligned}
	\widehat\xi_{\mathrm{un},p}(\lambda)\phi_0
	&=
	\frac{1}{j^p}
	\left[
	\lambda\phi_0
	-
	\left(
	j\Phi_0+\Phi_{W_j}
	+\frac{\lambda^j}{n_0j!}
	\right)d\lambda
	\right]-
	\frac{p}{j^{p+1}}
	\left(
	\lambda\,d\Phi_{W_j}
	-
	j\Phi_{W_j}\,d\lambda
	\right)\\
	&=\lambda\,d\widehat\Phi_{\mathrm{un},p}
	-
	\widehat\Phi_{\mathrm{un},p-1}\,d\lambda.
\end{aligned}
\]

\subsection{Flat vector fields for the intersection form}\label{sec-dulflatbasis}

We now show that the recursion families constructed above contain a complete set of flat vector fields for the intersection form.

Recall that Romano \cite{Romano2015} constructed an \(N\)-dimensional space \(H_0\) of degree-zero horizontal differentials, modulo covariantly constant differentials.
A basis of \(H_0\) consists of
\[
\begin{alignedat}{2}
	&\Phi_{q_a}^{(0)}, &&\qquad 2\le a\le d,\\
	&\Psi_{\infty_i}^{(0)}, &&\qquad 1\le i\le m,\\
	&\rho_i^{(0)},\ \omega_i, &&\qquad 1\le i\le g.
\end{alignedat}
\]
Here \(q_1,\ldots,q_d\) are the simple zeros of \(\lambda\). The first two families are normalized Abelian differentials of the third kind, with poles at \(q_1,q_a\) and \(\infty_0,\infty_i\), respectively. The differentials \(\rho_i^{(0)}\) have prescribed \(d\log\lambda\)-jumps along the \(\beta\)-cycles, while \(\omega_i\) form the normalized basis of holomorphic differentials.

We now identify a basis of \(H_0\) within our Cauchy construction. Let \(q_1,\ldots,q_d\) be the simple zeros of \(\lambda\), and let \(\infty_0,\ldots,\infty_m\) be its poles, of orders \(n_0,\ldots,n_m\). For each \(z\in\mathcal S\), let \(\delta_z\) be a positively oriented meridian around \(z\), based at \(P^\ast\) via a fixed connecting path. We also fix a canonical symplectic basis \(\alpha_1,\beta_1,\ldots,\alpha_g,\beta_g\) of \(H_1(C_u,\mathbb Z)\).

Choose zero-winding based contours \(\gamma_{q_a}\), \(\gamma_{\infty_i}\), \(\gamma_{\alpha_i}\), and \(\gamma_{\beta_i}\) in \(\mathcal G\), such that
\[
\begin{aligned}
	[\gamma_{q_a}]
	&=[\delta_{q_a}]-[\delta_{q_1}],
	&&2\le a\le d,\\
	[\gamma_{\infty_i}]
	&=[\delta_{\infty_i}]+n_i[\delta_{q_1}],
	&&1\le i\le m,
\end{aligned}
\]
in \(H_1(C_u\setminus\mathcal S,\mathbb Z)\), while
\[
[\gamma_{\alpha_i}]=[\alpha_i],
\qquad
[\gamma_{\beta_i}]=[\beta_i],
\qquad 1\le i\le g,
\]
in \(H_1(C_u,\mathbb Z)\). Here \([\delta_z]\) denotes the homology class of a positively oriented small loop around \(z\in\mathcal S\).

Consider the differentials
\[
\begin{alignedat}{2}
	&\widehat\phi_{\gamma_{q_a},0},
	&&\qquad 2\le a\le d,\\
	&\widehat\phi_{\gamma_{\infty_i},0},
	&&\qquad 1\le i\le m,\\
	&\widehat\phi_{\gamma_{\alpha_i},0},
	\quad
	\widehat\phi_{\gamma_{\beta_i},-1},
	&&\qquad 1\le i\le g.
\end{alignedat}
\]
For the first three families, we have
\[
\widehat\phi_{\gamma,0}
=
dC_\gamma(\log\lambda),
\]
while for the last family,
\[
\widehat\phi_{\gamma_{\beta_i},-1}
=
dC_{\gamma_{\beta_i}}(1)
=
\frac{1}{2\pi\mathrm i}dZ_i.
\]
Since \(dC_\gamma(1)=0\) for the first three families and \(\widehat\Phi_{\gamma,-2}=0\) for the last, the same argument as in Lemma~\ref{lem:dual-xi-Phi} shows that the corresponding initial recursion vector fields are \(\widehat\nabla\)-flat.

These differentials form a basis of Romano's space \(H_0\). Indeed, the first two families span the \(d+m-1\) independent third-kind directions modulo \(d\log\lambda\). The third family has independent \(d\log\lambda\)-jumps along the \(\beta\)-cycles, while the last family spans the \(g\)-dimensional space of holomorphic differentials. Thus they form a basis of \(H_0\), up to a constant linear transformation of Romano's basis. Consequently, the corresponding recursion vector fields form a complete local flat frame for the intersection form.

\subsection{Principal hierarchy}\label{dualprinH}

We now derive the principal hierarchy of the almost-dual structure \((\widehat H,g_{\phi_0},*,E)\).

Recall that, in the conformal case with \(\nu_0\neq0\) and in the non-conformal case, we have constructed a bi-infinite recursion family \(\{\widehat\xi_{\mathrm{un},p}\}_{p\in\mathbb Z}\) containing the unity vector field \(E\), together with the corresponding Abelian integrals \(\widehat\Phi_{\mathrm{un},p}\). When \(\nu_0=0\), the unity \(E\) is \(\widehat\nabla\)-flat and is a constant linear combination of the initial fields of the existing recursion families. The corresponding unity recursion family and its corrected Abelian integrals are constructed in Subsection \ref{sec-Erec}.

By Lemma \ref{cor:dual-Q-vector-recursion}, equalities \eqref{E-recursion-con} and \eqref{E-recursion-noncon}, the vector fields
$
\{\widehat\xi_I\}_{I\in\widehat{\mathcal I}},
$ and $
\{\widehat\xi_{\mathrm{un},p}\}_{p\geq0}
$
satisfy
$
\widehat\nabla_X\widehat\xi_{I^+}
=
X*\widehat\xi_I
$ and $
\widehat\nabla_X\widehat\xi_{\mathrm{un},p+1}
=
X*\widehat\xi_{\mathrm{un},p}
$
for every \(X\in\operatorname{Vect}(\widehat H)\), \(I\in\widehat{\mathcal I}\), and \(p\geq0\).
Moreover,  the recursion families contain a complete local frame of \(\widehat\nabla\)-flat vector fields. Consequently, these recursion families provide an extended calibration of the almost-dual structure.

Set
$
\widehat{\mathcal I}^{\mathrm{ext}}
=
\widehat{\mathcal I}
\sqcup
\{(\mathrm{un},p)\mid p\geq0\},
$
and write \(\widehat\xi_A\) and \(\widehat\Phi_A\) for the corresponding recursion fields and Abelian integrals, where \(A\in\widehat{\mathcal I}^{\mathrm{ext}}\).
The associated principal hierarchy consists of the flows on the loop space \(L\widehat H\) given by
\[
\frac{\partial v}{\partial\widehat T_A}
=
\widehat\nabla_{v_x}\widehat\xi_{A^+}
=
v_x*\widehat\xi_A,
\qquad
A\in\widehat{\mathcal I}^{\mathrm{ext}}.
\]
Here \(v=v(x)\) is a loop in \(\widehat H\), and \(v_x\) denotes its tangent vector along the loop.

By Lemma~\ref{lem:dual-xi-Phi} and \eqref{Erecpoisson}, these flows admit the dispersionless Lax representation
\[
\frac{\partial\lambda}{\partial\widehat T_A}\phi_0
=
\{\widehat\Phi_A,\lambda\},
\qquad
A\in\widehat{\mathcal I}^{\mathrm{ext}},
\]
where
$
\{f,g\}:=df\,\partial_xg-dg\,\partial_xf,
$
as in Subsection~\ref{prinH}.
In particular, since \(\widehat\xi_{\mathrm{un},0}=E\) is the unity of the almost-dual multiplication, the corresponding flow is the spatial translation
$
\frac{\partial}{\partial\widehat T_{\mathrm{un},0}}
=
\frac{\partial}{\partial x}.
$

\begin{rem}
	Romano \cite{Romano2015} constructed a Whitham-type hierarchy on double Hurwitz spaces using homogeneous horizontal differentials. In particular, for a horizontal differential \(\Omega\) satisfying
$
	D_E\Omega=k\Omega$ with $
k\in\mathbb Z\setminus\{0\},
$
	the associated deformed flat function
$
	h=\langle\phi_0,\Omega\rangle
$
	satisfies
	\[
	\widehat\nabla_X
	\bigl(g_{\phi_0}^{\sharp}(dh)\bigr)
	=
	kX*g_{\phi_0}^{\sharp}(dh).
	\]
	
This  deformed-flatness relation for the second-kind differentials in this construction can be recovered directly from our Cauchy representation.	
	Indeed, let \(I\) be an index for which
$
	D_E\phi_I=k\phi_I
$ with $k\ne 0$.
	Since \(Q_I\) depends only on \(\lambda\), we have
$
	\{Q_I,\lambda\}_X=0.
$
	Therefore, by \eqref{intdulconnformula},
	\[
	\bigl(\widehat\nabla_X(E\circ\xi_I)\bigr)
	(\lambda)\phi_0
	=
	\{C_{\gamma_I}(\lambda Q_I),\lambda\}_X.
	\]
For the corresponding homogeneous indices, the explicit contour weights and their Cauchy representations give
\[
C_{\gamma_I}(\lambda Q_I)=k\Phi_I
\]
on the chosen branches. This follows from \(\lambda Q_I=kQ_{I^+}\) whenever the successive indices have the same contour; the exceptional transition at \(I=(q,1)\) follows from the corresponding puncture Cauchy representations.

	Consequently,
\[
\begin{aligned}
	\bigl(\widehat\nabla_X(E\circ\xi_I)\bigr)
	(\lambda)\phi_0
	&=
	k\{\Phi_I,\lambda\}_X\\
	&=
	k(X\circ\xi_I)(\lambda)\phi_0\\
	&=
	k\bigl(X*(E\circ\xi_I)\bigr)(\lambda)\phi_0.
\end{aligned}
\]
Hence
\[
\widehat\nabla_X(E\circ\xi_I)
=
kX*(E\circ\xi_I).
\]
\end{rem}

\subsection{Open extension}\label{dual-open-extension}

We now construct the rank-one extension of the almost-dual structure
\((\widehat H,g_{\phi_0},*,E)\).
As in the construction for \(H\), we use the fibre coordinate determined by \(\phi_0\), with \(\lambda\) replaced by \(\log\lambda\).

Let \(\widehat t=(\widehat t^1,\ldots,\widehat t^N)\) be local flat coordinates for \(g_{\phi_0}\), and write
$
\partial_\alpha=\frac{\partial}{\partial\widehat t^\alpha}.
$
Fix a local inverse \(z=z(\widehat t,s)\) of
$
s=\Phi_0
$
on a domain where \(\phi_0\ne0\), \(\lambda\ne0\), and \(d\lambda\ne0\), together with a single-valued branch of \(\log\lambda\). Throughout this subsection, \(\lambda\), \(\log\lambda\), and the associated Abelian integrals are regarded as functions of \((\widehat t,s)\) through \(z=z(\widehat t,s)\).

For \(\widehat\nabla\)-flat vector fields \(\partial_\alpha,\partial_\beta\), we have
$
\widehat\nabla_{\partial_\alpha}\partial_\beta=0.
$
Hence the Levi--Civita formula \eqref{intconn2} yields
\[
\partial_\alpha\partial_\beta(\log\lambda)\phi_0
=
d\left(
\frac{
	\partial_\alpha(\log\lambda)
	\partial_\beta(\log\lambda)
	-
	(\partial_\alpha*\partial_\beta)(\log\lambda)
}{
	\partial_s\log\lambda
}
\right).
\]
Since \(d\) denotes differentiation along the spectral curve and \(d\Phi_0=\phi_0\), we have
$
df=(\partial_s f)\phi_0
$
for any locally defined function \(f=f(\widehat t,s)\).
Dividing the preceding identity by \(\phi_0\), we obtain
\[
\partial_\alpha\partial_\beta\log\lambda
=
\partial_s\left(
\frac{
	\partial_\alpha(\log\lambda)
	\partial_\beta(\log\lambda)
	-
	(\partial_\alpha*\partial_\beta)(\log\lambda)
}{
	\partial_s\log\lambda
}
\right).
\]
This is precisely the criterion \eqref{omegaequ} for the existence of a local solution of the open WDVV equations, with \(\log\lambda\) playing the role of the function in that criterion.

Consequently, there exists locally an open WDVV potential \(\widehat F^o\) satisfying
$
\partial_s\widehat F^o=\log\lambda
$
and
\[
\partial_\alpha\partial_\beta\widehat F^o
=
\frac{
	\partial_\alpha(\log\lambda)
	\partial_\beta(\log\lambda)
	-
	\widehat c_{\alpha\beta}^{\gamma}
	\partial_\gamma(\log\lambda)
}{
	\partial_s\log\lambda
}
\]
where \(\widehat c_{\alpha\beta}^{\gamma}\) are the structure constants of the almost-dual multiplication \(*\).
The corresponding rank-one extension carries the multiplication \(\widehat\star\), defined by
\[
\begin{aligned}
	X\,\widehat\star\,Y
	&=
	X*Y+
	\frac{
		X(\log\lambda)Y(\log\lambda)
		-
		(X*Y)(\log\lambda)
	}{
		\partial_s\log\lambda
	}\,\partial_s,\\
	X\,\widehat\star\,\partial_s
	&=
	X(\log\lambda)\,\partial_s,\\
	\partial_s\,\widehat\star\,\partial_s
	&=
	(\partial_s\log\lambda)\,\partial_s,
\end{aligned}
\]
for \(X,Y\in\operatorname{Vect}(\widehat H)\), viewed as vector fields on \(\widehat H\times U\) independent of \(s\).

We next construct the associated principal hierarchy. Let \(\widehat{\mathcal I}^{\mathrm{ext}}\) denote the index set of the calibration constructed in Subsection~\ref{dualprinH}, including the unity recursion family when present. Set
$
\widehat{\mathcal I}^{o}
=
\widehat{\mathcal I}^{\mathrm{ext}}
\sqcup
\{(s,p)\mid p\ge0\}.
$

For \(A\in\widehat{\mathcal I}^{\mathrm{ext}}\), define
$
\widehat\Theta_A
=
\widehat\xi_A
+
\widehat\Phi_{A^-}(s)\,\partial_s,
$
while for \(p\ge0\), set
$
\widehat\Theta_{(s,p)}
=
\frac{(\log\lambda)^p}{p!}\,\partial_s.
$
Here \(\widehat\Phi_{A^-}(s)\) denotes the Abelian integral evaluated at \(z=z(\widehat t,s)\), with the chosen local determination. We also define
$
(s,p)^+=(s,p+1).
$

Let \(\widehat\nabla^{\mathrm{ext}}\) denote the product flat connection on \(\widehat H\times U\), obtained by extending \(\widehat\nabla\) trivially in the \(s\)-direction. Thus, for vector fields \(X,Y\) on \(\widehat H\), regarded as independent of \(s\),
\[
\begin{aligned}
	\widehat\nabla^{\mathrm{ext}}_X Y
	&=\widehat\nabla_XY,\\
	\widehat\nabla^{\mathrm{ext}}_X\partial_s
	&=
	\widehat\nabla^{\mathrm{ext}}_{\partial_s}X
	=
	\widehat\nabla^{\mathrm{ext}}_{\partial_s}\partial_s
	=0.
\end{aligned}
\]
We claim that
\begin{equation}\label{eq:dual-open-calibration}
	\widehat\nabla^{\mathrm{ext}}_{\widetilde X}
	\widehat\Theta_{A^+}
	=
	\widetilde X\,\widehat\star\,\widehat\Theta_A,
	\qquad
	A\in\widehat{\mathcal I}^{o},
\end{equation}
for every vector field \(\widetilde X\) on \(\widehat H\times U\).

We first consider \(A\in\widehat{\mathcal I}^{\mathrm{ext}}\). Write
$
\widetilde X=X+a\partial_s,
$
where \(X\) is tangent to \(\widehat H\).
By the representation of the recursion vector fields,
\[
\widehat\xi_A(\lambda)\phi_0
=
\lambda\,d\widehat\Phi_A
-
\widehat\Phi_{A^-}\,d\lambda,
\]
we obtain
\begin{equation}\label{eq:dual-open-first-identity}
	\widehat\xi_A(\log\lambda)
	=
	\partial_s\widehat\Phi_A(s)
	-
	\widehat\Phi_{A^-}(s)
	\partial_s\log\lambda.
\end{equation}
On the other hand, the Poisson representation of the recursion relations gives
\[
(X*\widehat\xi_A)(\lambda)\phi_0
=
\{\widehat\Phi_A,\lambda\}_X
=
d\widehat\Phi_A\,X(\lambda)
-
X(\widehat\Phi_A)\,d\lambda.
\]
Dividing by \(\lambda\phi_0\), we find
\begin{equation}\label{eq:dual-open-second-identity}
	(X*\widehat\xi_A)(\log\lambda)
	=
	X(\log\lambda)\,
	\partial_s\widehat\Phi_A(s)
	-
	(\partial_s\log\lambda)\,
	X\bigl(\widehat\Phi_A(s)\bigr).
\end{equation}
Since
$
\widehat\Theta_{A^+}
=
\widehat\xi_{A^+}
+
\widehat\Phi_A(s)\,\partial_s,
$
the recursion relation
$
\widehat\nabla_X\widehat\xi_{A^+}
=
X*\widehat\xi_A
$
implies
\[
\begin{aligned}
	\widehat\nabla^{\mathrm{ext}}_{\widetilde X}
	\widehat\Theta_{A^+}
	&=
	X*\widehat\xi_A+
	\left[
	X\bigl(\widehat\Phi_A(s)\bigr)
	+
	a\,\partial_s\widehat\Phi_A(s)
	\right]\partial_s.
\end{aligned}
\]

We now compute the right-hand side of \eqref{eq:dual-open-calibration}. By the definition of \(\widehat\star\),
\[
\begin{aligned}
	\widetilde X\,\widehat\star\,\widehat\Theta_A
=	&
	(X+a\partial_s)
	\,\widehat\star\,
	\left(
	\widehat\xi_A+
	\widehat\Phi_{A^-}(s)\partial_s
	\right)\\
	=&
	X*\widehat\xi_A
	+
	\Biggl[
	\frac{
		X(\log\lambda)\widehat\xi_A(\log\lambda)
		-
		(X*\widehat\xi_A)(\log\lambda)
	}{
		\partial_s\log\lambda
	}
	+
	\widehat\Phi_{A^-}(s)X(\log\lambda)\\
	&+
	a\left(
	\widehat\xi_A(\log\lambda)
	+
	\widehat\Phi_{A^-}(s)\partial_s\log\lambda
	\right)
	\Biggr]\partial_s.
\end{aligned}
\]
Using \eqref{eq:dual-open-first-identity} and \eqref{eq:dual-open-second-identity}, we have
\[
\begin{aligned}
	&\frac{
		X(\log\lambda)\widehat\xi_A(\log\lambda)
		-
		(X*\widehat\xi_A)(\log\lambda)
	}{
		\partial_s\log\lambda
	}+
	\widehat\Phi_{A^-}(s)X(\log\lambda)
	=
	X\bigl(\widehat\Phi_A(s)\bigr),
\end{aligned}
\]
and
\[
\widehat\xi_A(\log\lambda)
+
\widehat\Phi_{A^-}(s)\partial_s\log\lambda
=
\partial_s\widehat\Phi_A(s).
\]
Consequently,
\[
\begin{aligned}
	\widetilde X\,\widehat\star\,\widehat\Theta_A
	&=
	X*\widehat\xi_A+
	\left[
	X\bigl(\widehat\Phi_A(s)\bigr)
	+
	a\,\partial_s\widehat\Phi_A(s)
	\right]\partial_s\\
	&=
	\widehat\nabla^{\mathrm{ext}}_{\widetilde X}
	\widehat\Theta_{A^+}.
\end{aligned}
\]
This proves \eqref{eq:dual-open-calibration} for \(A\in\widehat{\mathcal I}^{\mathrm{ext}}\).

Next, consider \(A=(s,p)\), where \(p\ge0\). By definition,
$
\widehat\Theta_{(s,p+1)}
=
\frac{(\log\lambda)^{p+1}}{(p+1)!}\partial_s.
$
Hence
\[
\begin{aligned}
	\widehat\nabla^{\mathrm{ext}}_{\widetilde X}
	\widehat\Theta_{(s,p+1)}
	&=
	\widetilde X\left(
	\frac{(\log\lambda)^{p+1}}{(p+1)!}
	\right)\partial_s\\
	&=
	\frac{(\log\lambda)^p}{p!}
	\left(
	X(\log\lambda)
	+
	a\,\partial_s\log\lambda
	\right)\partial_s.
\end{aligned}
\]
On the other hand, the multiplication formulas give
\[
\begin{aligned}
	\widetilde X\,\widehat\star\,
	\widehat\Theta_{(s,p)}
	&=
	(X+a\partial_s)
	\,\widehat\star\,
	\left(
	\frac{(\log\lambda)^p}{p!}\partial_s
	\right)\\
	&=
	\frac{(\log\lambda)^p}{p!}
	\left(
	X(\log\lambda)
	+
	a\,\partial_s\log\lambda
	\right)\partial_s.
\end{aligned}
\]
Therefore \eqref{eq:dual-open-calibration} also holds for the additional \(s\)-family.

For the initial contour indices belonging to the flat basis constructed in Subsection \ref{sec-dulflatbasis}, the fields \(\widehat\xi_A\) are \(\widehat\nabla\)-flat and the functions \(\widehat\Phi_{A^-}\) are constant, independent of both the spectral variable and the parameters on \(\widehat H\). Hence the corresponding fields \(\widehat\Theta_A\) are \(\widehat\nabla^{\mathrm{ext}}\)-flat. Together with
$
\widehat\Theta_{(s,0)}=\partial_s,
$
they form a local flat frame of the rank-one extension. Thus the fields constructed above define an extended calibration for \((\widehat H\times U,\widehat\nabla^{\mathrm{ext}},\widehat\star)\).

We finally derive the corresponding principal hierarchy. Let \((v(x),s(x))\) be a loop in \(\widehat H\times U\), and regard
$
\partial_x=v_x+s_x\partial_s
$
as its tangent vector. By the definition of the principal hierarchy and \eqref{eq:dual-open-calibration}, the flow associated with \(A\in\widehat{\mathcal I}^{o}\) is
\[
\frac{\partial(v,s)}{\partial\widehat T_A}
=
\widehat\nabla^{\mathrm{ext}}_{\partial_x}
\widehat\Theta_{A^+}
=
\partial_x\,\widehat\star\,\widehat\Theta_A.
\]

For \(A\in\widehat{\mathcal I}^{\mathrm{ext}}\), we have
$
\widehat\Theta_{A^+}
=
\widehat\xi_{A^+}
+
\widehat\Phi_A(s)\partial_s.
$
Therefore,
\[
\widehat\nabla^{\mathrm{ext}}_{\partial_x}
\widehat\Theta_{A^+}
=
\widehat\nabla_{v_x}\widehat\xi_{A^+}
+
\partial_x\widehat\Phi_A(s)\,\partial_s.
\]
The recursion relation gives
$
\widehat\nabla_{v_x}\widehat\xi_{A^+}
=
v_x*\widehat\xi_A.
$
Moreover, by the Poisson representation,
\[
(v_x*\widehat\xi_A)(\lambda)\phi_0
=
\{\widehat\Phi_A,\lambda\}.
\]
Thus the corresponding flows are
\begin{equation}\label{eq:dual-open-hierarchy}
	\begin{aligned}
		\frac{\partial\lambda}{\partial\widehat T_A}\phi_0
		=
		\{\widehat\Phi_A,\lambda\},\quad
		\frac{\partial s}{\partial\widehat T_A}
		=
		\partial_x\widehat\Phi_A(s),
	\end{aligned}
	\qquad
	A\in\widehat{\mathcal I}^{\mathrm{ext}}.
\end{equation}

For \(A=(s,p)\), \(p\ge0\), we have
\[
\widehat\nabla^{\mathrm{ext}}_{\partial_x}
\widehat\Theta_{(s,p+1)}
=
\partial_x\left(
\frac{(\log\lambda)^{p+1}}{(p+1)!}
\right)\partial_s,
\]
and the additional flows are
\begin{equation}\label{eq:dual-open-s-flows}
	\frac{\partial\lambda}{\partial\widehat T_{s,p}}=0,
	\qquad
	\frac{\partial s}{\partial\widehat T_{s,p}}
	=
	\partial_x\left(
	\frac{(\log\lambda)^{p+1}}{(p+1)!}
	\right),
	\qquad p\ge0.
\end{equation}

Finally, the unity recursion family is included, with
$
\widehat\Phi_{\mathrm{un},0}=\Phi_0+c$ with $
 c\in\mathbb C$.
Since \(\Phi_0(s)=s\), we have
$
\widehat\Phi_{\mathrm{un},0}(s)=s+c.
$
Hence the corresponding flow in \eqref{eq:dual-open-hierarchy} is
\[
\frac{\partial v}{\partial\widehat T_{\mathrm{un},0}}
=v_x,
\qquad
\frac{\partial s}{\partial\widehat T_{\mathrm{un},0}}
=s_x.
\]
Therefore, the \(\widehat T_{\mathrm{un},0}\)-flow coincides with spatial translation on the extended loop space.

Equations \eqref{eq:dual-open-hierarchy} and \eqref{eq:dual-open-s-flows} give the principal hierarchy of the rank-one extension.

\section{Shramchenko deformations}\label{sec:shramchenko}

In this section we explain how the constructions of the preceding sections
change under Shramchenko's deformation of the canonical bidifferential
\cite{shramchenko2005deformations}.

\subsection{Deformed bidifferential and Cauchy decompositions}

Assume first that $g\ge1$. 
Let $\mathbb B$ be the period matrix defined by
$
\mathbb B_{ij}
=\frac{1}{2\pi\mathrm i}\oint_{\beta_i}dZ_j.
$
Fix a constant symmetric $g\times g$ matrix
$\mathbf q=(q_{ij})$ and work on the open locus where
$
	\det(\mathbb B+\mathbf q)\ne0.
$
Set $M_{\mathbf q}=(\mathbb B+\mathbf q)^{-1}$.
Following Shramchenko, define the deformed
bidifferential by
\begin{equation}\label{eq:shram-B}
	B_{\mathbf q}(z,w)
	=
	B(z,w)
	-\frac{1}{2\pi\mathrm i}
	d\mathbf Z(z)^T M_{\mathbf q}d\mathbf Z(w),
\end{equation}
where $d\mathbf Z=(dZ_1,\ldots,dZ_g)^T$.

The correction in \eqref{eq:shram-B} is symmetric and holomorphic on
$\mathcal C_u\times\mathcal C_u$. Consequently, $B_{\mathbf q}$ has the
same double pole, biresidue, and vanishing residue on the diagonal as
$B$. Its periods, however, are
\begin{equation}\label{eq:shram-periods}
	\begin{aligned}
		\oint_{\alpha_i}B_{\mathbf q}(\cdot,w)
		&=-\sum_j(M_{\mathbf q})_{ij}\,dZ_j(w),\\
		\oint_{\beta_i}B_{\mathbf q}(\cdot,w)
		&=\sum_j(\mathbf qM_{\mathbf q})_{ij}\,dZ_j(w).
	\end{aligned}
\end{equation}
Thus, in place of the original vanishing $\alpha$-periods, we have
\begin{equation}\label{eq:shram-q-normalization}
	\oint_{\beta_i}B_{\mathbf q}(\cdot,w)
	+\sum_jq_{ij}\oint_{\alpha_j}B_{\mathbf q}(\cdot,w)=0.
\end{equation}

Let $\mathbf Z(z)=\int_{\infty_0}^z d\mathbf Z,$
with the usual determinations on the canonical polygon.
The deformed Cauchy kernel is
\begin{equation}\label{eq:shram-K}
	\begin{aligned}
		K_{\mathbf q}(z,w)
		&:=\int_{\infty_0}^z B_{\mathbf q}(\xi,w)\\
		&=K(z,w)-\frac{1}{2\pi\mathrm i}
		\mathbf Z(z)^T M_{\mathbf q}d\mathbf Z(w).
	\end{aligned}
\end{equation}
It has the same poles and residues in $w$ as
$K(z,w)$ and satisfies
\[
K_{\mathbf q}(\infty_0,w)=0,
\qquad
d_zK_{\mathbf q}(z,w)=B_{\mathbf q}(z,w).
\]
Its monodromies in the first variable are
\begin{equation}\label{eq:shram-K-monodromy}
	\begin{aligned}
		\Delta_{\alpha_i}K_{\mathbf q}
		&=-\sum_{j=1}^g(M_{\mathbf q})_{ij}dZ_j,\quad 
		\Delta_{\beta_i}K_{\mathbf q}
		=\sum_{j=1}^g
		(\mathbf qM_{\mathbf q})_{ij}dZ_j.
	\end{aligned}
\end{equation}
Consequently,
$
\Delta_{\beta_i}K_{\mathbf q}
+\sum_{j=1}^g q_{ij}
\Delta_{\alpha_j}K_{\mathbf q}=0.
$

Define the two Cauchy operators as in
Subsection~\ref{subseccauthy}:
\begin{equation}\label{eq:shram-Cauchy}
	\begin{aligned}
		C_\gamma^{\mathbf q}f(z)
		&:=\frac1{2\pi\mathrm i}\oint_\gamma
		f(w)K_{\mathbf q}(z,w),\\
		\mathcal C_\gamma^{\mathbf q}\varpi(w)
		&:=-\frac1{2\pi\mathrm i}\oint_\gamma
		K_{\mathbf q}(z,w)\varpi(z).
	\end{aligned}
\end{equation}
Since the correction to $K$ is holomorphic
along the diagonal, the local Cauchy jump
relations remain unchanged. The difference
lies in the period normalization. For the scalar Cauchy transform, one has
\[
\Delta_{\beta_i}C_\gamma^{\mathbf q}f
+\sum_{j=1}^g q_{ij}
\Delta_{\alpha_j}C_\gamma^{\mathbf q}f=0.
\]
Similarly, the differential Cauchy transform
satisfies
\[
\oint_{\beta_i}
\mathcal C_\gamma^{\mathbf q}\varpi
+\sum_{j=1}^g q_{ij}
\oint_{\alpha_j}
\mathcal C_\gamma^{\mathbf q}\varpi=0.
\]
Thus the Cauchy jump relations and the
corresponding additive decompositions of
Section~2 extend to the deformed kernel,
with the original $\alpha$-period
normalization replaced by the above
linear period condition.

The integration-by-parts identity retains
its full form:
\begin{equation}\label{eq:shram-integration-parts}
	\mathcal C_\gamma^{\mathbf q}(df)
	=
	dC_\gamma^{\mathbf q}(f)
	-\frac{f(P_*)}{2\pi\mathrm i}
	\Delta_\gamma K_{\mathbf q}(P_*,\cdot).
\end{equation}
In particular, the correction term vanishes
if $f(P_*)=0$ or if $\gamma$ is
null-homologous on the compact surface.
For general cycles, however, this term
must be retained.

When $g=0$, no holomorphic correction
is present. We set
$B_{\mathbf q}=B$ and $K_{\mathbf q}=K$,
so that the construction reduces to
the undeformed case.

\subsection{Extended primary differentials, metric, and multiplication}

We now recall  the deformed extended primary differentials introduced in \cite{shramchenko2005deformations}. Their local singularities and monodromies are prescribed exactly as in Subsection~2.2. The deformation changes only the global normalization.

For each corresponding index \(I\), let \(\phi_I^{\mathbf q}\) be the deformed differential. We require its derivative
$
\left.\partial_{u_k}\right|_\lambda\phi_I^{\mathbf q}
$
to be a single-valued meromorphic differential whose poles are confined to the ramification points of \(\lambda\), satisfying
\begin{equation}\label{eq:shram-primary-normalization}
	\oint_{\beta_i}
	\left.\partial_{u_k}\right|_\lambda\phi_I^{\mathbf q}
	+
	\sum_{j=1}^g q_{ij}
	\oint_{\alpha_j}
	\left.\partial_{u_k}\right|_\lambda\phi_I^{\mathbf q}
	=0,
	\qquad i=1,\ldots,g.
\end{equation}
We introduce the additional indices \((\alpha_i,-1)\), \(i=1,\ldots,g\), and define
\[
\phi_{\alpha_i,-1}^{\mathbf q}
=
-\frac1{2\pi\mathrm i}
\sum_{j=1}^g(M_{\mathbf q})_{ij}dZ_j.
\]
These differentials form a basis of the space of holomorphic differentials and satisfy
\[
\oint_{\beta_j}\phi_{\alpha_i,-1}^{\mathbf q}
+
\sum_{l=1}^gq_{jl}
\oint_{\alpha_l}\phi_{\alpha_i,-1}^{\mathbf q}
=-\delta_{ij}.
\]
The degree-zero \(\beta\)-type differentials are defined by
\[
\phi_{\beta_i,0}^{\mathbf q}
=
-2\pi\mathrm i
\sum_{j=1}^gq_{ij}\phi_{\alpha_j,-1}^{\mathbf q}
=
\sum_{j=1}^g(\mathbf qM_{\mathbf q})_{ij}dZ_j.
\]
For the remaining extended primary differentials, the holomorphic ambiguity is fixed by imposing the corresponding Euler conditions involving \(D_E\), together with the additional deformed period normalizations in the degree-zero logarithmic cases, as in Subsection~\ref{secextpri}.

Following Shramchenko's construction, we define the deformed Dubrovin primary differentials to be
\[
\begin{aligned}
	&\phi_{\infty_0,r}^{\mathbf q},
	\qquad 1\le r\le n_0-1,\\[3pt]
	&\phi_{\infty_i,r}^{\mathbf q},
	\qquad 1\le i\le m,\quad 0\le r\le n_i,\\[3pt]
	&\phi_{\alpha_i,-1}^{\mathbf q},\qquad
	\phi_{\beta_i,1}^{\mathbf q}
	+2\pi\mathrm i\sum\nolimits_{j=1}^g
	q_{ij}\phi_{\alpha_j,0}^{\mathbf q},
	\qquad 1\le i\le g.
\end{aligned}
\]
We denote by \(\mathcal D_u^{\mathbf q}\) their complex linear span. 
Fix a deformed extended primary differential \(\phi_0^{\mathbf q}\) and its normalized Abelian integral \(\Phi_0^{\mathbf q}\). A tangent vector \(X\in T_uH\) is represented by the differential
$
	X(\lambda)\phi_0^{\mathbf q}\in\mathcal D_u^{\mathbf q},
$
where \(X(\lambda)\) denotes differentiation along the \(\Phi_0^{\mathbf q}\)-horizontal lift.
The residue pairing retains its original form
\[
\langle\varpi_1,\varpi_2\rangle
=
\sum_{d\lambda=0}
\operatorname{Res}
\frac{\varpi_1\varpi_2}{d\lambda}.
\]
The deformed metric is defined by
\begin{equation}
	\eta_{\phi_0}^{\mathbf q}(X,Y)
	=
	\sum_{d\lambda=0}
	\operatorname{Res}
	\frac{
		X(\lambda)Y(\lambda)(\phi_0^{\mathbf q})^2
	}{
		d\lambda
	}.
\end{equation}
The multiplication, unity, and Euler vector field retain their canonical expressions:
\begin{equation}\label{eq:shram-canonical}
	\partial_{u_i}\circ\partial_{u_j}
	=
	\delta_{ij}\partial_{u_i},
	\qquad
	e=\sum_i\partial_{u_i},
	\qquad
	E=\sum_i u_i\partial_{u_i}.
\end{equation}

The Cauchy representations of the metric, Levi--Civita connection, and multiplication  extend naturally to the deformed setting.
More precisely, the metric dual of a covector represented by \(f\in\mathcal H_\gamma\) is given by
\begin{equation}\label{eq:shram-eta}
(\eta_\gamma^{\mathbf q})^\sharp(f)
=
-C_\gamma^{\mathbf q}(f)\,d\lambda
+
\mathcal C_\gamma^{\mathbf q}(f\,d\lambda).
\end{equation}

The Levi--Civita connection \(\nabla^{\mathbf q}\) of \(\eta_{\phi_0}^{\mathbf q}\) is represented by
\[
\bigl(\nabla_X^{\mathbf q}Y\bigr)(\lambda)\phi_0^{\mathbf q}
=
\mathcal P_+^{\mathbf q}
\left(
\left.X\right|_\lambda
\bigl(Y(\lambda)\phi_0^{\mathbf q}\bigr)
\right).
\]
Similarly, the contour representation of the multiplication is obtained from
\begin{equation}\label{eq:shram-Mul}
	M_{X,\gamma}^{\mathbf q}(f)
	=
	-C_\gamma^{\mathbf q}\bigl(fX(\lambda)\bigr)\,d\lambda
	+
	X(\lambda)\mathcal C_\gamma^{\mathbf q}(f\,d\lambda).
\end{equation}

Here \(X(\lambda)\) and \(Y(\lambda)\) are defined using the \(\Phi_0^{\mathbf q}\)-horizontal lift.

\subsection{Recursion vector fields}

We now construct the recursion vector fields associated with the Shramchenko-deformed structure. As in Subsection~3.1, we represent covectors by contour data and take their metric duals using the deformed Cauchy transforms.

For the pole-type indices, both the contours and weights remain unchanged. For the logarithmic indices, we retain the original contours and weight formulas, replacing the normalized third-kind Abelian integrals by their deformed counterparts.

For the cycle-type indices, we retain the original contours and weights for \((\alpha_i,p)\), \(p\geq 0\), and \((\beta_i,p)\), \(p\geq 1\). We also introduce the additional indices \((\alpha_i,-1)\), and choose the same contour for \((\alpha_i,-1)\) and \((\beta_i,0)\), namely, the positively oriented contour lying inside and sufficiently close to the boundary of the canonical polygon. Their weights are defined by
$
Q_{\alpha_i,-1}^{\mathbf q}
=
\frac{\phi_{\alpha_i,-1}^{\mathbf q}}{d\lambda},
$ and $
Q_{\beta_i,0}^{\mathbf q}
=
\frac{\phi_{\beta_i,0}^{\mathbf q}}{d\lambda}.
$

The deformed Cauchy transforms associated with these contour data reproduce the deformed extended primary differentials and their normalized Abelian integrals. More precisely, on the corresponding local branches, we have
\begin{equation}\label{eq:shram-primary-cauchy}
	C_{\gamma_I}^{\mathbf q}(Q_I^{\mathbf q})
	=\Phi_{I^-}^{\mathbf q},
	\qquad
	\mathcal C_{\gamma_I}^{\mathbf q}
	(Q_I^{\mathbf q}\,d\lambda)
	=\phi_I^{\mathbf q}.
\end{equation}
These identities follow from the same Cauchy jump arguments as in Subsection~3.1, with the deformed period normalization replacing the original one.

Fix \(\phi_0^{\mathbf q}=\phi_{I_0}^{\mathbf q}\). For each index \(I\), let \(\xi_I^{\mathbf q}\) be the metric dual of the covector represented by \((\gamma_I,Q_I^{\mathbf q})\). By \eqref{eq:shram-primary-cauchy} and \eqref{eq:shram-eta} the deformed metric-dual formula, we obtain
\begin{equation}\label{eq:shram-xi}
	\xi_I^{\mathbf q}(\lambda)\phi_0^{\mathbf q}
	=
	\phi_I^{\mathbf q}
	-
	\Phi_{I^-}^{\mathbf q}\,d\lambda.
\end{equation}
Furthermore, applying the deformed Cauchy projection formula for the Levi--Civita connection, the same argument as in Lemma~3.3 yields
\begin{equation}\label{eq:shram-xi-Phi}
	(\nabla_{\partial}^{\mathbf q}\xi_{I^+}^{\mathbf q})
	(\lambda)\phi_0^{\mathbf q}
	=
	\{\Phi_I^{\mathbf q},\lambda\}_{\partial}.
\end{equation}
Combining \eqref{eq:shram-xi}, \eqref{eq:shram-xi-Phi} and \eqref{eq:shram-Mul}, one has
\begin{equation}\label{eq:shram-recursion}
	\nabla_{\partial}^{\mathbf q}\xi_{I^+}^{\mathbf q}
	=
	\partial\circ\xi_I^{\mathbf q}.
\end{equation}

The flat sections of \(\nabla^{\mathbf q}\) can also be identified from \eqref{eq:shram-xi-Phi}. In particular, the initial recursion fields are flat. For the cycle-type indices, the relation
$
\Phi_{\beta_i,0}^{\mathbf q}
=
-2\pi\mathrm i\sum_{j=1}^g
q_{ij}\Phi_{\alpha_j,-1}^{\mathbf q}
$
implies
\[
\nabla_{\partial}^{\mathbf q}
\left(
\xi_{\beta_i,1}^{\mathbf q}
+
2\pi\mathrm i\sum_{j=1}^g
q_{ij}\xi_{\alpha_j,0}^{\mathbf q}
\right)=0.
\]
Hence the flat vector fields are represented by the differentials
\[
\phi_{\beta_i,1}^{\mathbf q}
+
2\pi\mathrm i\sum_{j=1}^g
q_{ij}\phi_{\alpha_j,0}^{\mathbf q},
\qquad i=1,\ldots,g,
\]
together with \(\phi_{\alpha_i,-1}^{\mathbf q}\) and the initial pole-type and logarithmic differentials. These constitute precisely the Shramchenko-deformed Dubrovin primary differentials.

The recursion relation \eqref{eq:shram-recursion} immediately yields the corresponding principal hierarchy, which admits the same Lax representation as in the undeformed case:
\[
\frac{\partial\lambda}{\partial T_I^{\mathbf q}}
\phi_0^{\mathbf q}
=
\{\Phi_I^{\mathbf q},\lambda\}.
\]
Moreover, the construction of the open extension carries over by the same arguments, with the original primary differentials, Abelian integrals, and recursion fields replaced by their deformed counterparts. Consequently, the corresponding open WDVV solutions and extended principal hierarchy are obtained in the same way.

\subsection{Almost-dual structure}

We now consider the Shramchenko deformation of the almost-dual structure. The intersection form \(g_{\phi_0}^{\mathbf q}\), its Levi--Civita connection \(\widehat\nabla^{\mathbf q}\), and the multiplication \(*\) are defined as in the undeformed case, with the corresponding deformed primary differential \(\phi_0^{\mathbf q}\).

We retain the zero-winding contours \(\gamma\in\mathcal G\) and the choices of \(P^\ast\) and \(\log\lambda\) from Subsection~\ref{almprisec}. The essential modification concerns the initial indices. Indeed, if
$
[\gamma]
=
\sum_{i=1}^g
\bigl(a_i[\alpha_i]+b_i[\beta_i]\bigr)$ 
 in $H_1(C_u,\mathbb Z),
$
then the period formulas for \(B_{\mathbf q}\) give
\[
dC_\gamma^{\mathbf q}(1)
=
\frac{1}{2\pi\mathrm i}
\sum_{i,j=1}^g
\bigl(
b_i(\mathbf qM_{\mathbf q})_{ij}
-a_i(M_{\mathbf q})_{ij}
\bigr)dZ_j.
\]
Thus, unlike the undeformed case, \(dC_\gamma^{\mathbf q}(1)\) may be nonzero even when \([\gamma]\) has no \(\beta\)-component.

Set
$
\mathcal G_{-1}^{\mathbf q}
=
\{\gamma\in\mathcal G\mid
dC_\gamma^{\mathbf q}(1)\neq0\},
$
and introduce
\[
\widehat{\mathcal I}^{\mathbf q}
=
\{(\gamma,p)\mid\gamma\in\mathcal G,\ p\ge0\}
\sqcup
\{(\gamma,-1)\mid\gamma\in\mathcal G_{-1}^{\mathbf q}\}.
\]
For \(I=(\gamma,p)\), we define the contour representatives and weights by
\[
\gamma_I=
\begin{cases}
	\gamma,&p\ge0,\\
	\partial\widehat C_u^\varepsilon,&p=-1,
\end{cases}
\] with \[
\widehat Q_I^{\mathbf q}=
\begin{cases}
	\displaystyle\frac{(\log\lambda)^p}{p!\lambda},
	&p\ge0,\\[3mm]
	\displaystyle\frac{dC_\gamma^{\mathbf q}(1)}{d\lambda},
	&p=-1.
\end{cases}
\]
Here \(\partial\widehat C_u^\varepsilon\) is the same polygon-boundary contour used in the undeformed construction.

For each \(I=(\gamma,p)\in\widehat{\mathcal I}^{\mathbf q}\), define \(I^\pm=(\gamma,p\pm1)\). We introduce the associated Abelian integrals and differentials by
\[
\widehat\Phi_{I^-}^{\mathbf q}
:=
C_{\gamma_I}^{\mathbf q}
(\lambda\widehat Q_I^{\mathbf q}),
\qquad
\widehat\phi_I^{\mathbf q}
:=
d\widehat\Phi_I^{\mathbf q}.
\]
Then
\[
\begin{aligned}
	C_{\gamma_I}^{\mathbf q}
	(\lambda\widehat Q_I^{\mathbf q})
=\widehat\Phi_{I^-}^{\mathbf q},\quad 
	\mathcal C_{\gamma_I}^{\mathbf q}
	(\widehat Q_I^{\mathbf q}\,d\lambda)
	=\widehat\phi_I^{\mathbf q}.
\end{aligned}
\]

Let \(\widehat\xi_I^{\mathbf q}\) be the \(g_{\phi_0}^{\mathbf q}\)-dual of the covector represented by \((\gamma_I,\widehat Q_I^{\mathbf q})\). The deformed intersection-metric dual formula then yields
\begin{equation}\label{eq:shram-dual-xi}
	\widehat\xi_I^{\mathbf q}(\lambda)\phi_0^{\mathbf q}
	=
	\lambda\widehat\phi_I^{\mathbf q}
	-
	\widehat\Phi_{I^-}^{\mathbf q}\,d\lambda.
\end{equation}
Applying the deformed Levi--Civita connection formula, the same arguments as in Lemmas~\ref{lem:dual-xi-Phi} and \ref{cor:dual-Q-vector-recursion} give
\begin{equation}\label{eq:shram-dual-recursion}
	\begin{aligned}
		(\widehat\nabla_{\partial}^{\mathbf q}
		\widehat\xi_{I^+}^{\mathbf q})(\lambda)\phi_0^{\mathbf q}
		&=
		\{\widehat\Phi_I^{\mathbf q},\lambda\}_{\partial},\\
		\widehat\nabla_{\partial}^{\mathbf q}
		\widehat\xi_{I^+}^{\mathbf q}
		&=
		\partial*\widehat\xi_I^{\mathbf q}.
	\end{aligned}
\end{equation}

We next identify a complete set of flat vector fields. For the cycle-type contours, the  connection computation yields
\[
\widehat\nabla_{\partial}^{\mathbf q}
\left(
\widehat\xi_{\gamma_{\beta_i},0}^{\mathbf q}
+
\sum_{j=1}^gq_{ij}
\widehat\xi_{\gamma_{\alpha_j},0}^{\mathbf q}
\right)=0.
\]
Together with the initial puncture-type fields and the holomorphic initial fields \(\widehat\xi_{\gamma_{\alpha_i},-1}^{\mathbf q}\), these vector fields form a complete local flat frame. Indeed, the corresponding differentials have independent third-kind singularities, cycle-type jumps, and holomorphic components, and their number equals \(\dim\widehat H\). They thus provide the deformed counterparts of Romano's degree-zero horizontal differentials.

The additional recursion family associated with the unity \(E\) is obtained as in the undeformed case, with the reference differential and Abelian integrals replaced by their deformed counterparts. Together with \eqref{eq:shram-dual-recursion}, this gives an extended calibration of the deformed almost-dual structure.

Consequently, the corresponding principal hierarchy admits the Lax representation
\[
\frac{\partial\lambda}{\partial\widehat T_I^{\mathbf q}}
\phi_0^{\mathbf q}
=
\{\widehat\Phi_I^{\mathbf q},\lambda\}.
\]
The construction of the rank-one open extension also carries over. In particular, the associated open WDVV solution satisfies
\[
\frac{\partial\widehat F^{o,\mathbf q}}{\partial s}
=
\log\lambda(s),
\]
and the extended principal hierarchy has the additional equation
\[
\frac{\partial s}{\partial\widehat T_I^{\mathbf q}}
=
\partial_x\widehat\Phi_I^{\mathbf q}(s).
\]

\bibliographystyle{unsrt}

\end{document}